\documentclass[a4 paper,12pt]{article}
\usepackage{graphicx}
\usepackage{amsmath}
\usepackage{authblk}
\usepackage[utf8]{inputenc}
\usepackage[english]{babel}
\usepackage{csquotes}
\usepackage{amssymb}
\usepackage{caption}
\usepackage[margin=2.5cm]{geometry}
\usepackage{setspace}
\usepackage{float}

\usepackage{algpseudocode}
\usepackage{algorithm}
\usepackage{algorithmicx} 
\usepackage{amsthm}
\usepackage{mathtools}
\usepackage[toc,page]{appendix}
\usepackage{tabulary}
\usepackage{booktabs}
\usepackage{subcaption}
\usepackage[authoryear, round]{natbib}
\usepackage{mathtools}

\usepackage[colorlinks = true, pdfstartview = FitV, linkcolor = blue, citecolor = blue, urlcolor = blue, bookmarks = false]{hyperref}
\usepackage{pdflscape}
\usepackage{afterpage}
\usepackage{capt-of}
\usepackage{bbm}
\usepackage{colortbl}
\usepackage{url}
\usepackage{bbold}
\usepackage{chngcntr}
\counterwithin{equation}{section}

\usepackage{authblk}
\usepackage[export]{adjustbox}
\usepackage{cleveref}
\usepackage[most]{tcolorbox}
\usepackage{enumitem}

\title{A Comprehensive Guide to Simulation-based Inference in Computational Biology}
\author{Xiaoyu Wang\textsuperscript{1,2*}, Ryan P. Kelly\textsuperscript{1,2}, Adrianne L. Jenner\textsuperscript{1,2}, David J. Warne\textsuperscript{1,2}, Christopher Drovandi\textsuperscript{1,2}
\\
\textbf{1} School of Mathematical Sciences, Queensland University of Technology, Brisbane, QLD, Australia
\\
\textbf{2} Centre for Data Science, Queensland University of Technology, Brisbane, QLD, Australia}

\begin{document}
\maketitle

\begin{abstract}
    Computational models are invaluable in capturing the complexities of real-world biological processes. Yet, the selection of appropriate algorithms for inference tasks, especially when dealing with real-world observational data, remains a challenging and underexplored area. This gap has spurred the development of various parameter estimation algorithms, particularly within the realm of Simulation-Based Inference (SBI), such as neural and statistical SBI methods. Limited research exists on how to make informed choices on SBI methods when faced with real-world data, which often results in some form of model misspecification. In this paper, we provide comprehensive guidelines for deciding between SBI approaches for complex biological models. We apply the guidelines to two agent-based models that describe cellular dynamics using real-world data. Our study unveils a critical insight: while neural SBI methods demand significantly fewer simulations for inference results, they tend to yield biased estimations, a trend persistent even with robust variants of these algorithms. On the other hand, the accuracy of statistical SBI methods enhances substantially as the number of simulations increases. This finding suggests that, given a sufficient computational budget, statistical SBI can surpass neural SBI in performance. Our results not only shed light on the efficacy of different SBI methodologies in real-world scenarios but also suggest potential avenues for enhancing neural SBI approaches. This study is poised to be a useful resource for computational biologists navigating the intricate landscape of SBI in biological modeling.
\end{abstract}

\section{Introduction}

Computational models have proven to be a powerful tool for gaining a deep understanding of biological processes by modeling their mechanisms \citep{walpole2013multiscale,brodland2015computational,metzcar2019review}. Continuum models, such as those governed by ordinary differential equations (ODEs) \citep{spencer2004ordinary,vanlier2013parameter,browning2022efficient} or partial differential equations (PDEs) \citep{leung2013systems,klowss2022stochastic,naevdal2023can}, are capable of describing dynamics at the population level. On the other hand, discrete models, like agent-based models \citep{an2009agent,hinkelmann2011mathematical,metzcar2019review}, focus on modeling individual-level dynamics and hence provide more detailed simulations but often require more computational effort than continuum models \citep{warne2022rapid}. 

An ongoing challenge in establishing stochastic models with good predictive performance is the accurate estimation of model parameters and quantification of associated uncertainty \citep{sun2011parameter,warne2019simulation,jorgensen2022efficient}. Bayesian inference is one of the most popular approaches for obtaining estimates for model parameters using data \citep{box2011bayesian}. In Bayesian inference, we update our initial beliefs with the likelihood of the model to obtain a posterior distribution. This means we adjust our initial understanding based on new data, leading to more informed estimates. When using Bayesian inference for parameter estimation, it is assumed that the models can replicate observed data. However, in real-world problems, this assumption does not always hold.

The inability of models to fully replicate observed data can stem from various factors. First, the model may inherently lack the capability to replicate the observed data \citep{box1979all,fisher2019all}. For example, if the model is designed to describe only linear growth, but the data exhibits exponential growth, then we would expect a mismatch between the model and the data. The second factor involves the possibility that the chosen inference algorithm might provide a biased estimation, preventing the model from accurately replicating the observed data \citep{black2011missing,wang2019variational,frazier2020model}.

The canonical approach for uncertainty quantification assumes that the likelihood function can be evaluated, and employs Markov Chain Monte Carlo (MCMC) for sampling \citep{sorensen2002likelihood,ghasemi2011bayesian,zucknick2014mcmc,valderrama2019mcmc}. It performs well for some continuum models, but this method has limitations. In many biological processes, accurately identifying the exact noise distribution is challenging, and often introduces errors \citep{lillacci2010parameter,tsimring2014noise}. Furthermore, for discrete models, the likelihood is often intractable, and the general practice is to approximate it with a continuum model \citep{browning2019bayesian,warne2019using,warne2022rapid}. Another popular approach when the likelihood is intractable is to employ a likelihood-free inference (LFI) \citep{sisson2018handbook} method, also known as simulation-based inference (SBI).  

One widely used SBI method is Approximate Bayesian Computation (ABC) \citep{csillery2010approximate,sunnaaker2013approximate,beaumont2019approximate}, which compares simulated data with observational data and accepts the simulation if the simulation data closely matches the observational data based on a certain distance metric. This method and its variants have been used in many biological applications such as \citet{beaumont2002approximate,toni2009approximate,liepe2014framework}. Another popular approach is Bayesian Synthetic Likelihood (BSL) \citep{wood2010statistical,price2018bayesian,frazier2023bayesian}, which approximates the likelihood by assuming its summary statistics likelihood density can be well approximated by a multivariate Gaussian distribution. Even through BSL makes strong assumption about the likelihood density, it has been used in many biological applications \citep{picchini2019bayesian,prescott2021quantifying,biswas2022hp1,morina2023estimated}. These LFI methods are theoretically grounded \citep{sunnaaker2013approximate,beaumont2009adaptive,beaumont2019approximate,frazier2024synthetic} and are also known as statistical SBI methods.

The statistical SBI methods necessitate a substantial amount of model simulations to provide a reasonable estimate, leading to computational inefficiency \citep{cranmer2020frontier}. As computational models become more complex, our computational budgets become more constrained, meaning we cannot afford a large number of simulations. To enhance efficiency, machine learning SBI methods have been introduced \citep{papamakarios2016fast,greenberg2019automatic,papamakarios2019sequential,wang2022adversarial}. These methods pair parameter values with their corresponding simulations to learn the mapping from simpler distributions, such as the standard Gaussian distribution, to the target distributions, i.e., the posterior distribution or likelihood. Neural network-based generative models, such as normalizing flows \citep{papamakarios2021normalizing}, are utilized to learn this mapping, and, as such, are often referred to as neural SBI. Compared to statistical SBI, machine learning SBI can require orders of magnitude fewer simulations, but the accuracy of these estimates is not guaranteed, particularly when fitting with real-world observational data.

It remains unknown how to select a suitable SBI algorithm for estimating model parameters in real-world settings. In the machine learning community, \citet{lueckmann2021benchmarking} present some benchmarking tasks and compare both statistical and neural SBI across various metrics. These tasks have ground-truth posteriors for model parameters, facilitating easy comparison, but such ground-truths are unavailable in real-world problems. \citet{jorgensen2022efficient} compare statistical SBI methods with neural SBI methods using two agent-based models. They use a synthetic dataset for parameter estimation, assuming the model can recover observation data. However, the application of real-world data complicates this. In such cases, a robust SBI algorithm is often required to address model misspecification problems when standard SBI approaches cannot provide reasonable accuracy. Development of robust SBI algorithms is an active research area in Bayesian statistics \citep{frazier2020model,frazier2024synthetic,kelly2024misspecification,ward2022robust,nott2023bayesian}.

For this purpose, we demonstrate the decision-making process for choosing SBI algorithms in two real-world applications of agent-based models in cellular biology: the biphasic tumour growth model \citep{jenner2020enhancing, wang2024calibration} and the stochastic cell invasion model \citep{carr2021estimating}. With the increase in computational power, agent-based models have become capable of describing complex biological processes at the individual level, even though their likelihoods are often intractable. For both models, we utilize synthetic data (generated from the models) and real-world observation data to infer model parameters and their associated uncertainties, employing both statistical and neural SBI algorithms. For synthetic data, where the models should be able to replicate the data, we do not use robust versions of the SBI algorithms. However, we apply robust versions of SBI when working with real-world observation data, due to the potential for model misspecification.  We then illustrate how to evaluate and validate the inference results from each SBI algorithm and compare their performances.

\begin{figure}[H]
\centering
  \includegraphics[width=1\linewidth]{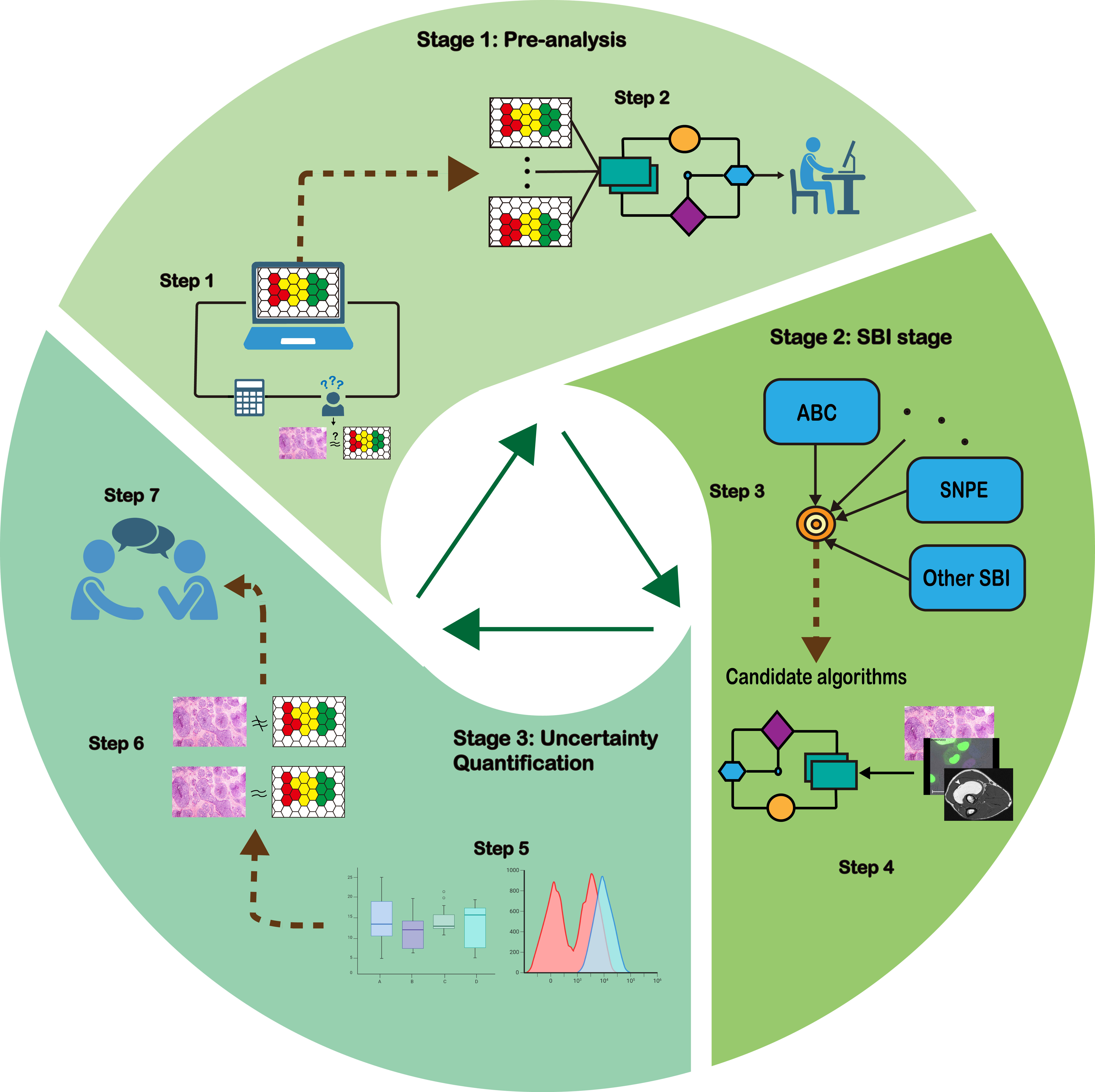}
  \caption{\textbf{The comprehensive guideline: } The guideline is divided into three stages: pre-analysis stage, SBI steps, and uncertainty analysis. The green arrows inside the cycle indicate that these three stages can be repeated. The important task of the first stage is to use SBI algorithms on synthetic datasets to check model compatibility and computational budget, allowing the selection of suitable candidate SBI algorithms to perform inference on real datasets, which leads to stage 2. Once inference results are obtained (stage 3), we need to gain a better understanding of the results, and if necessary, further improve the model, and return to stage 1 to use the newly extended model.}
\label{Workflow}
\end{figure}

In this paper, we present comprehensive guidelines for computational biologists and, more broadly, the mathematical biology community, on how to make decisions when using real-world observation data to infer model parameters and their associated uncertainties (the complete workflow is shown in Figure \ref{Workflow}). We implement both statistical and neural SBI methods, describing their performance in terms of efficiency and estimation accuracy, and provide a general discussion on these topics. While our focus is on two agent-based models of cellular dynamics, our approach can be extended to other models. Our goal is to contribute to the continuing discussion on how to efficiently and accurately calibrate complex biological models using real-world observational data.

\section{Methods and Models}
In this paper, we aim to use Bayesian inference for parameter estimation and uncertainty quantification. Our objective is to infer the posterior distribution of model parameters by considering:
\begin{equation}
p(\boldsymbol{\theta}|\boldsymbol{x}) \propto p(\boldsymbol{x}|\boldsymbol{\theta})p(\boldsymbol{\theta}),
\end{equation}
where $p(\boldsymbol{x}|\boldsymbol{\theta})$ represents the likelihood density, i.e., the probability distribution of the model simulation $\boldsymbol{x}$ given the model parameter values $\boldsymbol{\theta}$, and $p(\boldsymbol{\theta})$ is the prior distribution, which reflects our beliefs about the model parameters before incorporating the data. We distinguish between the likelihood density $p(\boldsymbol{x}|\boldsymbol{\theta})$ and the model $\mathcal{M}(\boldsymbol{\theta})$. For complex stochastic models, the former is often intractable, whereas the latter refers to the data generating process (DGP), which can often be simulated in a reasonable amount of time. The key idea of simulation-based inference is to approximate the posterior distribution while only relying on the ability to simulate from the DGP. 

When dealing with high-dimensional data, it is often necessary to use summary statistics to reduce the dimensionality of the data and avoid the curse of high dimensionality \citep{sisson2018handbook}. In this paper, we use $\boldsymbol{x}$ and $\boldsymbol{y}$ to refer to simulated and observed data, respectively. We denote the summary of simulated and observed data as $S(\boldsymbol{x})$ and $S(\boldsymbol{y})$, respectively, where $S(\cdot)$ represents the mapping from data to summaries.

In practice, limited or no information about $\boldsymbol{\theta}$ might lead us to use a vague prior distribution. We define a vague distribution as one with a wide range. For example, the Uniform distribution with a wide range, which we use as the prior for parameters in both models, is one such example. Notably, the Uniform distribution has a bounded parameter space. In SBI, and Bayesian inference more generally, it is often useful to transform this bounded space into an unbounded space. Here, we consider the logit transformation to facilitate this conversion. However, for neural SBI methods, a so-called leakage issue might occur \citep{greenberg2019automatic,durkan2020contrastive,wang2024preconditioned}, which we describe in detail in Section \ref{section: SBI}.

\subsection{Agent-based Models}
\begin{figure}[H]
\centering
  \includegraphics[width=.8\linewidth]{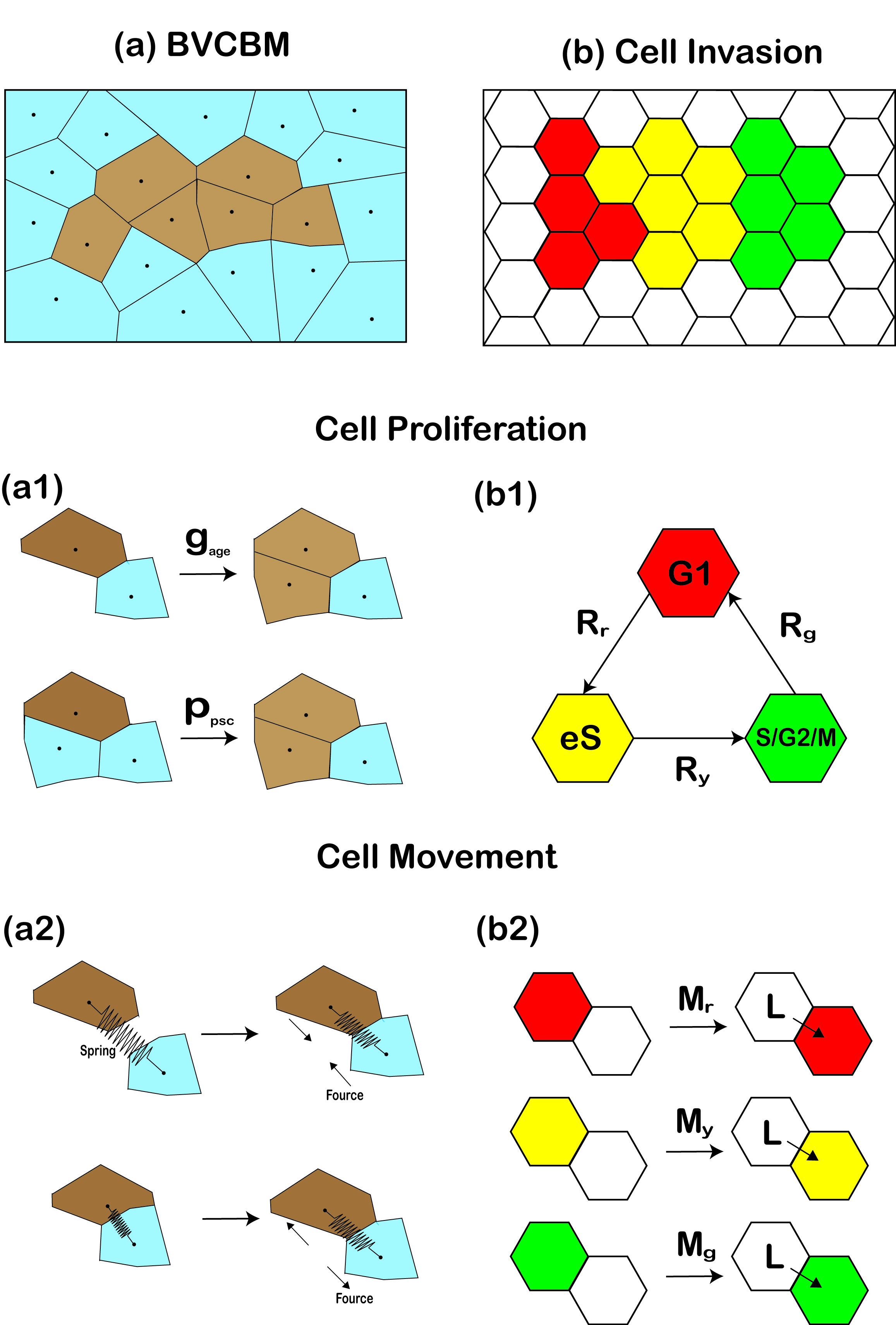}
  \caption{\textbf{Illustration plots for both agent-based models:} (a) and (b) depict the shape of cells, where (a) represents the off-lattice VCBM \citep{wang2024calibration} and (b) represents the on-lattice agent-based model \citep{carr2021estimating}. (a1) and (b1) describe cell proliferation in the BVCBM and the stochastic cell invasion model, respectively. (a1) illustrates (upper) a tumour cell (brown) dividing into two tumour cells, and (lower) cell invasion where a healthy cell (blue) becomes a tumour cell. (b1) illustrates the cell cycle, with the rate of change from one stage to another. (a2) and (b2) describe cell movement. (a2) describes how cells move according to Hooke's law: if two cells are far apart, they will attract each other to reach an equilibrium state; if too close, the cells will separate. (b2) describes how cells move in lattice space, where cells can only move to an empty lattice site according to a movement rate. The L represents the lattice space before the cell moves. }
\label{ABM examples}
\end{figure}

In this paper, we discuss two agent-based models as DGPs that describe complex cellular dynamics. The first model is an off-lattice agent-based model depicting tumour growth \citep{jenner2020enhancing,wang2024calibration}. In this model, cell centres are represented by points that move freely in the domain based on force-balance equations and cell edges are defined by a Voronoi tessellation of the set of cell centres. This approach introduces additional computational effort for cell movement, as cells can move in any direction. For this model, we do not employ summary statistics since the {\it in vivo} tumour volume data are single-trajectory time-series data \citep{wade2019fabrication}. The second model focuses on cell invasion, incorporating the cell cycle \citep{simpson2018stochastic,carr2021estimating}. It assumes all cells belong to one of three subpopulations corresponding to the cell cycle stages: G1, eS, and S/G2/M. The {\it in vivo} data for this model are image-based, generated using fluorescent ubiquitination-based cell cycle indicator (FUCCI) technology \citep{sakaue2008visualizing}. This necessitates the use of summary statistics to reduce data dimensionality. We utilize the same two sets of summary statistics as \citet{carr2021estimating}: cell density and cell trajectory data. We present the illustration for both models in Figure \ref{ABM examples}. \\

\subsubsection{Biphasic Voronoi Cell-based Model (BVCBM)} 

The BVCBM is an off-lattice agent-based model modeling the movement and proliferation of cells in a square domain. Initially the cells are placed in a hexagonal layout \citep{wang2024calibration}. The central cell is a cancer cell, and all others are healthy. The simulation runs until the tumour hits $100 {mm}^2$. At this size, the spatial configuration of healthy and cancerous cells is recorded and used as a base for further simulation of tumour growth. The growth is modeled by checking if a cancer cell divides, following the formula $p_d = p_0 \left( 1 - \frac{d}{d_{\mathrm{max}}}\right)$, where $p_d$ is division probability, $p_0$ initial division rate, $d$ is the Euclidean distance of a cell to the edge of the tumour and $d_{\mathrm{max}}$ is the maximum Euclidean distance a cell can be from the tumour edge and still proliferate. In a given time step, cancer cells have a probability $p_{\mathrm{psc}}$ of being invasive, meaning they can occupy a neighbouring healthy cells position. Cell positions, both healthy and cancerous, are updated using Hooke's law given by
\begin{align}
    \boldsymbol{r}_i(t+\Delta t) = \boldsymbol{r}_i(t) + \lambda \sum_{\forall j} \frac{\boldsymbol{r}_{i,j}(t)}{|\boldsymbol{r}_{i,j}(t)|}(s_{i,j}(t) - |\boldsymbol{r}_{i,j}(t)| ).
\end{align}
In this, $\boldsymbol{r}_i(t+\Delta t)$ is the updated position of cell $i$, $\mu$ the cell motility coefficient, $\boldsymbol{\mathrm{F}}_i(t)$ the force on cell $i$, $\lambda$ a mechanical interaction coefficient, $\boldsymbol{r}_{i,j}(t)$ the vector between cells $i$ and $j$, and $s_{i,j}(t)$ the natural spring length connecting them. Cell $j$ is considered in the neighbourhood of cell $i$ and contributing to its movement if it is connected to that cell in the Delaunay triangulation. Values for parameters like $\lambda$ and $\mu$ come from earlier studies \citep{meineke2001cell}. For more detailed simulation information, refer to \citet{jenner2020enhancing, jenner2023examining,wang2024calibration}.

In this paper, we focus on the biphasic BVCBM implemented in C++, where the tumour growth rate switches at time $\tau$ from one phase (usually slow growth) to another (usually fast growth). Following the approach of \citet{wang2024calibration}, we recognize that $g_{\mathrm{age}}$ is a key parameter in both phases of growth. The model parameters we aim to estimate are denoted as $\boldsymbol{\theta} = (g_{\mathrm{age}}^1, \tau, g_{\mathrm{age}}^2)$. We employ a uniform distribution constrained between 2 and 24$\times$ total experiment days for $(g_{\mathrm{age}}^1,g_{\mathrm{age}}^2)$, and 1 and total experiment days for $\tau$. To this end, we utilize three synthetic datasets generated with the same parameter sets as used in \citet{wang2024calibration}. This ensures that all algorithms can accurately replicate the synthetic datasets. Then we use the pancreatic tumour datasets \citep{jenner2023examining,wade2020dual} as our real-world observation data, chosen specifically for their display of biphasic behavior.\\

\subsubsection{Stochastic Cell Invasion Model} 

The stochastic cell invasion model proposed by \citet{carr2021estimating} is a two-dimensional hexagonal lattice agent-based model that simulates cell proliferation and movement to understand the mechanisms of cell invasion. This model assumes that the entire population of cells is divided into three subpopulations, each representing a different stage in the cell cycle: G1, early S (eS), and S/G2/M. The use of fluorescent ubiquitination-based cell cycle indicator (FUCCI) technology allows for the visualization of these cell cycle phases in distinct colors: red for G1, yellow for eS, and green for S/G2/M. To simulate the transitions through the cell cycle stages, \citet{carr2021estimating} define specific rates: a rate $R_r$ per hour for cells to transition from red (G1 phase) to yellow (eS phase), a rate $R_y$ per hour for transitioning from yellow to green (S/G2/M phase), and a rate $R_g$ per hour for cells to transition from green back to red (G1 phase). For modeling cell movement, \citet{carr2021estimating} propose using a nearest neighbor random walk, simulated through a Markov process employing the Gillespie algorithm. This approach involves defining movement rates, $M_r$, $M_y$, and $M_g$, which represent the hourly moving rates for cells in the red, yellow, and green phases, respectively. Under this model, a cell can randomly move to an adjacent vacant position. The time interval between each movement is characterized by an exponential distribution, and the total measurement time for the experiment is 48 hours.

The experimental datasets derived from FUCCI technology consist of image data. This necessitates the use of summary statistics to reduce the dimensionality of the observational data. \citet{carr2021estimating} address this by counting the number of cells in each phase of the cell cycle at 48 hours, using this count as the summary statistics for the cell cycle transition rates. In this paper, we use the count of each cell type for the transition rates and denote it as $({\mathrm{sx}}_1, {\mathrm{sx}}_2, {\mathrm{sx}}_3)$ for number of cells in each stage of cell cycle. In terms of summarizing the cell movement rate, \citet{carr2021estimating} propose two summary statistics: cell density and cell trajectory. The first, cell density, refers to the median position and interquartile range of each cell type on either side of the scratched region. The second, cell trajectory, also known as cell track, is the average distance that cells travel during each phase of the cell cycle, either until they return to the G1 phase or the simulation terminates.

In this paper, we implement the model proposed by \citet{simpson2018stochastic,carr2021estimating} using MATLAB, and utilize the \texttt{ImageJ} package in R for image analysis. We have developed a wrapper for the MATLAB code, enabling us to perform inference tasks in Python. The model parameters we aim to estimate are denoted as $\boldsymbol{\theta} = (R_r, R_y, R_g, M_r, M_y, M_g)$. For all these parameters, we employ a uniform distribution constrained between 0 and 1 for $(R_r, R_y, R_g)$, and 0 and 10 for $(M_r, M_y, M_g)$.

\subsection{Simulation-based Inference}\label{section: SBI}
We define Simulation-Based Inference (SBI) as an inference technique that does not explicitly evaluate the likelihood function, but rather relies on the ability to simulate from the DGP. As we mentioned in the previous section, two popular approaches for SBI are statistical and neural SBI methods. In this paper, we focus on ABC and BSL to represent statistical SBI methods. ABC and its variants are the most popular methods in biology \citep{beaumont2002approximate,toni2009approximate,beaumont2010approximate,liepe2014framework,vo2015quantifying,ross2017using}. They approximate the likelihood non-parametrically by using a kernel, such as indicator or Gaussian kernel, to help determine if the parameter values used to generate simulated data can be accepted. This kernel depends on a tolerance value that is chosen by the user. BSL, on the other hand, is a parametric approach that approximates the likelihood density by assuming the summary statistics generated from the DGP follow a multivariate Gaussian distribution. 

Generally, statistical SBI approaches such as ABC and BSL require a large number of model simulations, since many model simulations are wasted as they are associated with parameter proposals that are rejected.  Thus for complex simulators, statistical SBI approaches can be highly computationally intensive.

In order to work with a much smaller number of model simulations, neural SBI approaches have been developed. The key idea is to use a Conditional Neural Density Estimator (CNDE) to learn a mapping from a simple and tractable distribution to the target distribution. Neural SBI approaches only require the availability of the $N$ pairs of ${\{\boldsymbol{\theta}_i,\boldsymbol{x}_i\}}_{i=1}^N$ to train the CNDE for the posterior (NPE) or likelihood (NLE) density.  Such methods can also be performed sequentially, called SNPE and SNLE, by updating the proposal distribution of the parameter values based on previous approximations. The sequential versions can be more computationally efficient when analysing a single dataset, particularly if there is a large discrepancy between the prior and the posterior. 

Model misspecification presents a significant challenge when dealing with real-world observational data. A DGP can be considered as misspecified if it fails to replicate the observed data. This indicates that the likelihood density might not be correctly approximated. In this paper, we focus on two robust SBI algorithms for handling real-world observational data: robust BSL (RBSL) \citep{frazier2021robust} and robust SNLE (RSNLE) \citep{kelly2024misspecification}. Detailed descriptions of these algorithms and their hyperparameters are presented in Section \textcolor{blue}{A} of the Supplementary document. 

\subsubsection{Statistical SBI: Approximate Bayesian Computation (ABC)} 

ABC is categorized as a {\it statistical} SBI algorithm for directly approximating the posterior distribution. Intuitively, the ABC method compares simulated data with observed data and accepts the parameter values if the simulated data is sufficiently close to the observed data according to some distance metric, $\rho(\cdot,\cdot)$.  If a uniform kernel is selected, which is a popular choice in the literature, then the parameter proposal is accepted if it produces simulation data $\boldsymbol{x}$ such that $\rho(\boldsymbol{y},\boldsymbol{x}) < \epsilon$, where $\epsilon$ is the user-specified ABC threshold. ABC methods aim to approximate the posterior distribution by
\begin{equation}
    p_{\epsilon}(\boldsymbol{\theta}|\boldsymbol{y}) \propto p(\boldsymbol{\theta})\int_{\boldsymbol{x}} \mathbb{1}\{\rho(\boldsymbol{y},\boldsymbol{x}) < \epsilon\}p(\boldsymbol{y}|\boldsymbol{\theta}) d\boldsymbol{x},
\end{equation}
where $\mathbb{1}$ is an indicator kernel and $p_{\epsilon}(\boldsymbol{\theta}|\boldsymbol{y})$ denotes the approximate posterior.  When the data dimension is high, it can be inefficient to compare $\boldsymbol{y}$ and $\boldsymbol{x}$ directly.  In these cases, the discrepancy function $\rho\big(S(\boldsymbol{y}),S(\boldsymbol{x}) \big)$ can compare summary statistics of the observed and simulated data. There are many algorithms for sampling the ABC posterior.  The most fundamental algorithm is ABC rejection \citep{beaumont2002approximate,beaumont2009adaptive}, which generates parameter proposals from the prior. This can be inefficient when the prior and posterior differ substantially.  A more efficient approach, called Sequential Monte Carlo (SMC) ABC (e.g. \cite{toni2009approximate,drovandi2011estimation}), defines a sequence of decreasing ABC thresholds, and updates the parameter proposal at each iteration. See \citet{sisson2018handbook} for detailed descriptions of ABC algorithms.

We selected the adaptive SMC ABC algorithm proposed by \citet{drovandi2011estimation} because it has been used to calibrate both models demonstrated in this paper, providing reasonable estimations with real-world observational datasets \citep{wang2024calibration,carr2021estimating}. We briefly describe the adaptive SMC ABC algorithm that we use in this paper and provide a detailed description, along with pseudocode, in Section \textcolor{blue}{A1} of the Supplementary document.

The adaptive SMC ABC algorithm generates $N$ samples from a sequence of ABC posteriors, utilizing a series of decreasing ABC thresholds, denoted as $\epsilon_1 > \dots > \epsilon_T$, where $\epsilon_T = \epsilon$ is the target ABC threshold. In each iteration of the algorithm, the sequence of tolerances is adaptively determined. This is done by discarding a proportion, $a\cdot N$, of the samples that exhibit the highest discrepancy, where $a$ is a tuning parameter. Subsequently, the sample population is rejuvenated through a process of resampling followed by a move step. During the move step, an MCMC ABC kernel is used to diversify particles and simultaneously maintain the distribution of particles consistent with the current tolerance level. The number of MCMC steps, $R_t$, applied to each particle is adaptively determined based on the overall MCMC acceptance rate. This is calculated as $R_t = \left\lceil \frac{\log(c)}{\log(1-p_t^{\mathrm{acc}})}\right\rceil$, where $c$ is a tuning parameter and $p_t^{\mathrm{acc}}$ is the estimated MCMC acceptance probability at the $t$th SMC iteration. The algorithm terminates when the MCMC acceptance rate becomes intolerably low or when $\epsilon_t = \epsilon$.  \\

\subsubsection{Statistical SBI: Bayesian Synthetic Likelihood (BSL)} 

BSL \citep{price2018bayesian,frazier2023bayesian}  is an alternative {\it statistical} SBI algorithm that approximates the likelihood density parametrically. It assumes that the likelihood density follows a multivariate Gaussian distribution, and it samples from the approximate posterior by implementing a sampling algorithm such as MCMC. BSL employs summary statistics when dealing with high-dimensional data  $\boldsymbol{y}$. This results in the likelihood density being approximated by the summary statistics likelihood as $p(S(\boldsymbol{y})|\boldsymbol{\theta})$, which is assumed to follow a multivariate Gaussian distribution:
\begin{equation}
p(S(\boldsymbol{y})|\boldsymbol{\theta}) \approx \mathcal{N}(S(\boldsymbol{y}); \mu(\boldsymbol{\theta}), \Sigma(\boldsymbol{\theta})).
\end{equation}
The auxiliary parameters $\mu(\boldsymbol{\theta})$ and $\Sigma(\boldsymbol{\theta})$ are typically unknown and need to be estimated using simulated data. Specifically, we use $m$ simulated datasets of summary statistics generated from the DGP conditional on $\theta$, ${\{S(\boldsymbol{x_i})\}}_{i=1}^m$, to estimate these parameters by the following equations:
\begin{align}
\Hat{\mu}(\boldsymbol{\theta}) &= \frac{1}{m} \sum_{i=1}^{m} S(\boldsymbol{x}_i),\\
\Hat{\Sigma}(\boldsymbol{\theta}) &= \frac{1}{m-1} \sum_{i=1}^{m} (S(\boldsymbol{x}_i) - \mu(\boldsymbol{\theta}))(S(\boldsymbol{x}_i) - \mu(\boldsymbol{\theta}))^T.
\end{align}
Subsequently, MCMC can be used to sample from the approximate posterior distribution, which is given by:
\begin{equation}
p(\boldsymbol{\theta}|S(\boldsymbol{y})) \propto p(\boldsymbol{\theta})\mathcal{N}(S(\boldsymbol{y}); \Hat{\mu}(\boldsymbol{\theta}), \Hat{\Sigma}(\boldsymbol{\theta})).
\end{equation}
It is clear that the choice of the number $m$ is crucial in BSL methodology, as it influences the mixing of the MCMC chain. According to empirical evidence presented in \citet{price2018bayesian}, the most efficient results occur when $m$ is tuned such that the standard deviation of the log synthetic likelihood estimator (estimated at some central parameter value) is between 1 and 2.

In this paper, we use standard BSL during the pre-analysis step to achieve two objectives: firstly, to understand how BSL performs on synthetic datasets, and secondly, to select a suitable value for $m$. It is also important to note that the covariance matrix of the random walk proposal for the MCMC implementation of BSL requires tuning.  This can be tuned based on several pilot runs of MCMC. The output from SMC ABC could be used to help form the initial random walk covariance matrix. 

\subsubsection{Neural SBI: Neural Posterior Estimation (NPE)} 
NPE utilizes state-of-the-art deep generative models to learn a sequence of transformations that aim to transport a simple distribution to the posterior distribution, classifying NPE as a {\it neural} SBI algorithm \citep{papamakarios2021normalizing}. Unlike ABC methods, which can require a vast number of simulated datasets, NPE employs a fixed number $n$ of parameter values and its associated simulated data, forming a training dataset ${\{\boldsymbol{\theta}_i, \boldsymbol{x}_i\}}_{i=1}^n$. This dataset is used to train a neural conditional density estimator (NCDE), $q_{F(\boldsymbol{\psi})}(\boldsymbol{\theta}|\boldsymbol{x})$. The aim is to learn the mapping from $p(\boldsymbol{z}|\boldsymbol{x})$ to $p(\boldsymbol{\theta}|\boldsymbol{x})$. Here, $F$ represent a neural network with hyperparameter $\boldsymbol{\psi}$, and $\boldsymbol{z} \sim p(\boldsymbol{z}|\boldsymbol{x})$ signifies a simple and tractable density, such as a multivariate Gaussian distribution. To simulate the posterior distribution, observational data $\boldsymbol{y}$ is inputted into the trained neural conditional density estimator, leading to $\boldsymbol{\theta}_j \sim q_{F(\hat{\boldsymbol{\psi}})}(\boldsymbol{\theta}|\boldsymbol{y})$ where $\hat{\boldsymbol{\psi}}$ is the learned hyperparameter and $j=1,\ldots,M$ where $M$ is the desired number of samples from the approximated posterior.

The NPE method may not perform well if many of the training datasets are not close to the observed dataset, which can occur when using a vague prior distribution and when the parameter dimension is increased.  Sequential NPE (SNPE) has been developed to improve estimation accuracy for a given observed dataset. This approach involves using the NPE approximation from the current round as the proposal distribution $\Tilde{p}(\boldsymbol{\theta})$ for the next round. However, when training $q_F$ with parameter values drawn from $\Tilde{p}(\boldsymbol{\theta})$ (i.e.\ the proposal distribution which is not necessarily the prior), the samples generated from the NCDE do not converge to the true posterior distribution. Instead, it approaches the distribution defined by:
\begin{equation}
\label{npe equation}
\Tilde{p}(\boldsymbol{\theta}|\boldsymbol{y}) \propto p(\boldsymbol{\theta}|\boldsymbol{y})\frac{\Tilde{p}(\boldsymbol{\theta})}{p(\boldsymbol{\theta})}.
\end{equation}
Various SNPE methods have been suggested such as \citet{papamakarios2016fast,lueckmann2017flexible,greenberg2019automatic}. In this paper, we use the SNPE method proposed in \citet{greenberg2019automatic}, which, according to \citet{lueckmann2021benchmarking}, significantly outperforms other methods.

Generally, both NPE and SNPE operate under the assumption that the model $\mathcal{M}(\theta)$ can accurately replicate observational data. However, this assumption may not always be valid, particularly when dealing with real-world data. Recent research has begun to explore methods for handling model misspecification in NPE techniques \citep{ward2022robust,glockler2023adversarial}. Despite these efforts, it remains unclear how to address this issue both efficiently and reliably. 

Another challenge faced by SNPE is the ``leakage" issue, which we elaborate on here. The leakage problem can occur when there is no regularization in the loss function acting on the NCDE to constrain the density within the prior support, or when the NCDEs ignore some extreme or invalid data to stabilize the training, leading to unexplored areas of the parameter space. \citet{durkan2020contrastive, deistler2022truncated, wang2024preconditioned} identify this problem in SNPE-C \citep{greenberg2019automatic} and have proposed methods such as using transformations to map parameters from bounded space to unbounded space or employing a preconditioning step to improve the quality of the training dataset. To address this issue, we have applied the truncated SNPE (TSNPE) method as proposed by \citet{deistler2022truncated}.

In this paper, we use normalizing flows as the NCDE and we briefly introduce the details of how to train the normalizing flows in Section \textcolor{blue}{B 1.3} and \textcolor{blue}{B 1.4} of the Supplementary document. We use neural spline flow (NSF) \citep{durkan2019neural} as our flow model. \\

\subsubsection{Neural SBI: Neural Likelihood Estimation (NLE)} 

NLE is similar to BSL, but it learns the likelihood density using NCDE, as discussed in \citet{papamakarios2019sequential}, and is a {\it neural} SBI algorithm. This approach renders NLE more flexible than BSL by not making any parametric assumptions about the form of the likelihood density. NLE trains the NCDE model on pairs from a training dataset ${\{\boldsymbol{\theta}_i, \boldsymbol{x}_i\}}_{i=1}^n$. This training is aimed at approximating the likelihood density $q_{G(\boldsymbol{\phi})}(\boldsymbol{\theta}|\boldsymbol{x})$, where $G(\boldsymbol{\phi})$ in NLE is distinct from $F(\boldsymbol{\psi})$ used in NPE. Subsequently, sampling algorithms like MCMC can be employed to draw samples from the approximate posterior distribution by using $q_{G(\hat{\boldsymbol{\phi}})}(\boldsymbol{\theta}|\boldsymbol{y})$ as the likelihood density, where $\hat{\boldsymbol{\phi}}$ is the learned hyperparameter. In a manner similar to SNPE, Sequential NLE (SNLE) can enhance estimation accuracy for a given observed dataset.  As in SNPE, a new training dataset are generated from the current NLE approximation, and then the NCDE can be re-trained using all the model simulations produced thus far.  This process is repeated for a given number of rounds.

\subsection{Model Misspecification}

In practice, it is generally infeasible to develop a model that can perfectly recover real datasets. Since our focus is on fitting real data with models, model misspecification will always exist, potentially resulting in the model being unable to recover such datasets. \citet{marin2014relevant, frazier2020model} use the term compatible to describe when there exists a parameter value of the model that can recover observed summary statistics as the sample size diverges.  When the model is unable to recover the observed summaries, the model is said to be incompatible, and represents a particular form of model misspecification.

We follow the definition of the model misspecification problem from \citet{frazier2020model} and apply it in the context of systems biology. Since it is implausible to develop a model that can perfectly recover real datasets, we define a model as well-specified or compatible if the real datasets lie reasonably within the predictive interval of the posterior predictive distribution, and as misspecified or incompatible if they do not. We also want to clarify the distinction between an algorithm performing poorly in inference and a model being unable to recover the observational datasets. The former indicates that the algorithm is unsuitable for the specified task, while the latter refers to the model misspecification problem that is the focus of this paper.

\citet{frazier2020model} has shown that ABC and its variants are partly robust to model misspecification, in the sense that the ABC posterior concentrates onto the pseudo-true parameter value as the sample size increases.  Here the pseudo-true parameter is defined as the parameter value that minimises the discrepancy between the model summary and the observed summary, as the sample size diverges. Therefore, we treat the adaptive SMC ABC algorithm used in this paper as a robust algorithm for handling model misspecification problems. 

Under incompatibility, DGPs might not be able to recover the observation datasets accurately. Methods like BSL or SNLE might lead to poor approximations. Consequently, many approaches have been developed to improve the robustness of BSL or SNLE. For an introduction to some popular approaches, see \citet{nott2023bayesian}. In this paper, we focus on adjustment methods that have been widely used to enhance robustness in both BSL \citep{frazier2020robust, frazier2024synthetic, frazier2021robust, frazier2023bayesian} and SNLE \citep{kelly2024misspecification}. This approach suggests using a vector of free parameters to adjust either the mean of the simulated summaries or the variance of the simulated summaries to mitigate the effects of model misspecification. Specifically, in the robust BSL (RBSL) context, the adjustment method introduces a free parameter vector, $\Gamma = [\gamma_1, \dots, \gamma_{\mathrm{d}}]$, where $\mathrm{d}$ is the dimension of the summary statistic vector, to adjust the sample mean or sample variance of the summary statistics. 

For the mean adjustment method, the simulated means of the summary statistics become:
\begin{equation}
\phi_n(\xi) = \mathrm{\mu}_n(\boldsymbol{\theta}) + \mathrm{diag}(\mathrm{\Sigma}_n^{\frac{1}{2}}(\boldsymbol{\theta}))\Gamma.
\end{equation}
The parameters $\boldsymbol{\theta}$ and $\Gamma$ are considered independent with a new prior $p(\xi) = p(\boldsymbol{\theta})p(\mathrm{\Gamma})$, where \citet{frazier2021robust} recommend a Laplace prior for $\Gamma$. \citet{kelly2024misspecification} use a similar idea as the mean adjustment method to adjust the surrogate likelihood in SNLE, $q_{g(\psi, \boldsymbol{\theta})}(S(\boldsymbol{y}) - \Gamma)$ to learn the joint posterior,
\begin{equation}
    p(\boldsymbol{\theta}, \Gamma|S(\boldsymbol{y})) \propto q_{g(\mathrm{\psi}, \boldsymbol{\theta})}(S(\boldsymbol{y}) - \Gamma) p(\boldsymbol{\theta})p(\Gamma).
\end{equation}
\citet{kelly2024misspecification} propose to use a data-driven prior for $p(\mathrm{\Gamma})$, which has following form:
\begin{equation}
p(\gamma_i) = \mathrm{Laplace}(0, \lambda = |\tau \tilde{S}^i(\boldsymbol{y})|) = \frac{1}{2\lambda} \exp\left( -\frac{|\gamma_i|}{\lambda} \right),
\end{equation}
where $\tilde{S}^i(\boldsymbol{y})$ is the $i$-th standardised observed summary.

For the variance adjustment method, the introduced vector $\Gamma = [\gamma_1, \dots, \gamma_{\mathrm{d}}]$ is used to inflate the variance of the simulated summary statistics. When the values for $m$ and $n$ are large enough, if the simulated and observed summaries differ by more than a few standard deviations, the summaries can be considered misspecified. The inflated variance has the form:
\begin{equation}
    V_n(\xi) = \mathrm{\Sigma}_m(\boldsymbol{\theta}) + \begin{bmatrix}
\Sigma_n(\boldsymbol{\theta})_{11} \gamma_1^2 & 0 & \cdots & 0 \\
0 & \Sigma_n(\boldsymbol{\theta})_{22} \gamma_2^2 & \cdots & 0 \\
\vdots & \vdots & \ddots & \vdots \\
0 & \cdots & \cdots & \Sigma_n(\boldsymbol{\theta})_{\mathrm{d}\mathrm{d}} \gamma_\mathrm{d}^2
\end{bmatrix},
\end{equation}
where a Laplace prior can again be used for $\Gamma$.

\subsection{Guidelines}

In this section, we detail the proposed guidelines as three main stages so that the readers can easily follow and find the potential solution to improve the inference performance. As shown in Figure \ref{Workflow}, the three stages are (1) pre-analysis stage, (2) SBI stage and (3) uncertainty analysis stage. 

\subsubsection{Stage 1: pre-analysis stage}
The first stage is the pre-analysis stage. In this stage, the initial step is to estimate the computational cost. Typically, agent-based models demand more computational time than continuous models when describing the same phenomena. For some cases the simulation time can depend greatly on where the parameter value lies in the parameter space. That is, different regions of the parameter space can result in vastly different simulation times, for example parameter values that lead to large numbers of cells. Thus, the pre-analysis step is crucial because it provides a rough indication of whether the statistical SBI is computationally feasible or prohibitive. We propose the first guideline below:

\begin{enumerate}
    \item \textbf{Know the computational cost for simulation. }\\
    If the model describes complex phenomenon, it is a good idea to repeatedly generate parameter samples from the prior distribution and simulate datasets from the model to determine if it is computationally expensive.
\end{enumerate}

If the computational cost for the model to simulate datasets is reasonable, then we can calibrate the model with synthetic datasets to investigate parameter identifiability and sensitivity. However, it is necessary to inform the reader that directly applying a certain SBI algorithm might not work well, which means we need to perform some diagnostic tests to check if the candidate algorithm is suitable for the task. For example, BSL can use the marginal distribution of model summaries to check the normality assumption by repeatedly simulating datasets from a fixed set of parameter values. When we treat synthetic datasets as real datasets, the true parameter values are known, allowing us to easily check if the model parameters are identifiable. If the posterior distribution of a parameter overlaps with the prior distribution, then we can say the dataset may not be informative for this parameter. If a parameter is sensitive, a small perturbation in its value will generate very different simulated datasets, meaning a highly concentrated posterior distribution indicates sensitivity. Therefore, we propose the final guideline for the pre-analysis stage as follows:

\begin{enumerate}
    \setcounter{enumi}{1} 
    \item \textbf{Perform inference on synthetic datasets. }\\
    The approximate posterior distributions on synthetic datasets can be a good indicator to select candidate algorithms.
\end{enumerate}

\subsubsection{Stage 2: SBI stage}

Checking for potential model misspecification is also important. As discussed in the previous section, model misspecification is inevitable, and we can only aim to reduce its effects. Ideally, we hope that the model can recover important features of the observed data, leading to model compatibility. However, if the model is incompatible, we still want to extract some information about the parameters during inference. For example, if tumour growth is exponential, a quadratic model may be incompatible. Prior predictive checks are a widely used tools to empirically determine if the model is capable of recovering the observed dataset. A recently developed test for detecting model misspecification, proposed by \citet{ramirez2024testing}, provides a more rigorous approach to identifying misspecification. This leads to the next guideline:

\begin{enumerate}
    \setcounter{enumi}{2}
    \item \textbf{Check model suitability.}\\
    It is important to check if a model has the ability to recover the real-world dataset. 
\end{enumerate}

The inference results on synthetic dataset are a good indicator of how well the inference will perform on real datasets, leading to the second stage: the SBI stage. In this stage, we need to carefully choose from a set of candidate algorithms to ensure that computational resources are not wasted. If the computational time to simulate a single dataset is long—say, it takes 10 minutes—then statistical SBI algorithms might not be suitable, as they would require a significant amount of time due to the need for millions of simulations. On the other hand, if the dataset contains a lot of noise, the prior distributions are vague, or the model itself is highly nonlinear, neural SBI algorithms might provide inaccurate estimates. Hence, we propose the next guideline as: 

\begin{enumerate}
    \setcounter{enumi}{3} 
    \item \textbf{Understand how to choose suitable SBI algorithms. }\\
    Choose a suitable SBI algorithm based on inference results on synthetic dataset and the advantages and disadvantages for candidate SBI algorithms. 
\end{enumerate}

\subsubsection{Stage 3: uncertainty analysis stage}

The two stages outlined above can provide reliable inference results, which leads to the next stage of performing uncertainty analysis and extending the model based on these results. The first step at this stage is to conduct a posterior predictive check. This is the most important step because it first assesses how well the model recovers the actual observations and, second, ensures that the posterior distributions are not estimated to be overconfident. In practice, most models have two or more parameters, making it necessary to plot the marginal posterior distributions for each parameter. Additionally, bivariate plots of the posterior distributions can help in understanding the correlation between parameters. Depending on the specified task, further analysis of the inference results can be performed. Therefore, we propose the following guidelines:

\begin{enumerate}
    \setcounter{enumi}{4} 
    \item \textbf{Understand the inference results on real-world datasets. }\\
    Understanding the univariate and bivariate posterior distribution of model parameters and the corresponding posterior predictive distribution can provide a clear path to further improving the model in more realistic settings.
\end{enumerate}

The inference results often provide suggestions related to model misspecification. The posterior predictive distribution is an empirical approach to visualize how close the simulated datasets generated based on posterior samples are to the actual observations. An indication that a model and corresponding inference algorithm performs well is that the posterior predictive distribution would cover most of the observations.  For example, for a nominal 95$\%$ posterior predictive interval (formed by the 2.5$\%$ and 97.5$\%$ quantiles of the posterior predictions) would contain roughly 95$\%$ of the data. However, it is possible that the inference algorithm performs poorly or fails to work, see Section \ref{BVCBM example} for an example. If most of the actual observations lie outside the posterior predictive distribution, then the model is incompatible. Therefore, the following guideline can be used to investigate the effects of model misspecification:

\begin{enumerate}
    \setcounter{enumi}{5} 
    \item \textbf{Model misspecification? }\\
    It is always a good idea to check whether the actual observations lie within the posterior predictive distribution. Ideally, for example, we hope the 95$\%$ posterior predictive interval can contain roughly 95$\%$ of the data. 
\end{enumerate}

We should highlight that understanding the inference results is not the endpoint of the calibration process. The key to performing calibration correctly is to improve the model so that it can describe the target phenomenon better or more realistically. Since we focus on biological models in this paper, we should collaborate with biologists to understand how to improve the model for a more accurate description of the phenomenon.

\begin{enumerate}
    \setcounter{enumi}{6} 
    \item \textbf{Work with biologists to improve the model.}\\
    It is important to work with biologists (or, more generally, domain experts) to gain a better understanding of biological processes and to improve the model in light of the inference results.
\end{enumerate}

\section{Results}
In this section, we use SBI algorithms to perform inference tasks on two agent-based models by using real-world observations. We outline the workflow for conducting these inference tasks through a series of steps: initially, we undertake a pre-analysis step, corresponding to part (a) in Figure \ref{Workflow}. The primary objective at this stage is to evaluate whether the SBI algorithms can accurately replicate synthetic datasets generated from the model. If the SBI algorithms are unable to replicate these synthetic datasets, it is likely to encounter difficulties with real-world datasets, indicating that it may not be the most suitable choice for our purposes. The pre-analysis step is crucial for selecting the appropriate algorithm and provides a detailed guideline for its implementation during the inference process. 

We present and compare the results of the prior predictive check and other findings from the pre-analysis step. The goal is to highlight the differences in the approximate posterior distributions provided by each candidate SBI algorithm and to establish a quantifiable method for selecting the most appropriate algorithm. Based on the results of the pre-analysis, we select the most suitable SBI algorithms from the following options: SMC ABC, RBSL, SNPE, and RSNLE. We employ a robust version of the likelihood approximation method when it is identified as the most appropriate algorithm.

For both examples, we select candidate SBI algorithms based on the trade-off between computational cost (measured by the total number of simulations used by an SBI algorithm) and estimation accuracy. If an SBI algorithm requires a significantly larger number of simulations compared with others, it will not be considered, even if it provides the most accurate estimation. We then show the posterior predictive distribution provided by the most suitable candidate SBI algorithms on real-world observations, which we select based on the pre-analysis step.

\subsection{Implementation}
The implementation details for each of the algorithms is as follows: For SMC ABC, we used $N = 1000$ samples, set the tuning parameters to $a = 0.5$ and $c = 0.01$, and stop the algorithm when the MCMC acceptance rate of the move step drops below 1\%. For BSL, we set the total number of MCMC iterations to $N = 10000$ and chose the values for $m$ based on the estimated standard deviation of the log synthetic likelihood. \citet{price2018bayesian} suggests that these values should be between 1 and 2, and we followed this recommendation. For BVCBM, we used $m = 300$ for the synthetic datasets, and $m = 150, 250, 300$ for pancreatic tumour growth datasets with 19, 26, and 32 measurement days, respectively. For the stochastic cell invasion model, we used $m = 200$ and $300$ for cell trajectories and cell density as summary statistics. For both SNPE and SNLE, we ran 10 rounds, with each round simulating 10,000 datasets for training. We used SNPE for synthetic datasets in both examples but opted for TSNPE for BVCBM in pancreatic tumour growth datasets to avoid leakage issues.

In the implementation of neural SBI, we utilized the \texttt{SBI} package \citep{tejero2020sbi} for NPE, SNPE, and TSNPE. For NLE, SNLE, and RSNLE, the \texttt{JAX} package \citep{frostig2018compiling} was employed to expedite the MCMC step, given the slower performance of the SNLE implementation within the \texttt{SBI} package. Neural SBI was facilitated through the use of normalizing flows, which were trained utilizing the Adam optimizer. The training settings included a learning rate of $5 \times 10^{-4}$ and a batch size of 256. Details of the implementation is provided in Section \textcolor{blue}{C1} of the Supplementary document.

We outline the pre-analysis steps and show the corresponding results for each example. In this stage, we use synthetic datasets because the true model parameter values are available, making it possible to quantify the uncertainty of each parameter analytically. For each of the agent-based models we used for demonstration, we use multiple ground truth model parameter values to build synthetic datasets from the DGPs. In Section \textcolor{blue}{B2} of the Supplementary document , we list all the ground truth parameter values we used to generate the synthetic datasets for BVCBM.

\subsection{Example 1: Biphasic Voronoi Cell-based Model}\label{BVCBM example}

\subsubsection{Stage 1: pre-analysis stage}
\begin{figure}[H]
\begin{subfigure}{1\textwidth}
\caption{ }
    \begin{subfigure}{.32\textwidth}
        \includegraphics[width=1\linewidth]{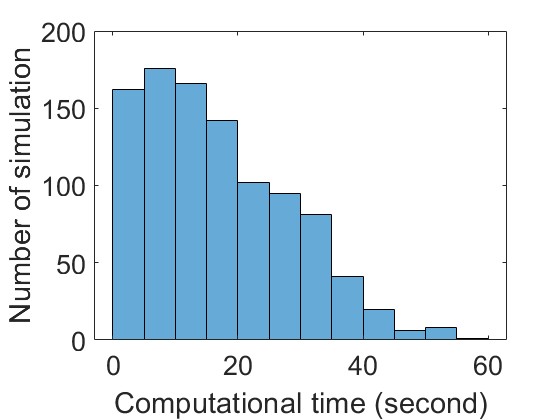}
    \end{subfigure}
    \hfill
    \begin{subfigure}{.32\textwidth}
        \includegraphics[width=1\linewidth]{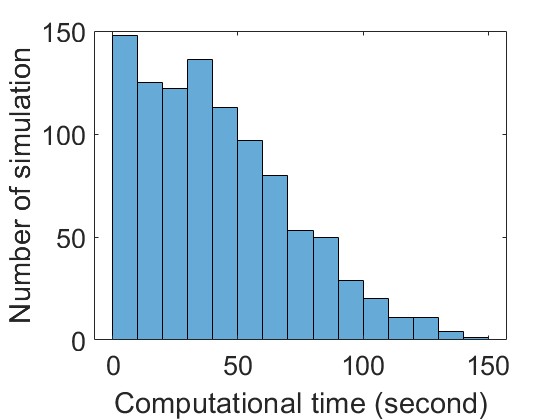}
    \end{subfigure}
    \hfill
    \begin{subfigure}{.32\textwidth}
        \includegraphics[width=1\linewidth]{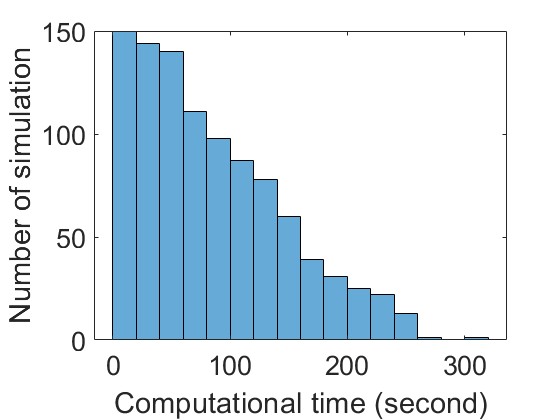}
    \end{subfigure}
    \label{bvcbm computational analysis}
\end{subfigure}
\vfill
\begin{subfigure}{1\textwidth}
\caption{ }
    \begin{subfigure}{.32\textwidth}
        \includegraphics[width=1\linewidth]{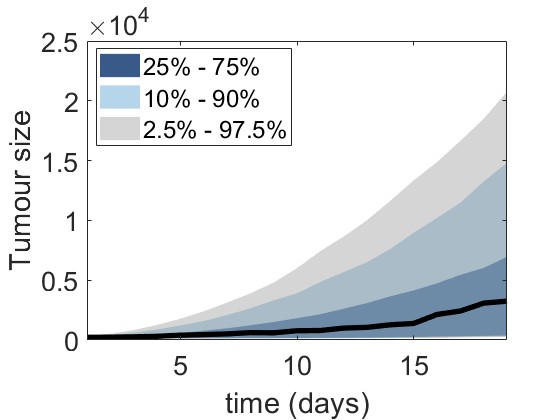}
    \end{subfigure}
    \hfill
    \begin{subfigure}{.32\textwidth}
        \includegraphics[width=1\linewidth]{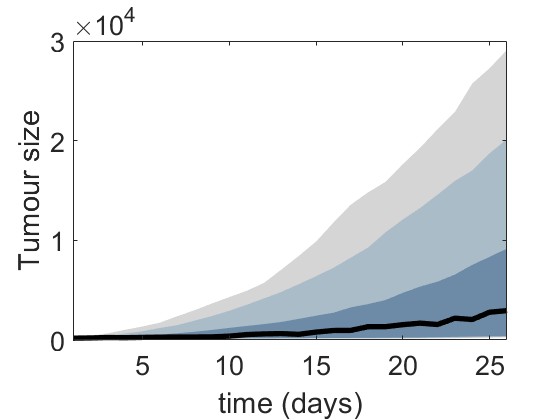}
    \end{subfigure}
    \hfill
    \begin{subfigure}{.32\textwidth}
        \includegraphics[width=1\linewidth]{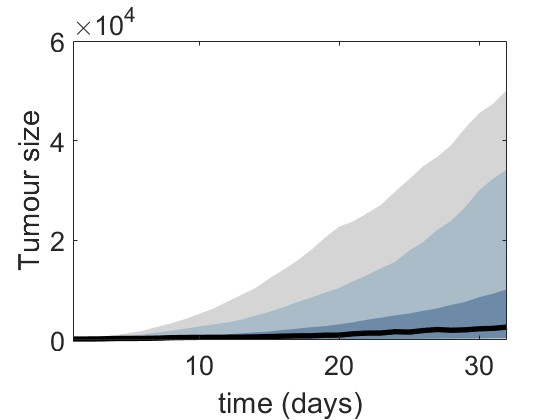}
    \end{subfigure}
    \label{bvcbm prior predictive check}
\end{subfigure}
\caption{ \textbf{Pre-analysis steps for BVCBM example:} (a) The histogram of computational time for BVCBM based on 1000 simulated datasets from the prior predictive distribution with measurement length 19, 26 and 32 days. (b) Prior predictive check for three real pancreatic tumour growth datasets with measurement length 19, 26 and 32 days, which are inside the $(10\%-90\%)$ predictive intervals.}
\end{figure}

The computational cost for inference needs to be considered before implementation. As the first step in performing pre-analysis, it is essential to estimate the average computational time for single simulation dataset from DGPs and check if the model is capable of recovering the actual observations through prior predictive check. We generate $N = 1000$ samples from the prior distribution and simulate its associated simulated data from the model to obtain the prior predictive distribution. In Figure \ref{bvcbm computational analysis}, we show the histogram of computational costs of 1000 simulated datasets for BVCBM. We report a computational cost range of 0.19 to 319.50 seconds per simulation for BVCBM depending on the length of tumour growth time series datasets. It is evident that increases in the length of time-series datasets will lead to longer simulation time. 

The prior predictive checks in Figure \ref{bvcbm prior predictive check} confirm that the model may be capable of recovering the actual observations, which led us to apply the four candidate SBI algorithms to perform the inference task with synthetic datasets to check parameters identifibility and parameters sensitivity, as proposed in guideline 3. 

\begin{figure}[h]
\begin{subfigure}{1\textwidth}
    \begin{subfigure}{.32\textwidth}
        \includegraphics[width=1\linewidth]{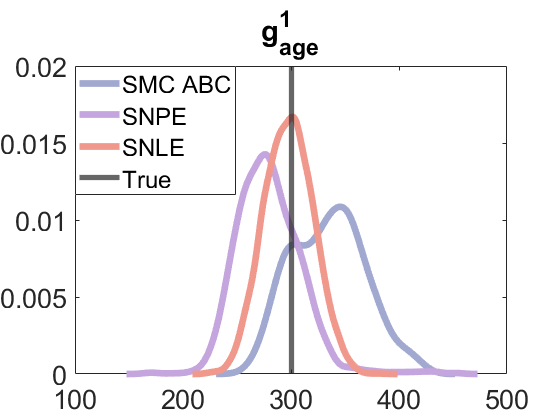}  
    \end{subfigure}
    \hfill
    \begin{subfigure}{.32\textwidth}
        \includegraphics[width=1\linewidth]{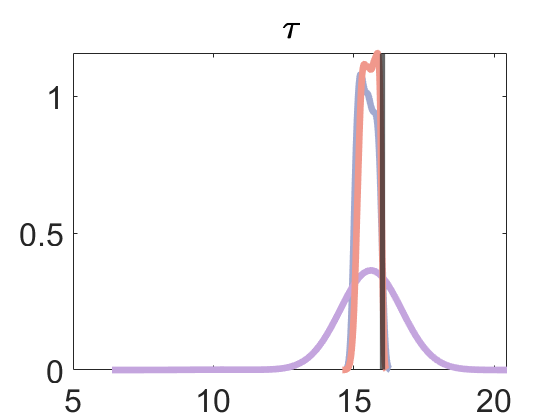}  
    \end{subfigure}
    \hfill
    \begin{subfigure}{.32\textwidth}
        \includegraphics[width=1\linewidth]{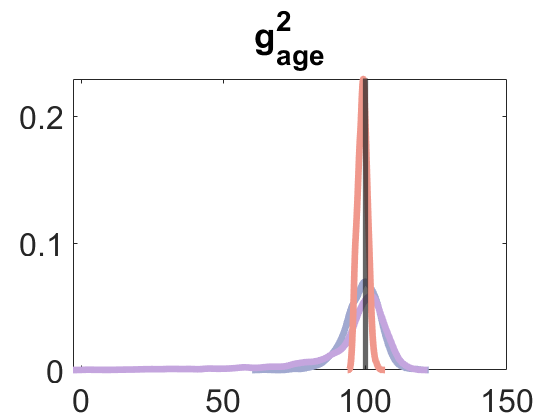}  
    \end{subfigure}
\end{subfigure}
\caption{\textbf{Univariate posterior estimates of the BVCBM parameters for a synthetic dataset:} The plots represent the estimated posterior distributions of parameters \(g^1_{\text{age}}\),  \(\tau\) and \(g^2_{\text{age}}\) across different methods. The violet lines show the SMC ABC's approximate posterior distributions, the purple lines show the approximations by SNPE and the orange lines show the approximations by SNLE. The black vertical lines are the true parameter values used to generate this synthetic dataset.}
\label{bvcbm vaidate synthetic dataset}
\end{figure}

Figure \ref{bvcbm vaidate synthetic dataset} shows the approximated marginal posterior distributions for three key model parameters, $g_{\mathrm{age}}^1$, $\tau$ and $g_{\mathrm{age}}^2$, from each SBI algorithm on a synthetic dataset. The full results for all synthetic datasets we used are presented in Section \textcolor{blue}{C1} of the Supplementary document. It is evident that most algorithms perform well for BVCBM. 

For BSL, we find the algorithm fail to work. To investigate whether the normality assumption of BSL may be reasonable for this example, we generate 100,000 simulations from the model at a parameter value favourable for dataset 1 and 3 (taken from \citet{wang2024calibration} and listed in Section \textcolor{blue}{B2} of the Supplementary document).  The marginal distributions for some of the summaries as shown in Figure \ref{BVCBM model summaries} and for all the summaries in Section \textcolor{blue}{B2} of the Supplementary document.  It can be seen that the model summaries are highly non-normal for dataset 3 (Figure \ref{BVCBM model summaries 3}), with several of the summaries showing strong multimodality. For dataset 1, a few of the model summaries show strong multimodality (Figure \ref{BVCBM model summaries 1}). For this reason, BSL cannot be expected to perform well in this example, so we do not consider it. There are various extensions of BSL that aim to relax the normality assumption \citep{fasiolo2018extended,an2020robust} that could be explored for this example, but that is not within the scope of this present paper.

\begin{figure}[h]
\begin{subfigure}{1\textwidth}
    \caption{ }
    \begin{subfigure}{.32\textwidth}
        \includegraphics[width=1\linewidth]{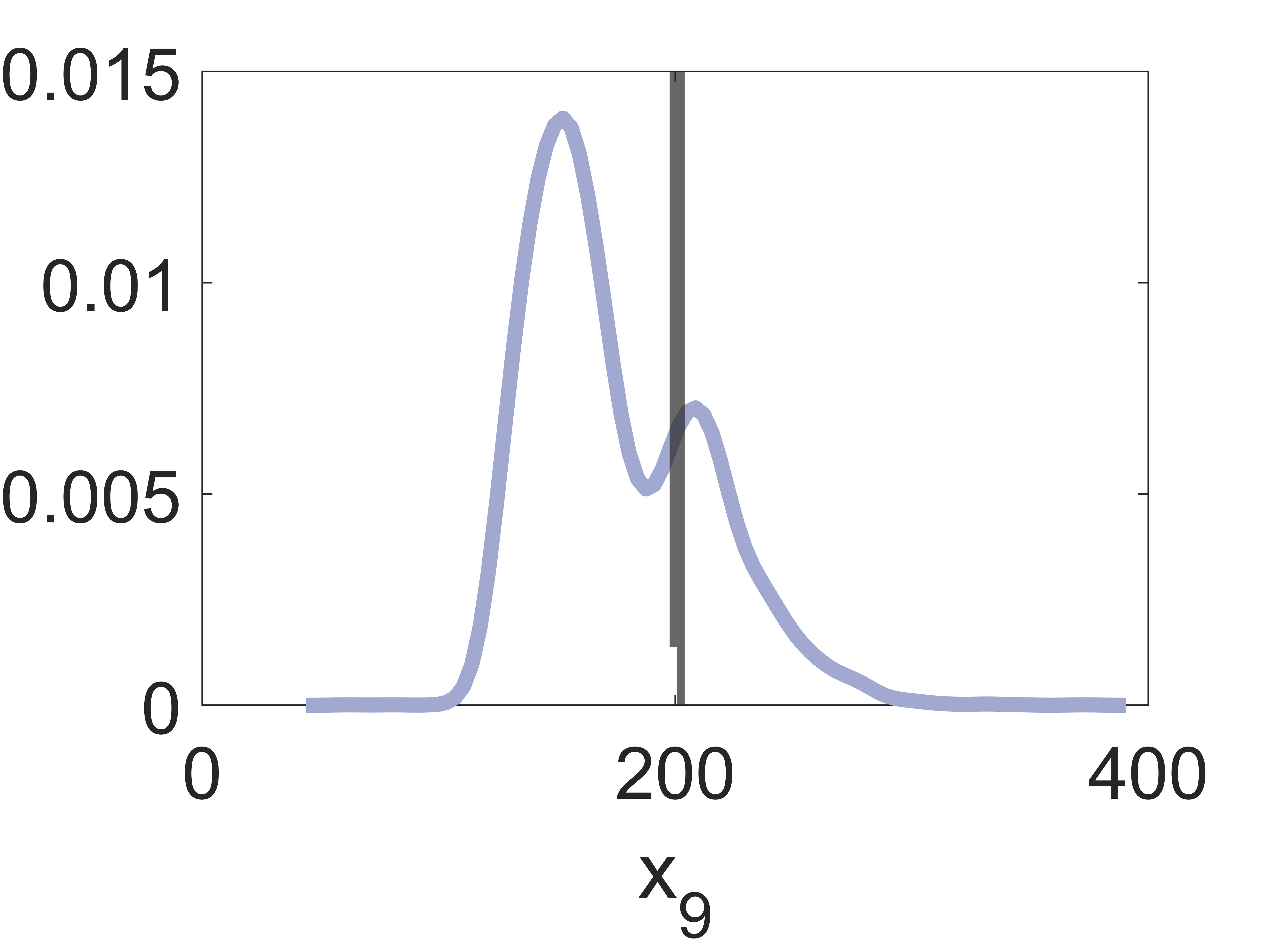}
    \end{subfigure}
    \hfill
    \begin{subfigure}{.32\textwidth}
        \includegraphics[width=1\linewidth]{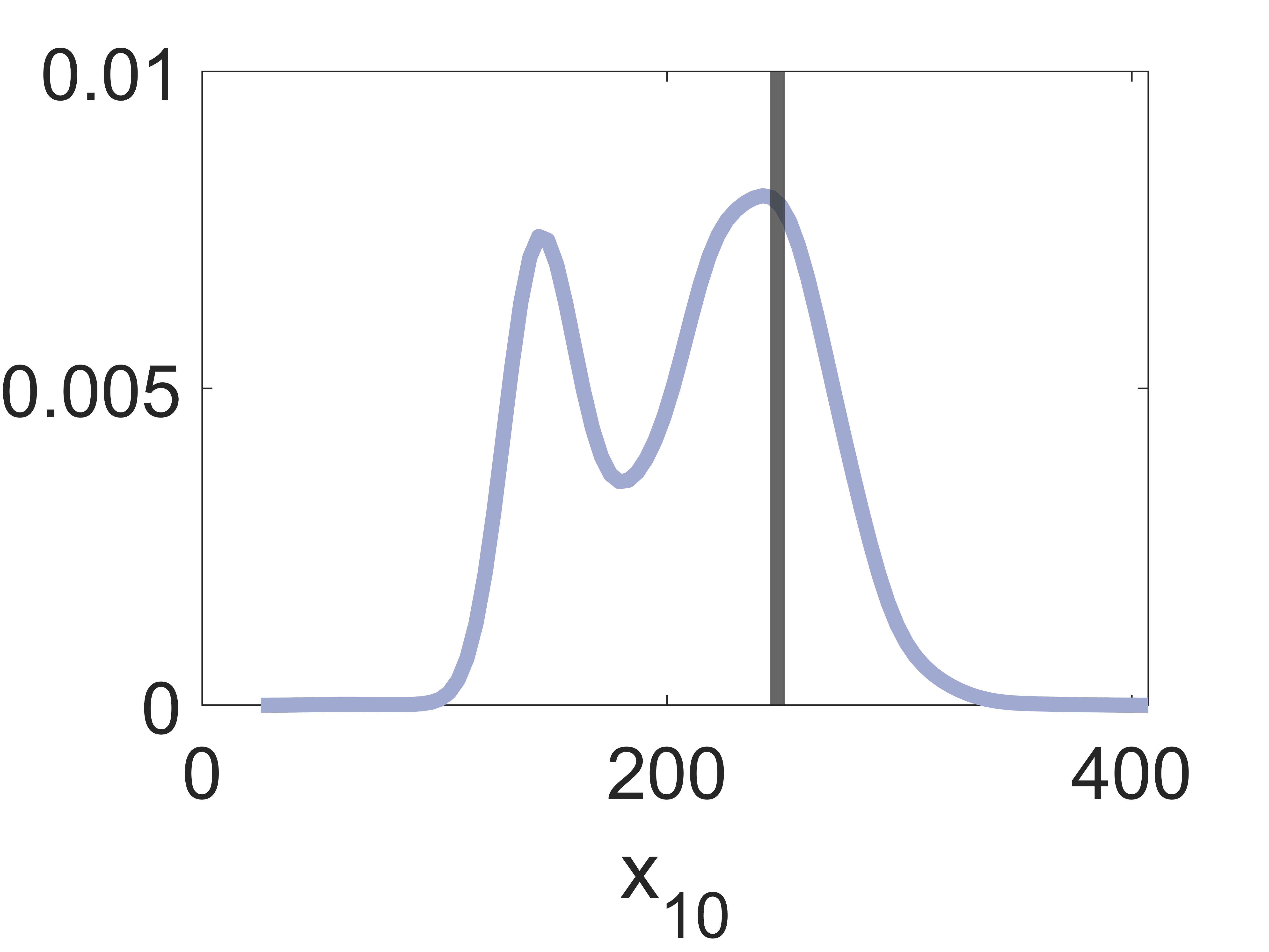}
    \end{subfigure}
    \hfill
    \begin{subfigure}{.32\textwidth}
        \includegraphics[width=1\linewidth]{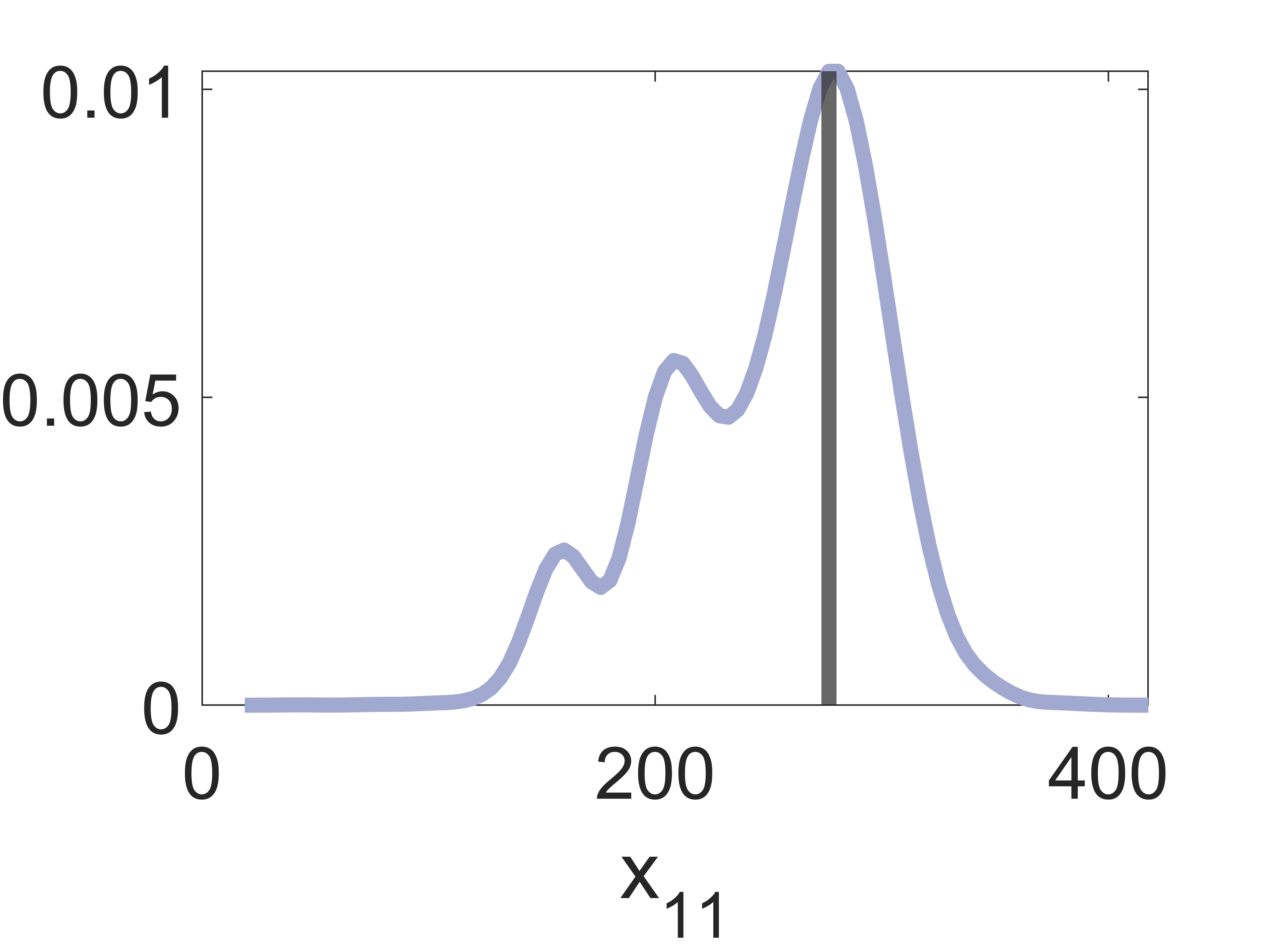}  
    \end{subfigure}
    \label{BVCBM model summaries 1}
\end{subfigure}
\vfill
\begin{subfigure}{1\textwidth}
\caption{}
    \begin{subfigure}{.32\textwidth}
        \includegraphics[width=1\linewidth]{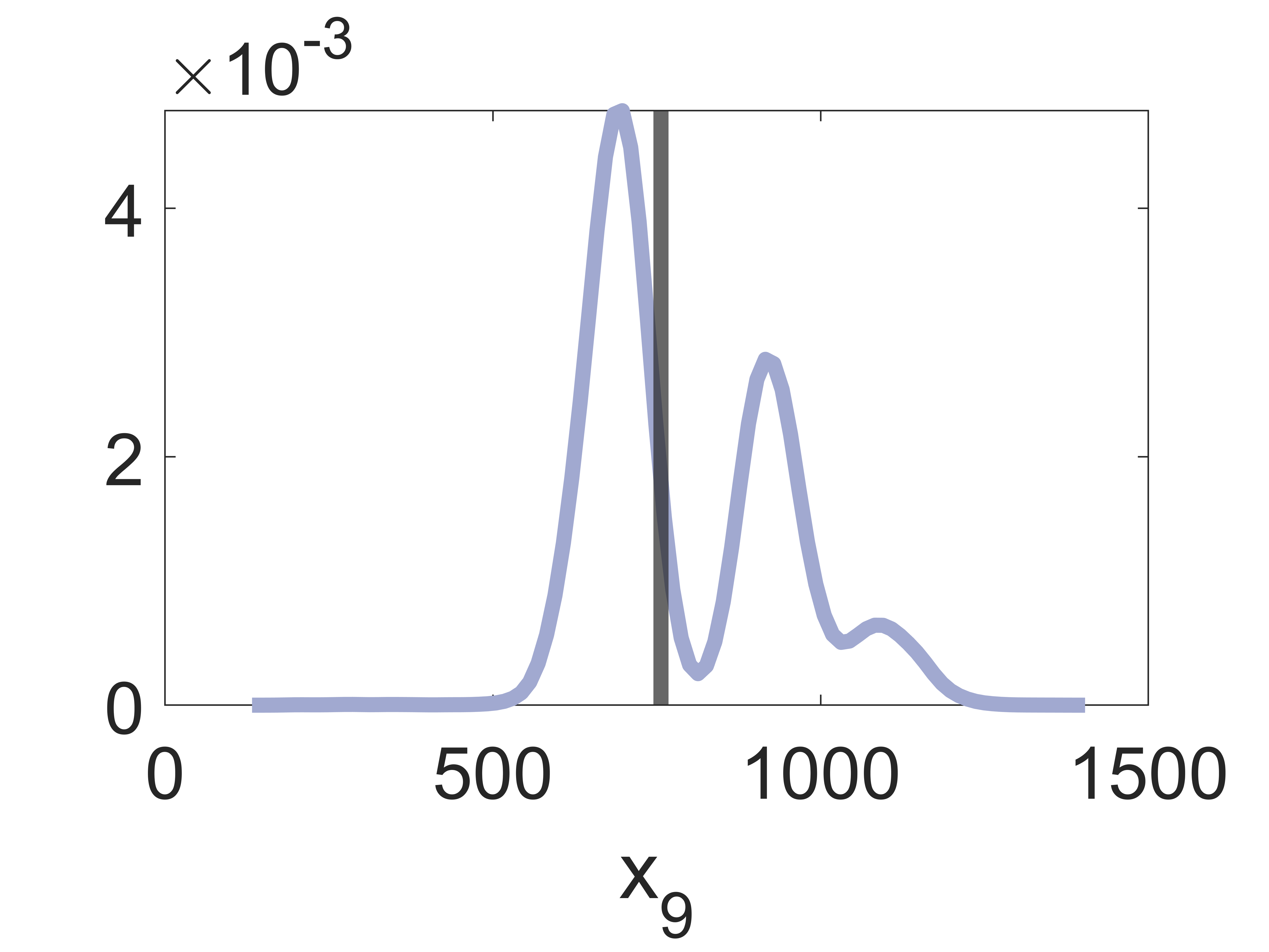}
    \end{subfigure}
    \hfill
    \begin{subfigure}{.32\textwidth}
        \includegraphics[width=1\linewidth]{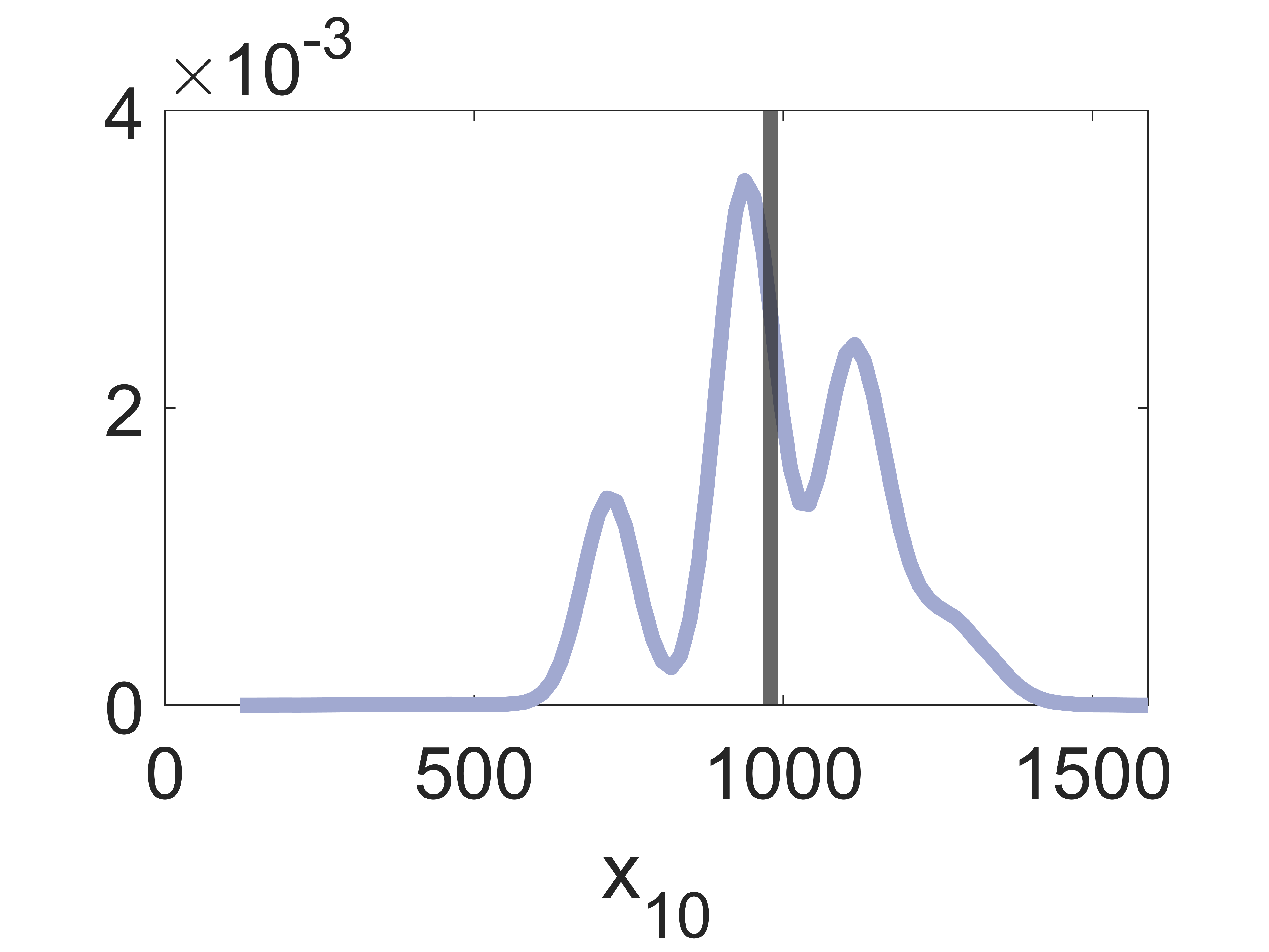}
    \end{subfigure}
    \hfill
    \begin{subfigure}{.32\textwidth}
        \includegraphics[width=1\linewidth]{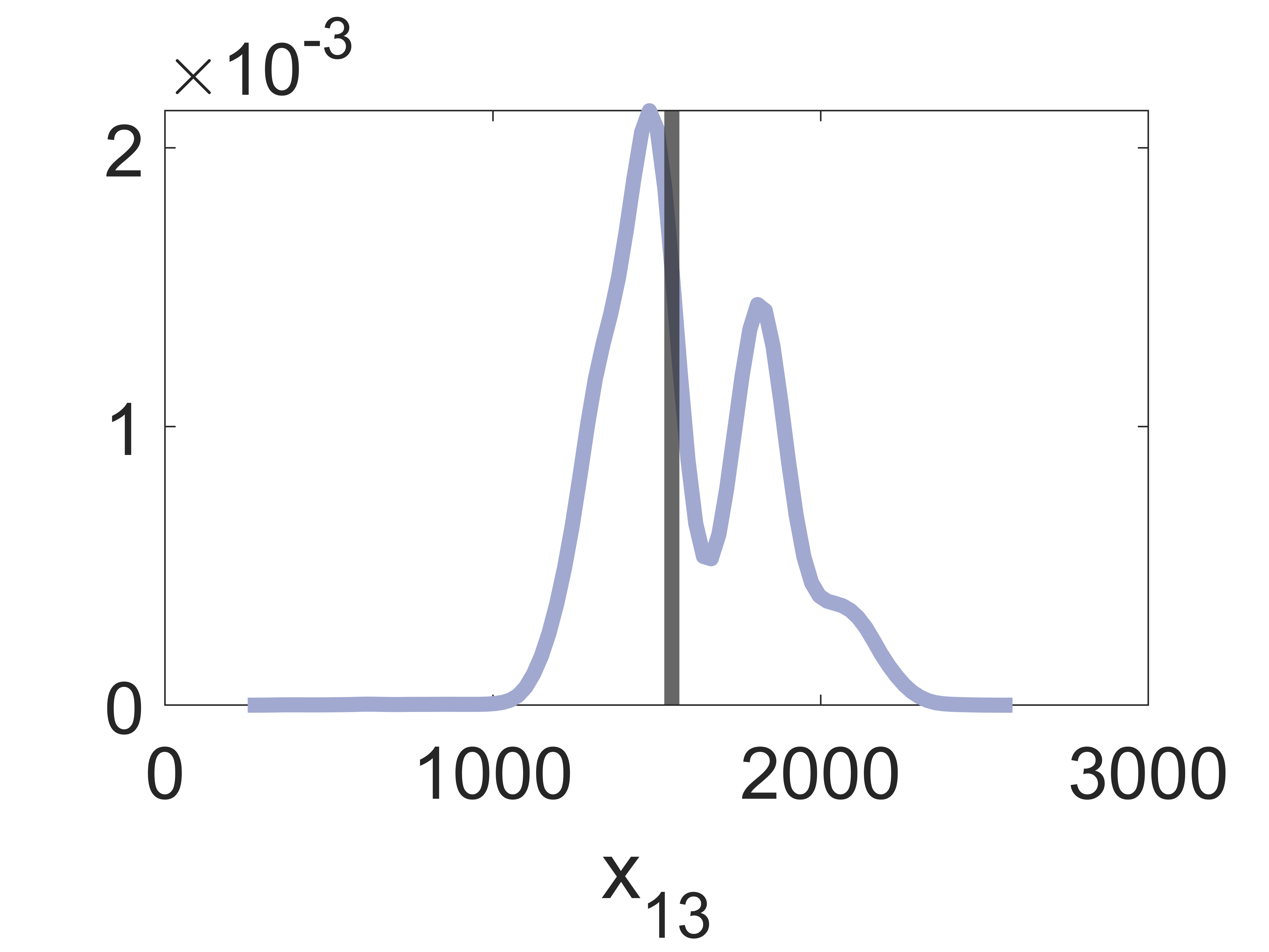}  
    \end{subfigure}
    \label{BVCBM model summaries 3}
\end{subfigure}
\caption{ \textbf{Marginal distribution for some of the model summaries of two datasets generated from BVCBM:} (a) three marginal distributions of model summaries in dataset 1, and, (b) three marginal distributions of model summaries in dataset 3.}
\label{BVCBM model summaries}
\end{figure}

\begin{figure}[H]
\begin{subfigure}{1\textwidth}
    \begin{subfigure}{.32\textwidth}
    \caption{ }
        \includegraphics[width=1\linewidth]{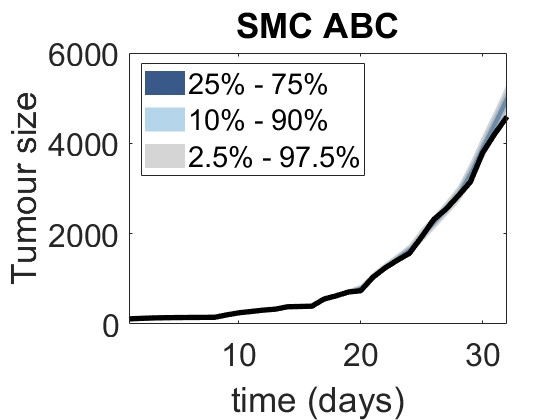}
    \end{subfigure}
    \hfill
    \hfill
    \begin{subfigure}{.32\textwidth}
    \caption{ }
        \includegraphics[width=1\linewidth]{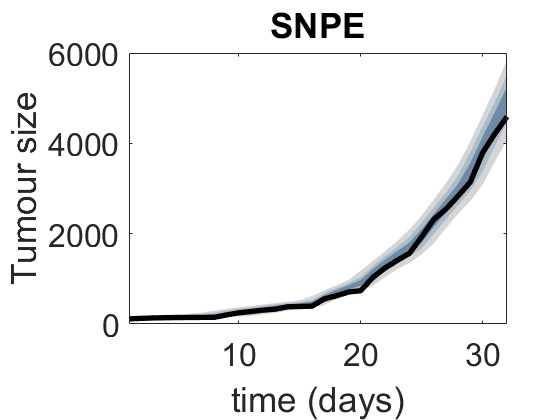}  
    \end{subfigure}
    \hfill
    \begin{subfigure}{.32\textwidth}
    \caption{ }
        \includegraphics[width=1\linewidth]{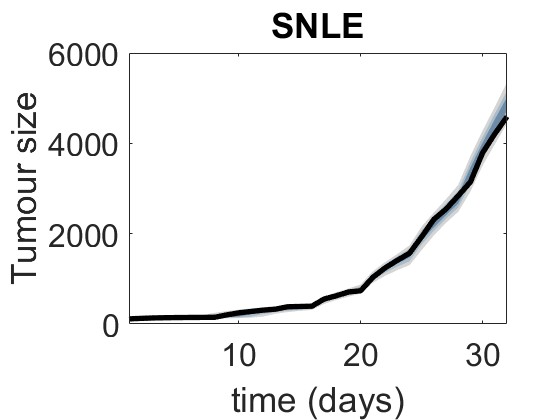}  
    \end{subfigure}
\end{subfigure}
\caption{ \textbf{The posterior predictive plots for three SBI algorithms on synthetic dataset $y$:} (a)-(c) refers to SMC ABC, SNPE and SNLE results on synthetic dataset with 32 measurement days and parameters values $(g^1_{\text{age}},\tau,g^2_{\text{age}}) = (300,16,100)$. }
\label{BVCBM synthetic data calibration}
\end{figure}

In Figure \ref{BVCBM synthetic data calibration}, we present the posterior predictive distribution for the synthetic dataset $y$ based on all three SBI algorithms. It is evident that SMC ABC and SNL perform better than SNPE for all synthetic datasets (SNPE perform poorly in Figure \textcolor{blue}{S 11} in Section \textcolor{blue}{C 1.1} of the Supplementary document). Although the observation dataset lies within the $(25\%, 75\%)$ predictive interval for SNPE for synthetic dataset 1 and 2, it fails to work for synthetic dataset 3. 

\subsubsection{Stage 2: SBI stage}

We recorded the computational cost for each algorithm on synthetic dataset for BVCBM. All neural SBI algorithms (SNPE and SNLE) used 100k total simulations. SMC ABC required around 500k to 700k total simulations for the three synthetic datasets in BVCBM. Considering both the posterior predictive distribution and computational cost, we chose SMC ABC and SNL as the candidate SBI algorithms for BVCBM with real-world pancreatic tumour growth datasets.

For the BVCBM with pancreatic tumour growth datasets, the robust versions of these algorithms should be implemented to reduce the effect of model misspecification. Based on \citet{frazier2020model}, ABC and its variants are robust, so we use them directly. For SNL, we implement the robust version (RSNL), where the adjustment method is used for correcting any potential misspecification. The inference results can be obtained by running the candidate SBI algorithms on real datasets.

\subsubsection{Stage 3: uncertainty analysis stage}

\begin{figure}[H]
\begin{subfigure}{1\textwidth}
    \begin{subfigure}{.32\textwidth}
    \caption{ }
        \includegraphics[width=1\linewidth]{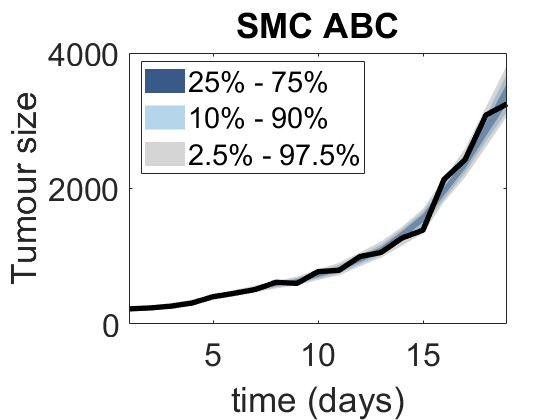}  
    \end{subfigure}
    \hfill
    \begin{subfigure}{.32\textwidth}
    \caption{ }
        \includegraphics[width=1\linewidth]{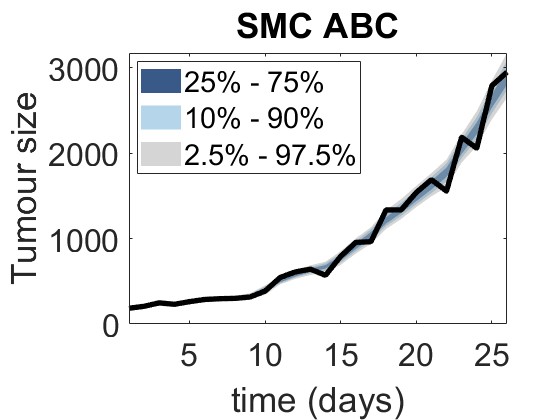}  
    \end{subfigure}
    \hfill
    \begin{subfigure}{.32\textwidth}
    \caption{ }
        \includegraphics[width=1\linewidth]{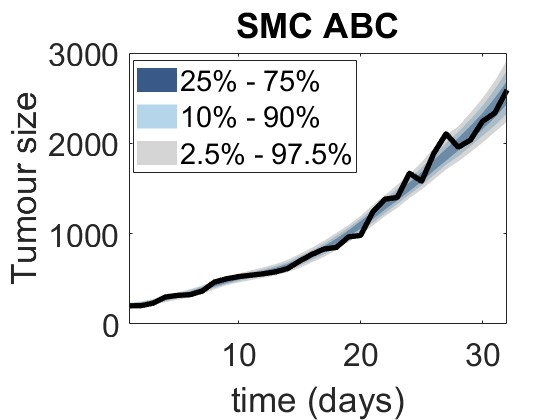}  
    \end{subfigure}
\end{subfigure}
\vfill
\begin{subfigure}{1\textwidth}
    \begin{subfigure}{.32\textwidth}
    \caption{ }
        \includegraphics[width=1\linewidth]{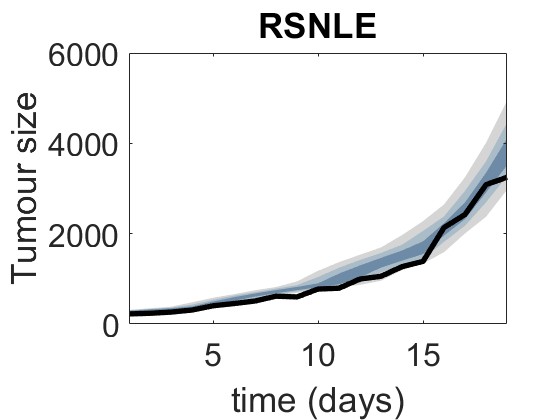}  
    \end{subfigure}
    \hfill
    \begin{subfigure}{.32\textwidth}
    \caption{ }
        \includegraphics[width=1\linewidth]{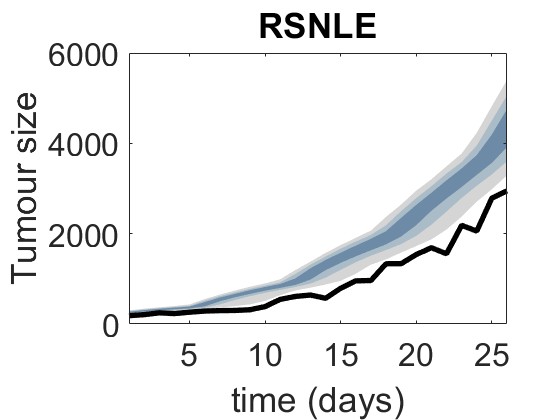}  
    \end{subfigure}
        \hfill
    \begin{subfigure}{.32\textwidth}
    \caption{ }
        \includegraphics[width=1\linewidth]{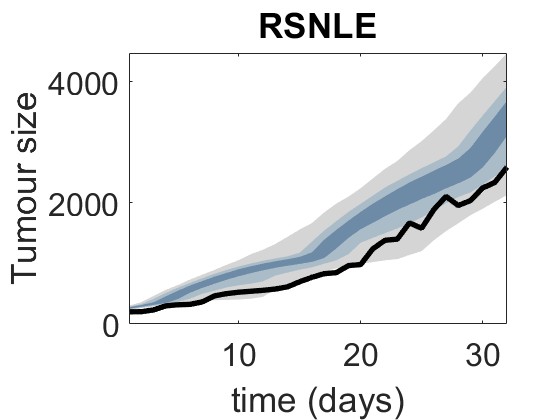}  
    \end{subfigure}
\end{subfigure}
\caption{ \textbf{The posterior predictive plots for candidate SBI algorithms on the pancreatic tumour datasets:} (a)-(c) refers to SMC ABC results on pancreatic tumour datasets with 19, 26 and 32 days, respectively. (e)-(f) refers to RSNL results on pancreatic tumour datasets with 19, 26 and 32 days, respectively.}
\label{BVCBM real data calibration}
\end{figure}

For BVCBM, the real-world pancreatic tumour growth datasets contain noise, meaning the growth data do not always increase consistently. Although cell proliferation and movement are described by stochastic processes, the model only considers exponential growth, which can lead to potential model misspecification. In Figure \ref{BVCBM real data calibration}\textcolor{blue}{(a)-(c)}, SMC ABC performs reasonably well for the first two pancreatic datasets but slightly worse on the third one, as the third dataset does not exhibit strong biphasic growth. As suggested by \textbf{Guideline 6}, the results of SMC ABC indicate that the BVCBM is compatible with all three pancreatic tumour growth datasets.

We note that neural SBI methods may not perform well under model misspecification, particularly when the observed dataset displays features that are different to the simulated datasets used to train the NCDE.  The real tumour growth datasets exhibit substantially more noise than what is generated from the BVCBM.  As can be seen from Figures \ref{BVCBM real data calibration}\textcolor{blue}{(e)} and \ref{BVCBM real data calibration}\textcolor{blue}{(f)}, even the robust version of SNLE (i.e. the version that includes adjustment parameters that aim to soak up the misspecification) fails to produce accurate predictions. The first pancreatic tumour growth dataset in \ref{BVCBM real data calibration}\textcolor{blue}{(e)} is smoother than the other two datasets, allowing RSNL to provide a better fit. 

We highlight the drawbacks of the SNPE method and explain why we did not choose it for inference tasks on real datasets, even though SNPE performs well on synthetic datasets. NPE is an amortized method that fully depends on neural networks and performs sampling based on trained feedforward networks. In such cases, it is sensitive to the values of the random seed and often requires averaging the results based on multiple random seed values, which significantly increases computational resource requirements. Additionally, neural networks tend to struggle when the training datasets are highly noisy, which might cause SNPE to fail in learning. For more details on the limitations of SNPE, we refer to \citet{wang2024preconditioned}.

\subsection{Example 2: Stochastic Cell Invasion Model}

\subsubsection{Stage 1: pre-analysis stage}
\begin{figure}[H]
    \begin{subfigure}{.49\textwidth}
    \caption{Cell Trajectory}
        \includegraphics[width=1\linewidth]{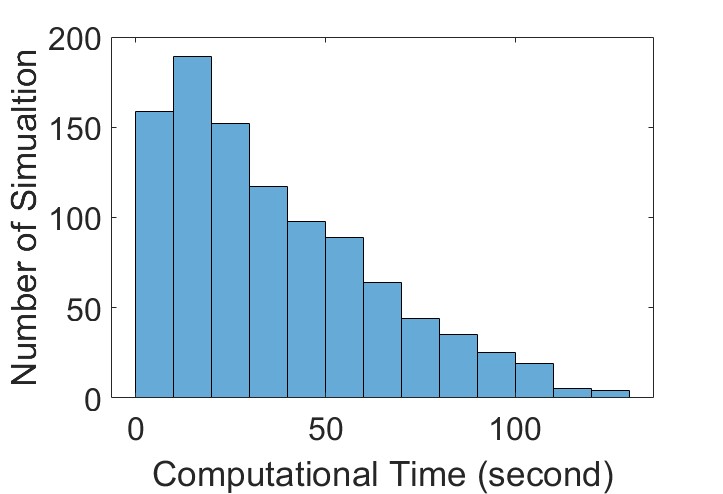}
    \end{subfigure}
    \hfill
    \begin{subfigure}{.49\textwidth}
    \caption{Cell Density}
        \includegraphics[width=1\linewidth]{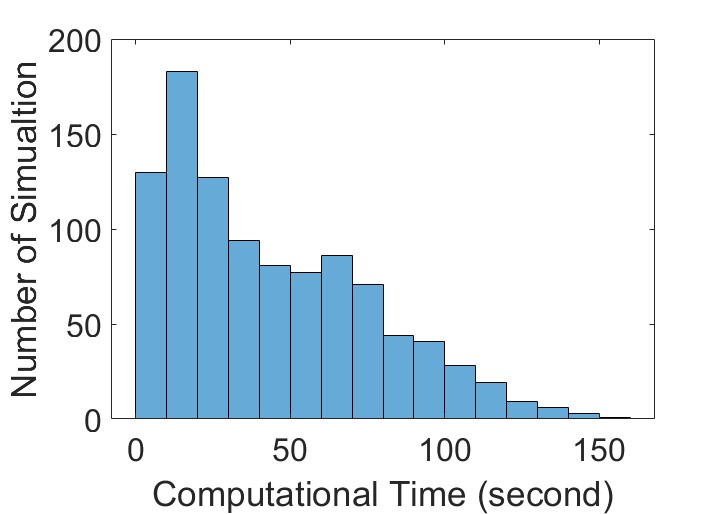}  
    \end{subfigure}
    \caption{\textbf{Computational time analysis for stochastic cell invasion model}. (a) The histogram of computational time of 1000 simulated summary statistics of cell trajectories; (b) The histogram of computational time of 1000 simulated summary statistics of cell density.}
\label{cim computational analysis}
\end{figure}

We use the same configuration for implementation as in Example \ref{BVCBM example}, but with the summary statistics of the real datasets. Following \textbf{Guideline 1}, we first investigate the computational cost. In Figure \ref{cim computational analysis}, we present the histogram of computational costs for 1000 simulated datasets using the stochastic cell invasion model. We report a computational cost ranging from 1.40 to 129.22 seconds per simulation when using cell trajectory as summary statistics, and from 0.94 to 168.91 seconds per simulation when using cell density as summary statistics for cell movement. The summary statistics for cell proliferation are the counts of cells in each cell cycle stage. For convenience, we refer to the summary statistics vector that combines the summaries of cell proliferation and cell movement using cell trajectories as cell trajectory. Similarly, we refer to the summary vector that combines the summaries of cell proliferation and cell movement using cell densities as cell density.

\begin{figure}[H]
\begin{subfigure}{1\textwidth}
\caption{}
    \begin{subfigure}{.32\textwidth}
        \includegraphics[width=1\linewidth]{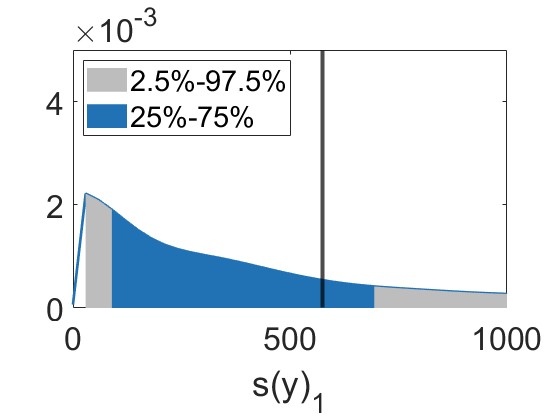}
    \end{subfigure}
    \hfill
    \begin{subfigure}{.32\textwidth}
        \includegraphics[width=1\linewidth]{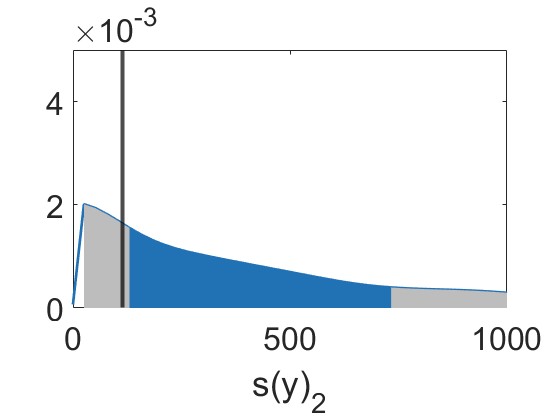}  
    \end{subfigure}
    \hfill
    \begin{subfigure}{.32\textwidth}
        \includegraphics[width=1\linewidth]{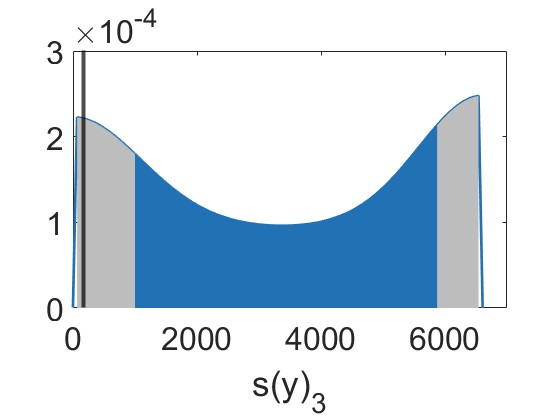}  
    \end{subfigure}
    \vfill
    \begin{subfigure}{.32\textwidth}
        \includegraphics[width=1\linewidth]{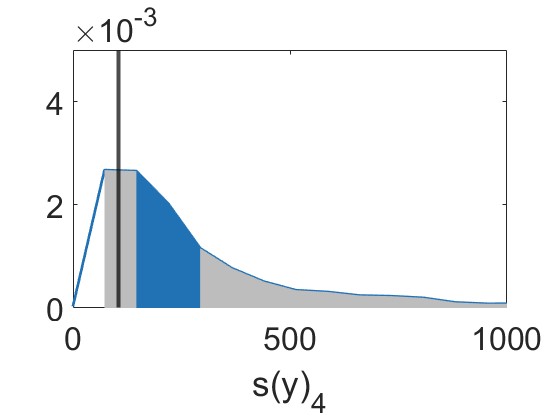}  
    \end{subfigure}
    \hfill
    \begin{subfigure}{.32\textwidth}
        \includegraphics[width=1\linewidth]{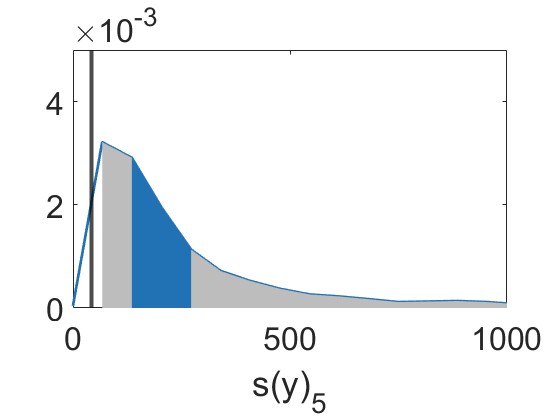}  
    \end{subfigure}
    \hfill
    \begin{subfigure}{.32\textwidth}
        \includegraphics[width=1\linewidth]{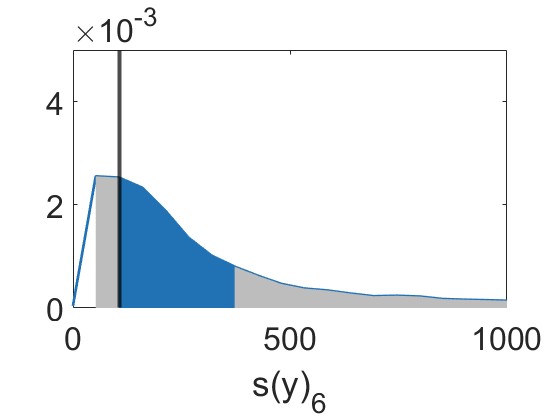}  
    \end{subfigure}
    \label{cim prior predictive check tracking}
\end{subfigure}
\vfill 
\begin{subfigure}{1\textwidth}
\caption{}
    \begin{subfigure}{.19\textwidth}
        \includegraphics[width=1\linewidth]{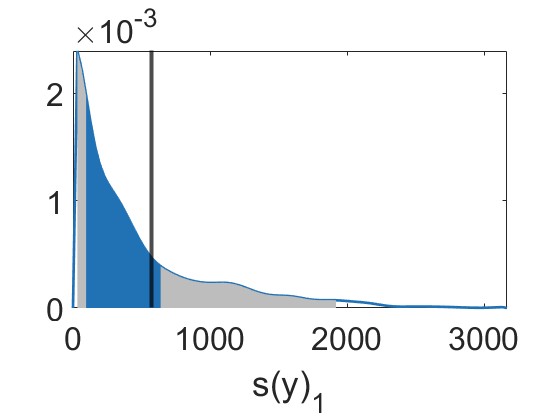}
    \end{subfigure}
    \hfill
    \begin{subfigure}{.19\textwidth}
        \includegraphics[width=1\linewidth]{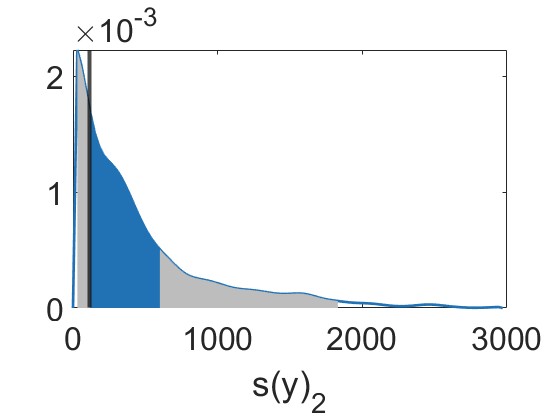}  
    \end{subfigure}
    \hfill
    \begin{subfigure}{.19\textwidth}
        \includegraphics[width=1\linewidth]{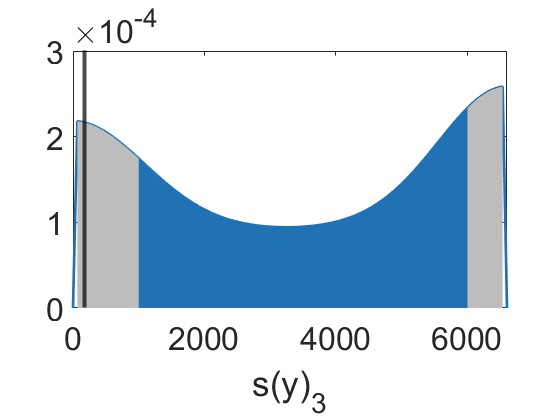}  
    \end{subfigure}
    \hfill
    \begin{subfigure}{.19\textwidth}
        \includegraphics[width=1\linewidth]{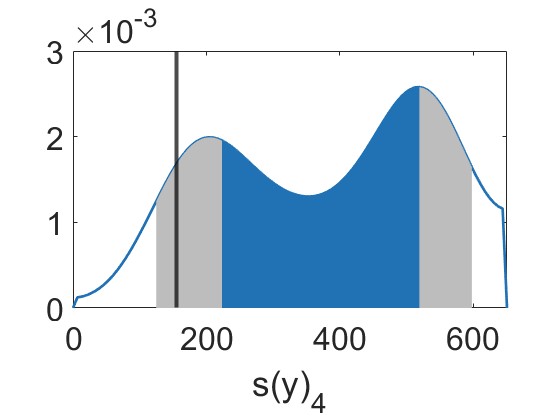}  
    \end{subfigure}
    \hfill
    \begin{subfigure}{.19\textwidth}
        \includegraphics[width=1\linewidth]{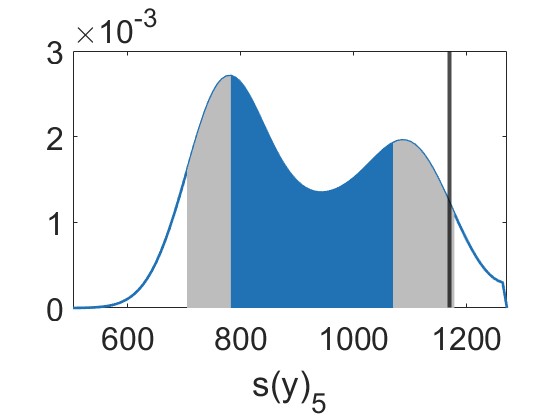}  
    \end{subfigure}
    \vfill
    \begin{subfigure}{.19\textwidth}
        \includegraphics[width=1\linewidth]{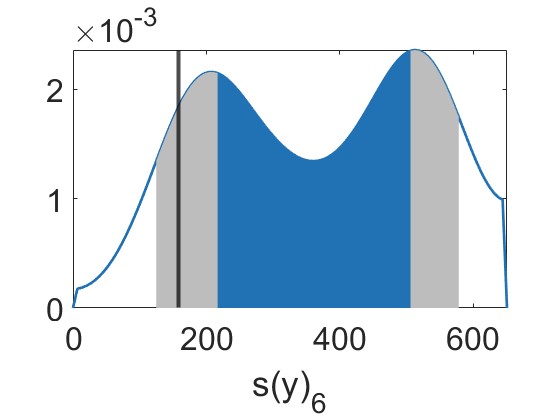}  
    \end{subfigure}
    \hfill
    \begin{subfigure}{.19\textwidth}
        \includegraphics[width=1\linewidth]{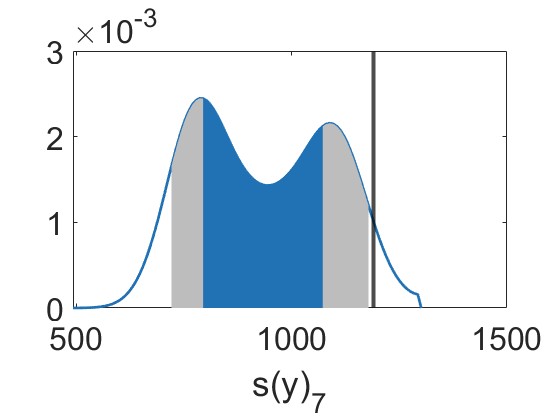}  
    \end{subfigure}
    \hfill
    \begin{subfigure}{.19\textwidth}
        \includegraphics[width=1\linewidth]{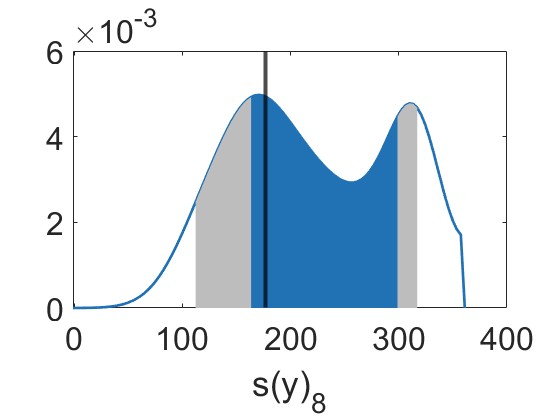}  
    \end{subfigure}
    \hfill
    \begin{subfigure}{.19\textwidth}
        \includegraphics[width=1\linewidth]{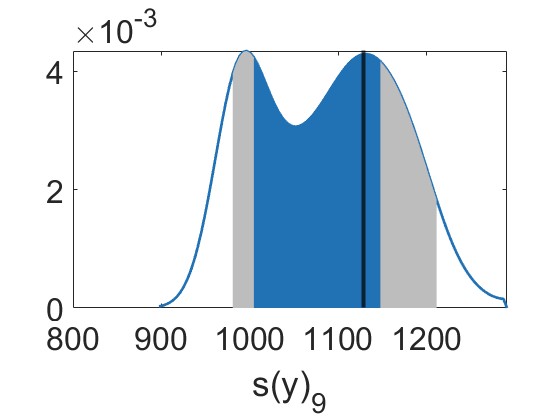}  
    \end{subfigure}
    \hfill
    \begin{subfigure}{.19\textwidth}
        \includegraphics[width=1\linewidth]{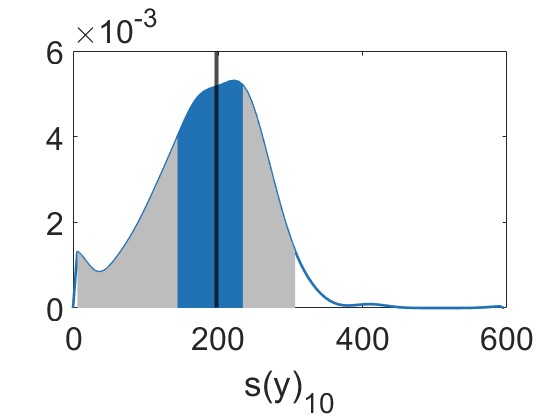}  
    \end{subfigure}
    \vfill
    \begin{subfigure}{.19\textwidth}
        \includegraphics[width=1\linewidth]{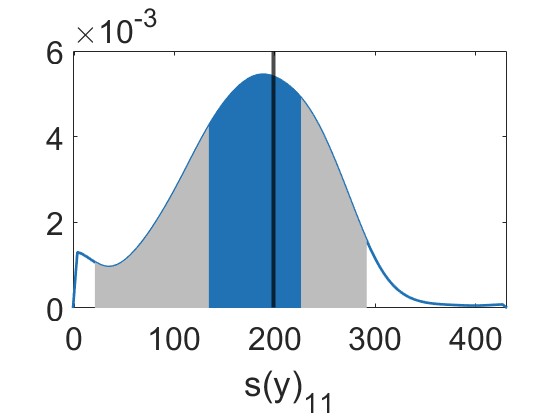}  
    \end{subfigure}
    \hfill
    \begin{subfigure}{.19\textwidth}
        \includegraphics[width=1\linewidth]{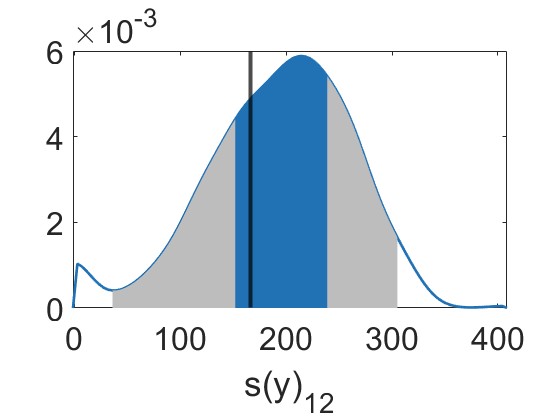}  
    \end{subfigure}
    \hfill
    \begin{subfigure}{.19\textwidth}
        \includegraphics[width=1\linewidth]{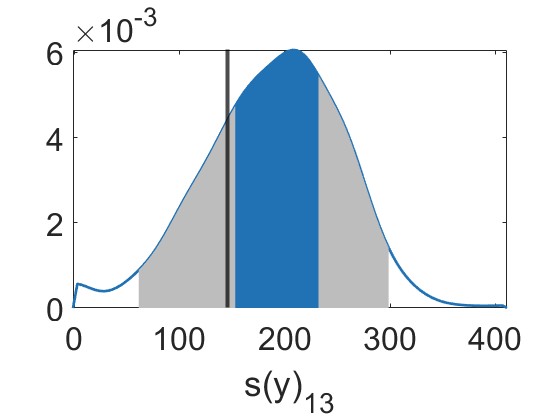}  
    \end{subfigure}
    \hfill
    \begin{subfigure}{.19\textwidth}
        \includegraphics[width=1\linewidth]{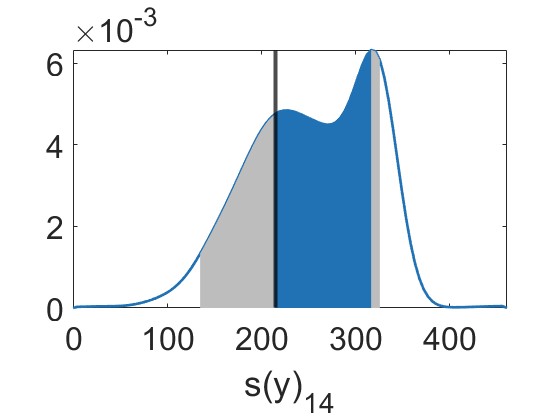}  
    \end{subfigure}
    \hfill
    \begin{subfigure}{.19\textwidth}
        \includegraphics[width=1\linewidth]{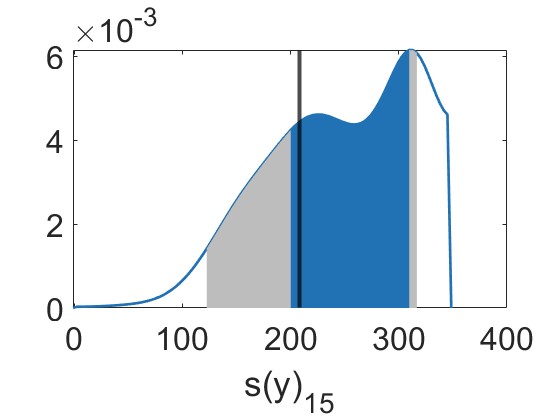}  
    \end{subfigure}
    \label{cim prior predictive check density}
\end{subfigure}
\caption{\textbf{Prior predictive check for stochastic cell invasion model}. The summary statistics of (a) cell trajectories, and (b) cell density for the real datasets. The blue areas represent the $(25\%-75\%)$ predictive interval and grey areas represent the $(2.5\%-97.5\%)$ predictive interval. The black vertical lines are the true values. }
\label{cim prior predictive check}
\end{figure}

Next, we perform prior predictive checks for both cell trajectory and cell density of the real datasets, as this can provide an initial indication of model misspecification, as stated in \textbf{Guideline 2}. In Figure \ref{cim prior predictive check}, the observed values of the summary statistics for cell trajectories (represented by black vertical lines) mostly lie within the 2.5\%-97.5\% predictive interval (Figure \ref{cim prior predictive check tracking}), as do the summary statistics for cell density (Figure \ref{cim prior predictive check density}).

\begin{figure}[H]
\begin{subfigure}{1\textwidth}
\caption{}
    \begin{subfigure}{.32\textwidth}
        \includegraphics[width=1\linewidth]{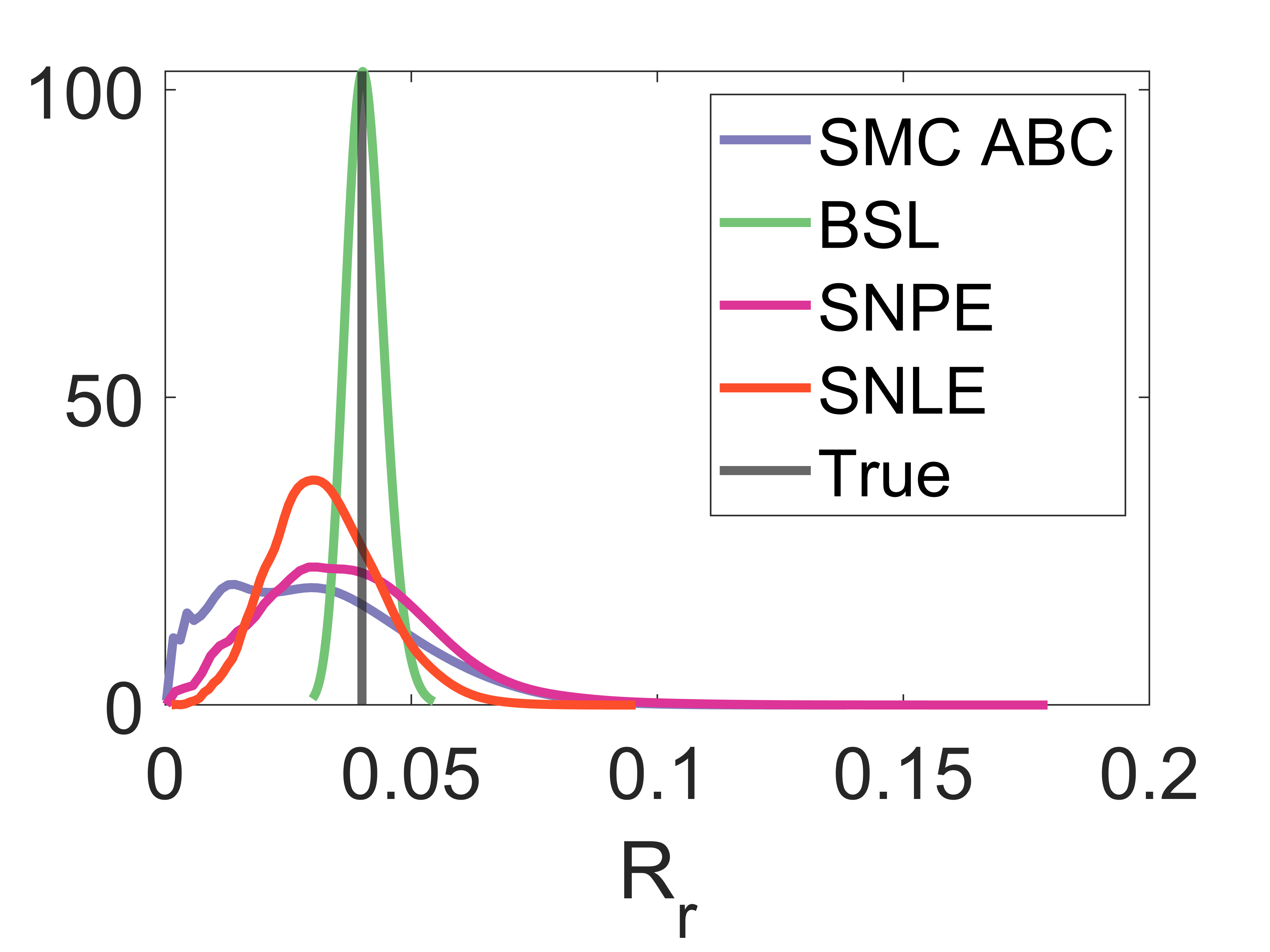}
    \end{subfigure}
    \hfill
    \begin{subfigure}{.32\textwidth}
        \includegraphics[width=1\linewidth]{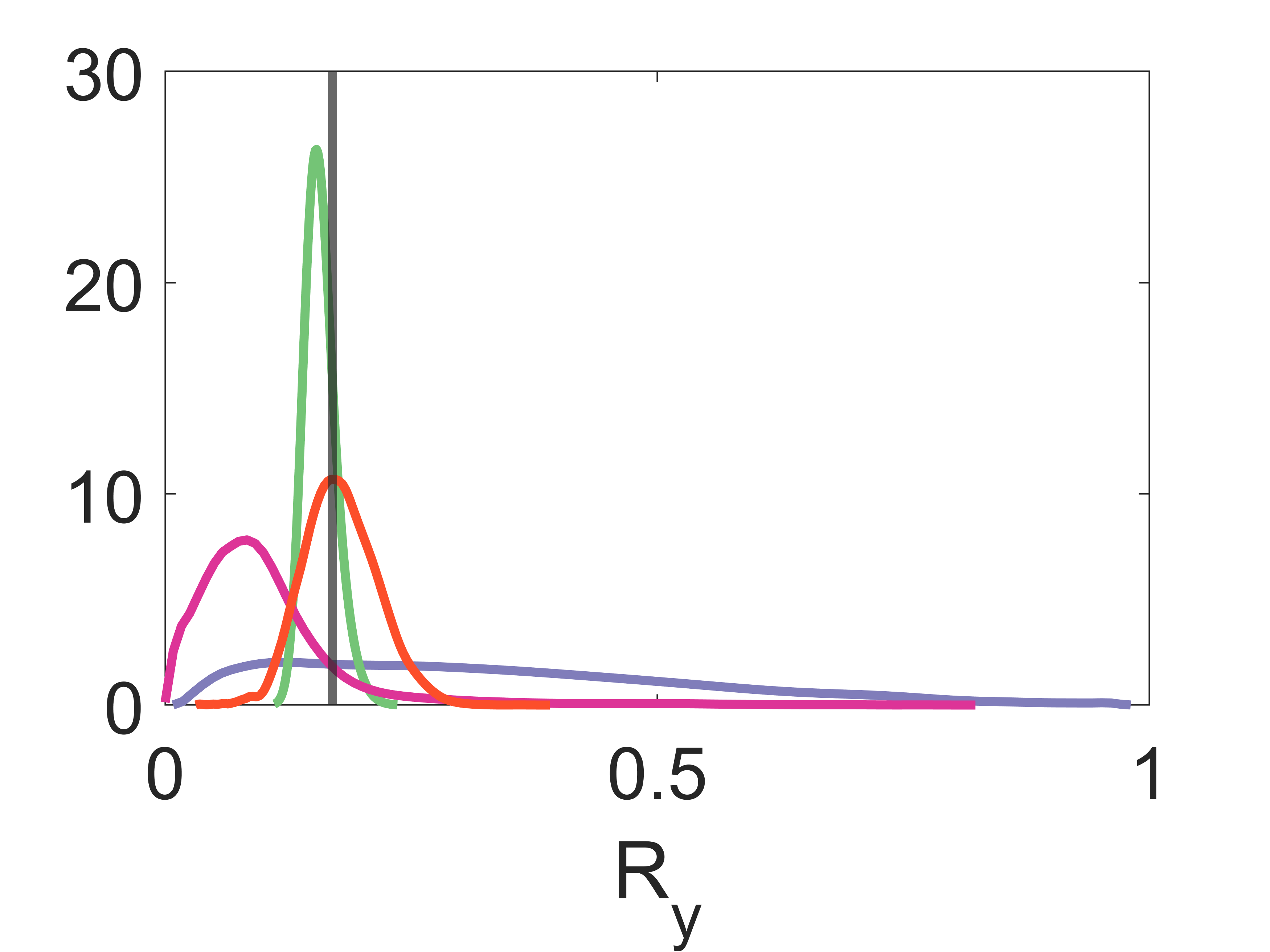}  
    \end{subfigure}
    \hfill
    \begin{subfigure}{.32\textwidth}
        \includegraphics[width=1\linewidth]{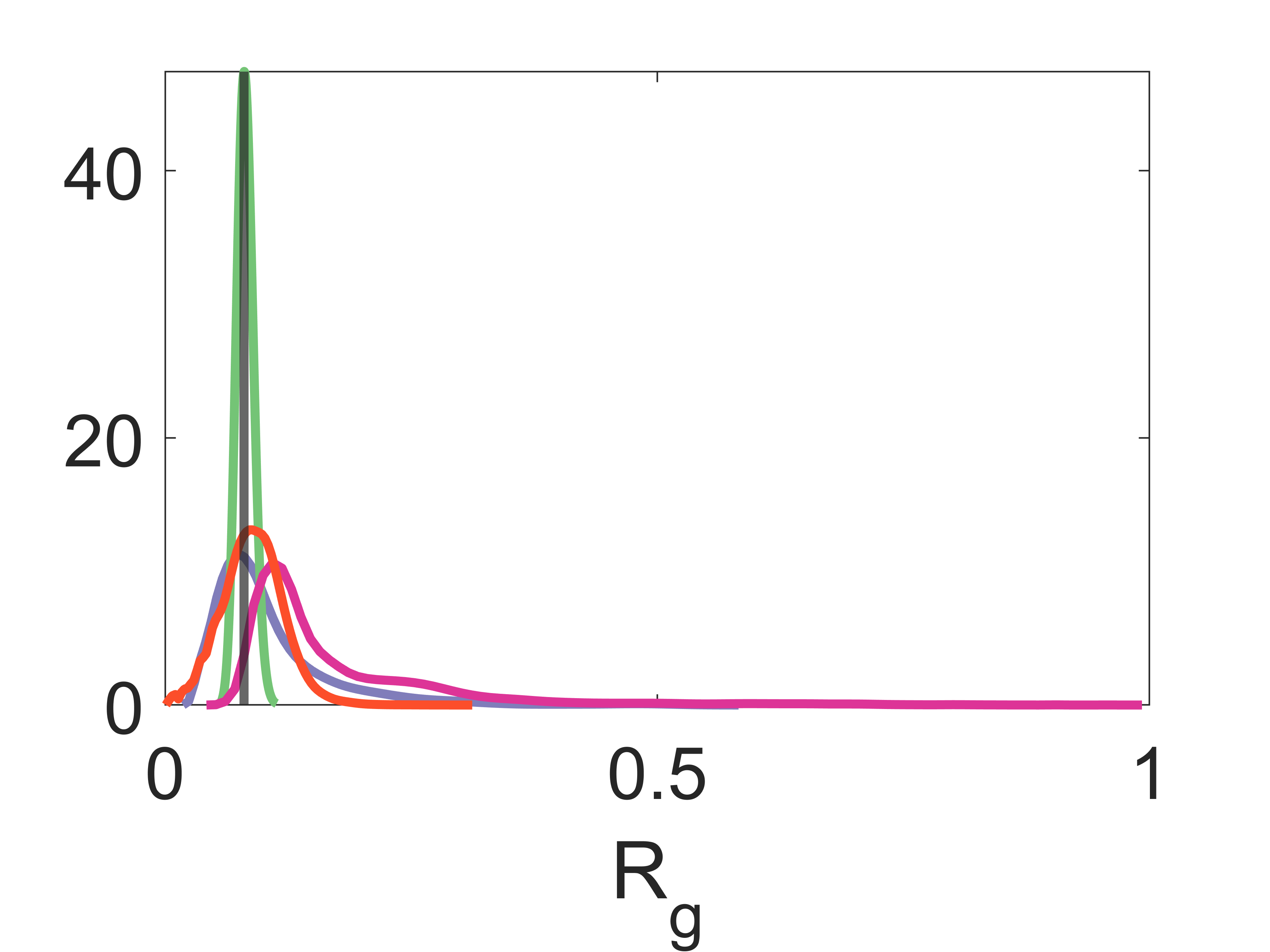}  
    \end{subfigure}
    \vfill
    \begin{subfigure}{.32\textwidth}
        \includegraphics[width=1\linewidth]{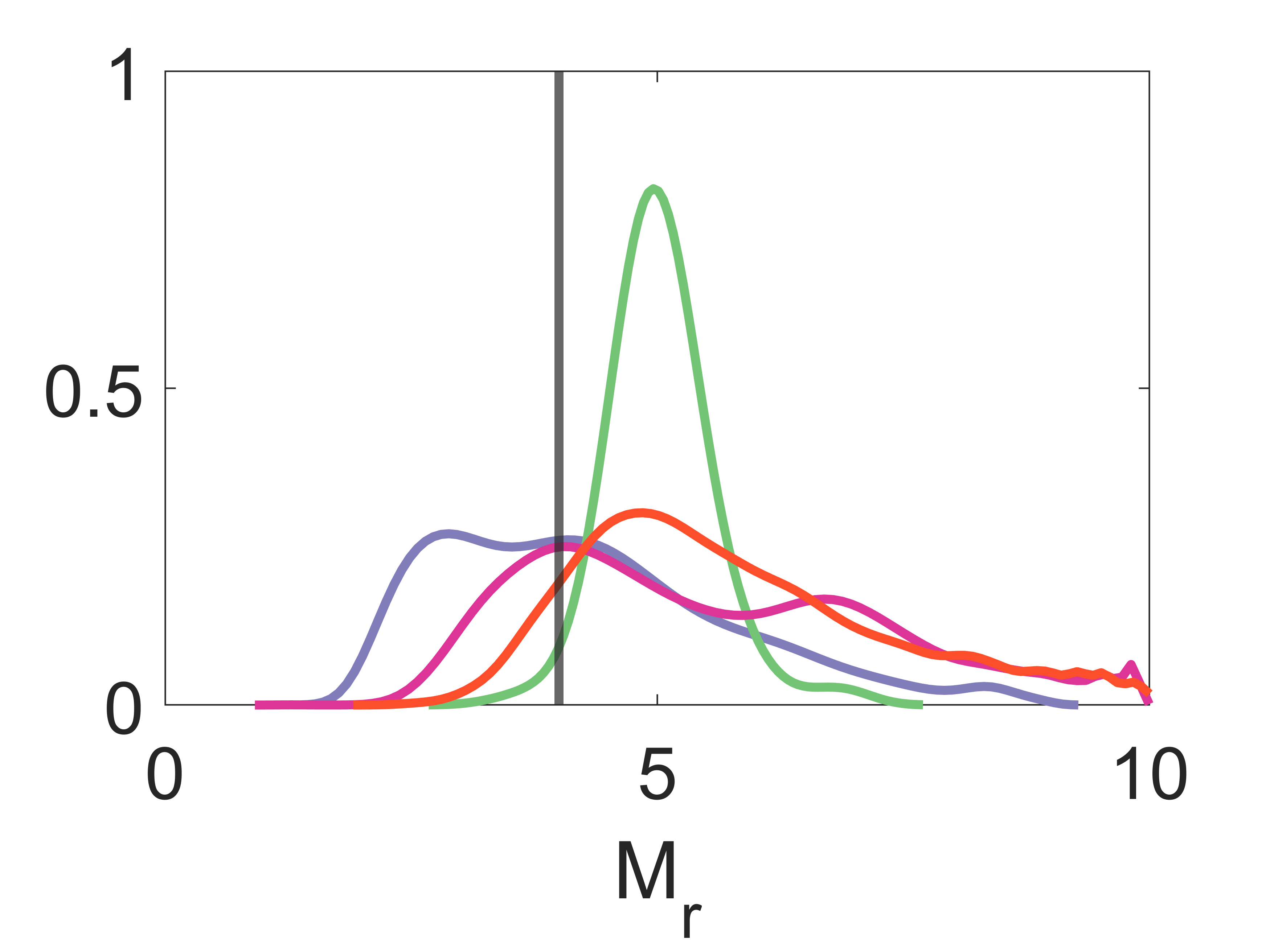}  
    \end{subfigure}
    \hfill
    \begin{subfigure}{.32\textwidth}
        \includegraphics[width=1\linewidth]{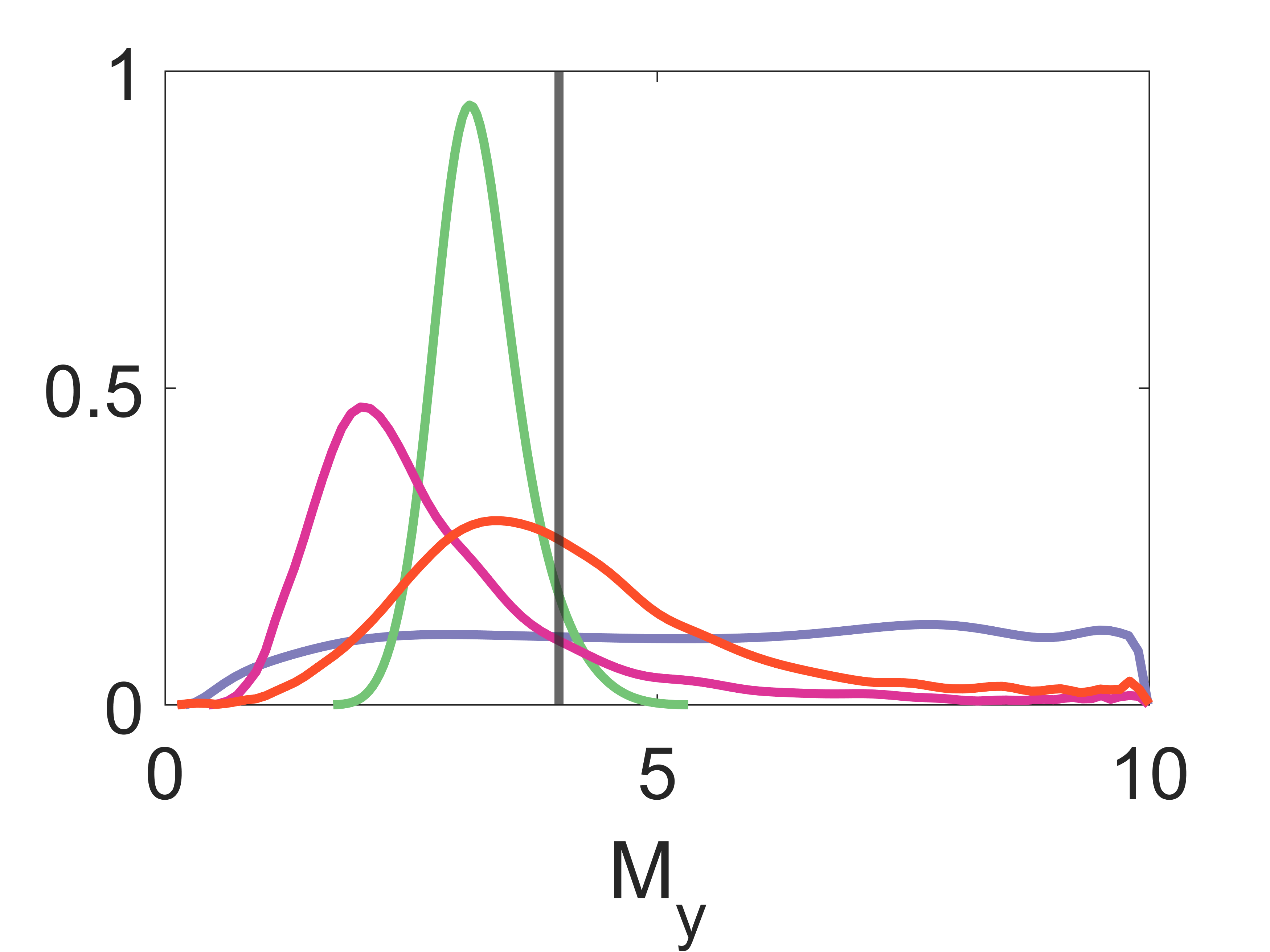}  
    \end{subfigure}
    \hfill
    \begin{subfigure}{.32\textwidth}
        \includegraphics[width=1\linewidth]{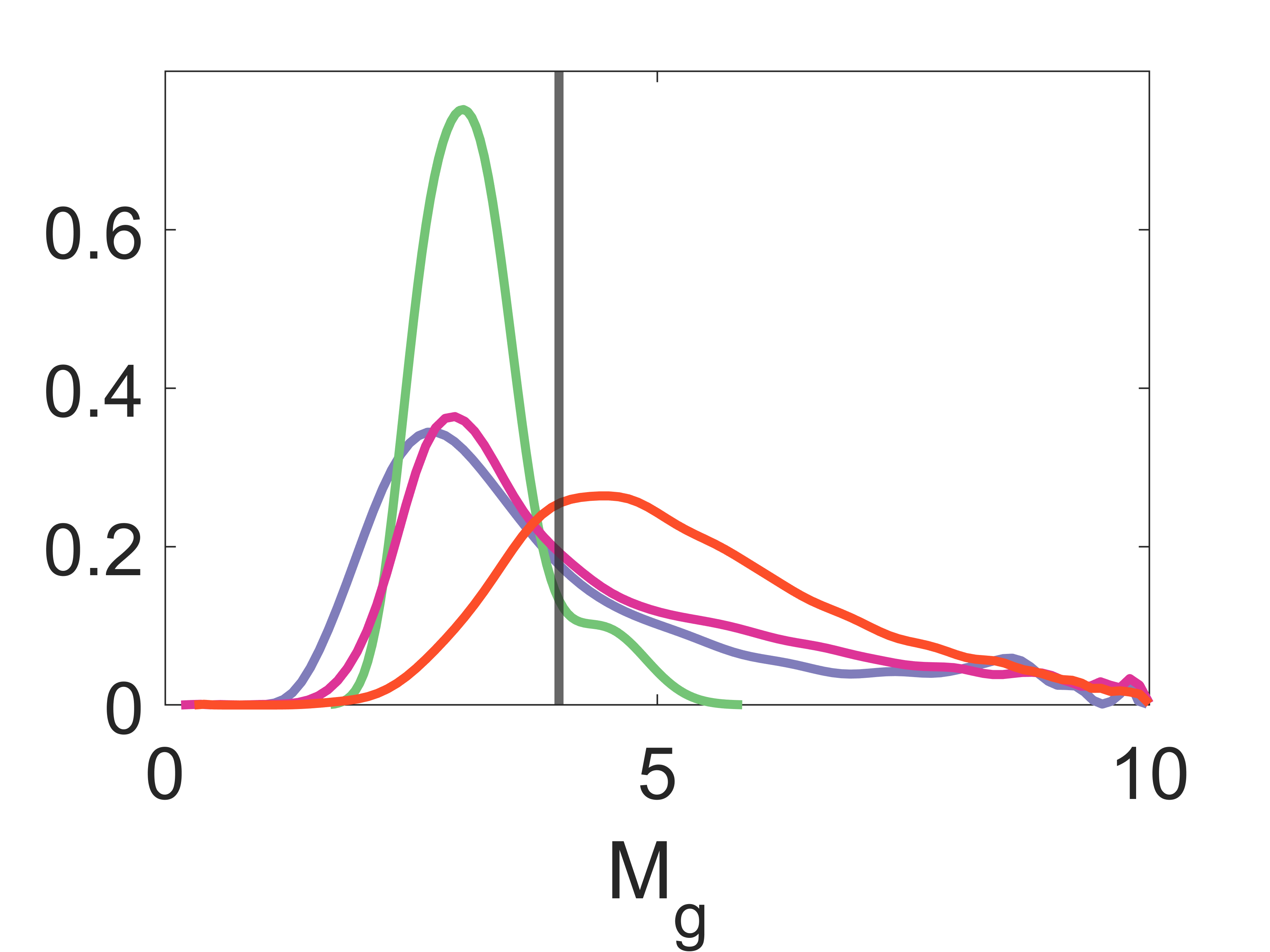}  
    \end{subfigure}
    \label{tracking synthetic marginal posterior}
\end{subfigure}
\vfill
\begin{subfigure}{1\textwidth}
\caption{}
    \begin{subfigure}{.32\textwidth}
        \includegraphics[width=1\linewidth]{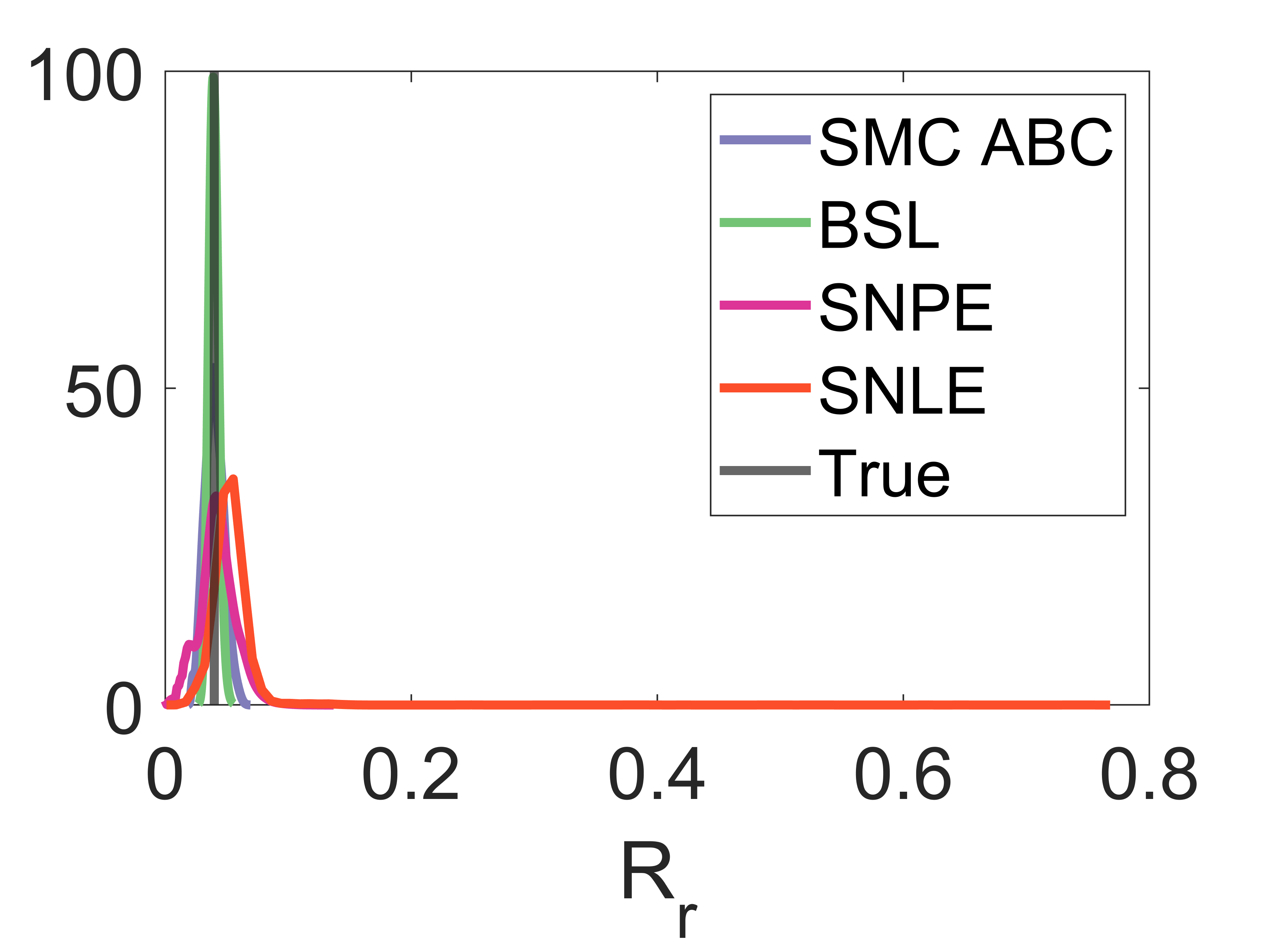}  
    \end{subfigure}
    \hfill
    \begin{subfigure}{.32\textwidth}
        \includegraphics[width=1\linewidth]{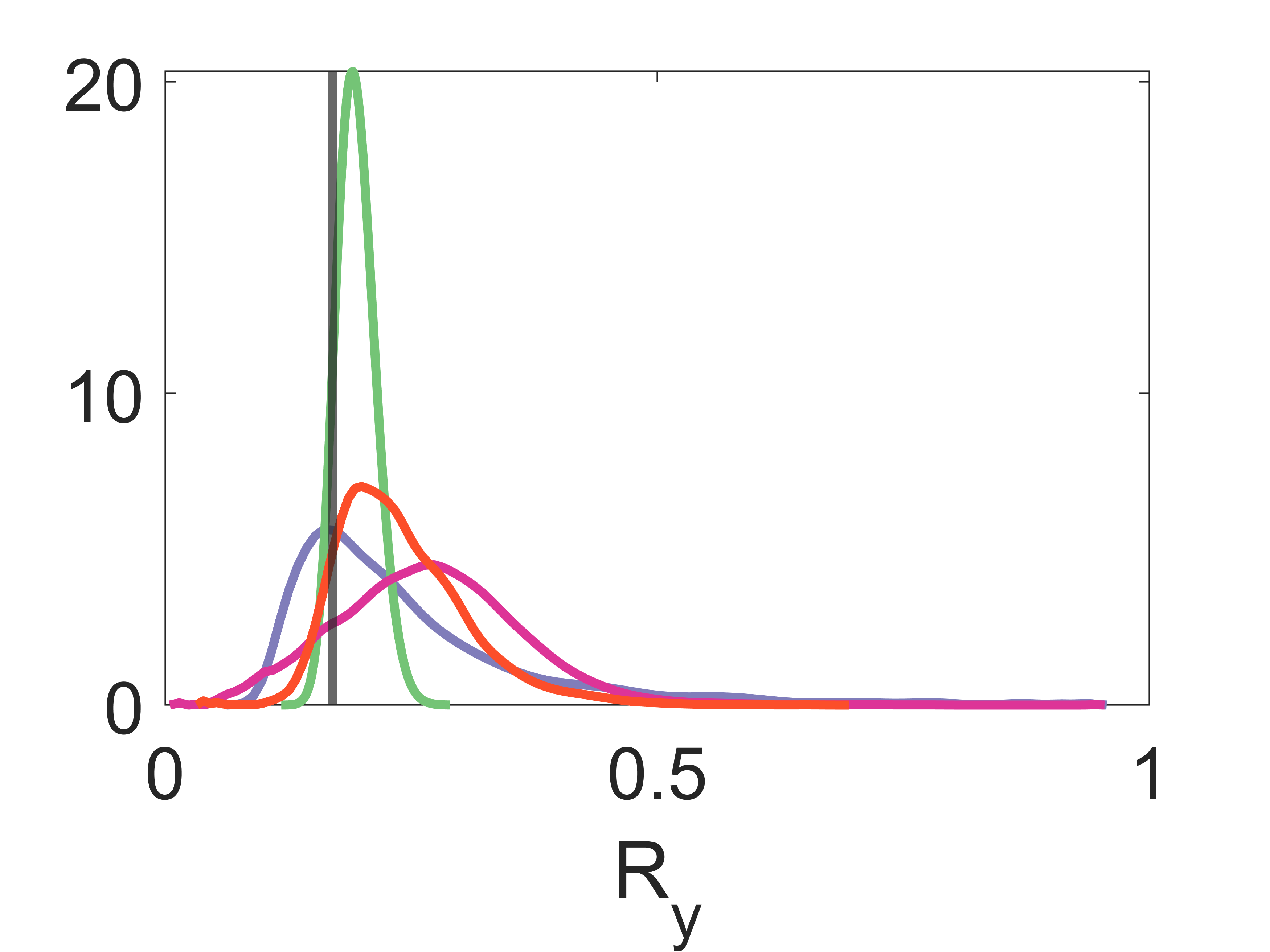}  
    \end{subfigure}
    \hfill
    \begin{subfigure}{.32\textwidth}
        \includegraphics[width=1\linewidth]{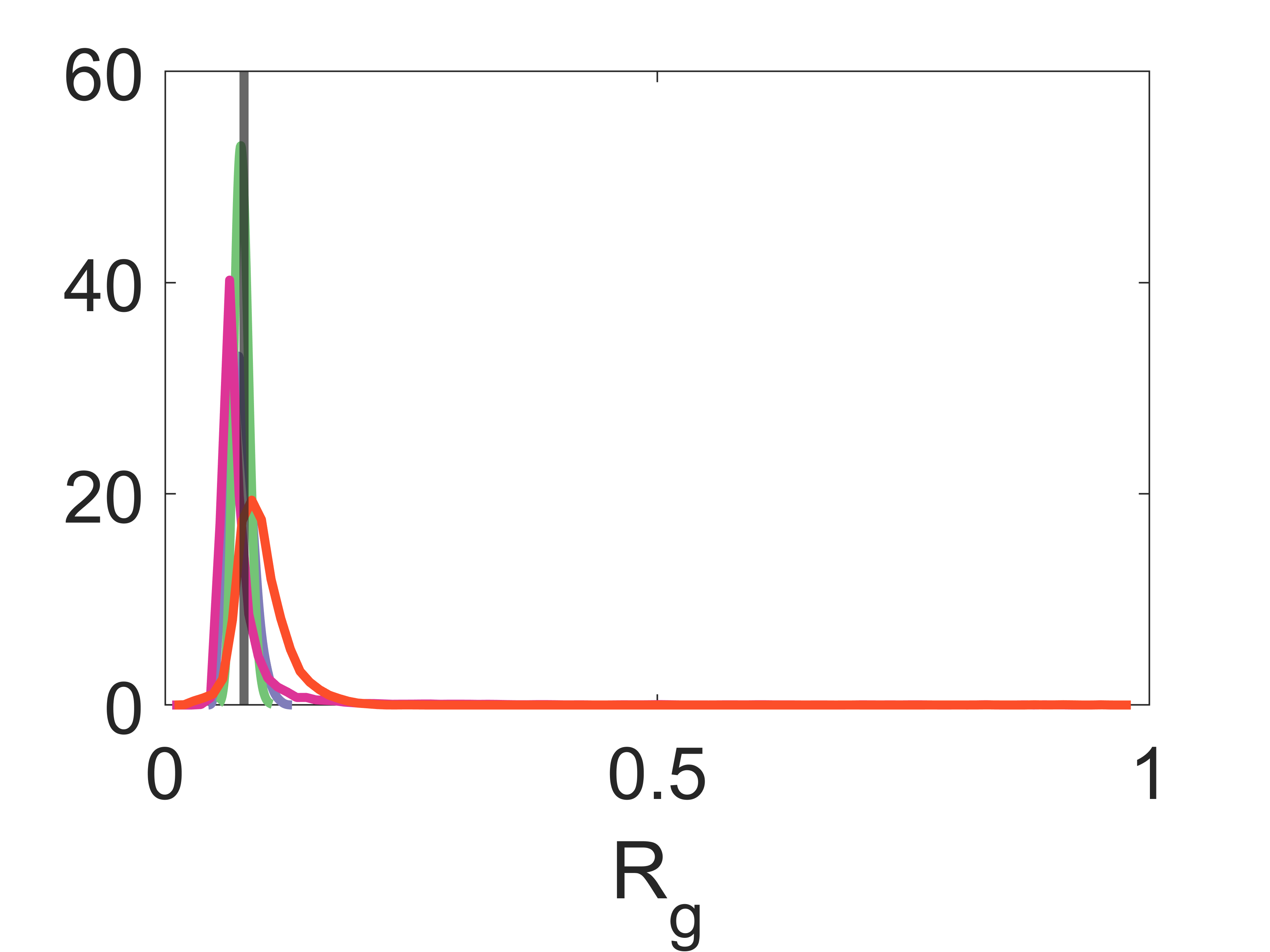}  
    \end{subfigure}
    \vfill
    \begin{subfigure}{.32\textwidth}
        \includegraphics[width=1\linewidth]{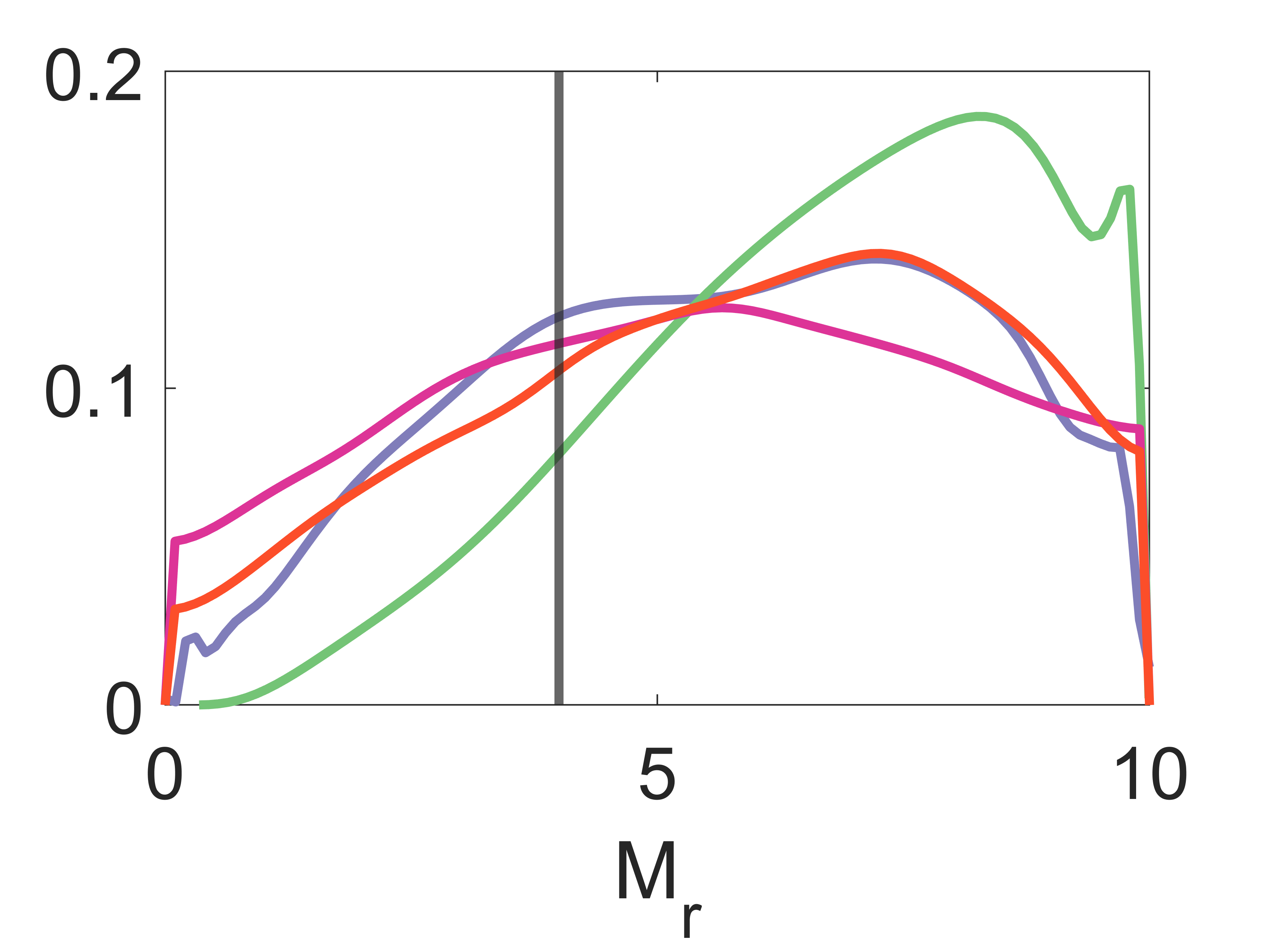}  
    \end{subfigure}
    \hfill
    \begin{subfigure}{.32\textwidth}
        \includegraphics[width=1\linewidth]{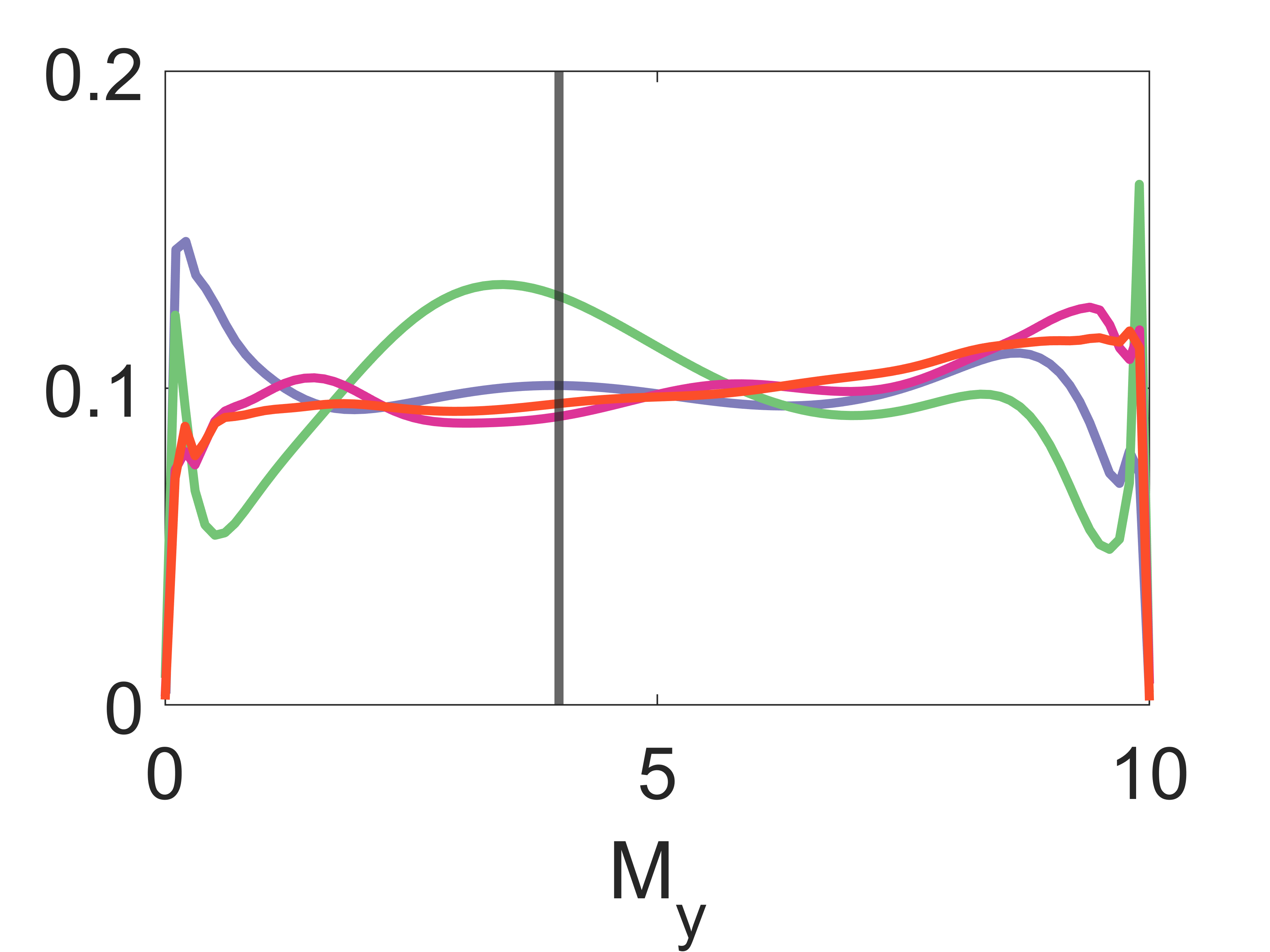}  
    \end{subfigure}
    \hfill
    \begin{subfigure}{.32\textwidth}
        \includegraphics[width=1\linewidth]{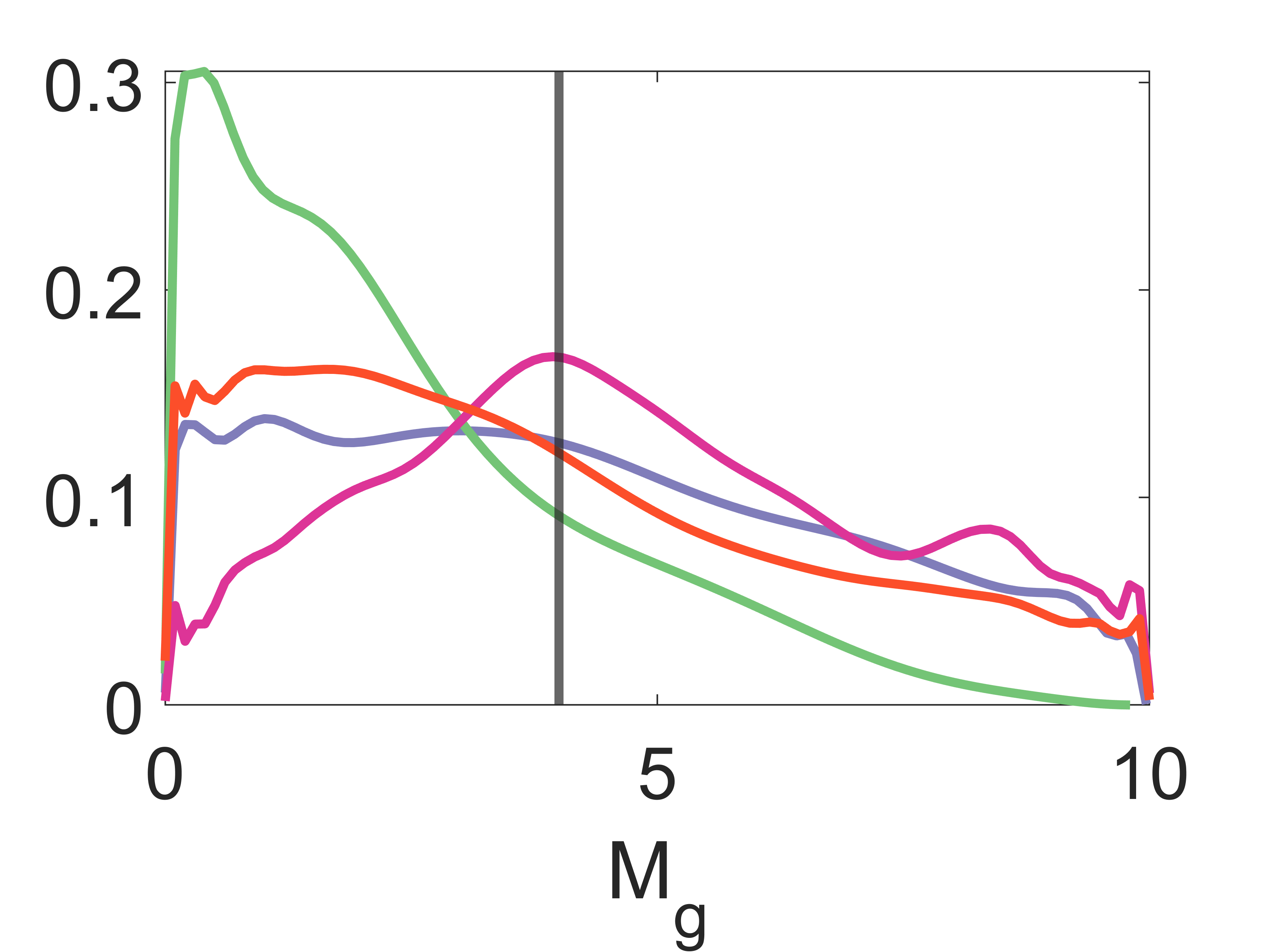}  
    \end{subfigure}
    \label{density synthetic marginal posterior}
\end{subfigure}
\caption{\textbf{Univariate posterior estimates of the stochastic cell invasion model's parameters for synthetic dataset 1 with (a) cell trajectories and (b) cell density as summary statistics for cell movement:} The plots represent the estimated posterior distributions of parameters $(R_r, R_y, R_g, M_r, M_y, M_g)$ across different methods. The violet solid lines show the SMC ABC's approximate posterior distributions, the green solid lines show those from the BSL method, the purple solid lines show the approximations by SNPE and the orange solid lines show the approximations by SNLE. The black vertical lines are the true parameter values used to generate this synthetic dataset.}
\label{cim synthetic marginal posterior}
\end{figure}

We then run four candidate SBI algorithms on synthetic datasets. The three parameters sets used to generate the synthetic datasets are 
\begin{align*}
    \theta = \{& (0.04,0.17,0.08,4,4,4),(0.25,0.15,0.22,4,4,4),\\
    &(0.12,0.07,0.03,4,4,4)\},
\end{align*}
which are the same as suggested in \citet{carr2021estimating}. Figures \ref{cim synthetic marginal posterior} shows the estimated marginal posterior distributions from each SBI algorithm on the synthetic dataset with true values $\theta = (0.04,0.17,0.08,4,4,4)$ for both cell trajectory (Figure \ref{tracking synthetic marginal posterior}) and cell density (Figure \ref{density synthetic marginal posterior}). The full results for all synthetic datasets we used are presented in Section \textcolor{blue}{C 2.1} of the Supplementary document. 

For both cell trajectory and cell density, SMC ABC, SNPE, and SNLE show similar performance across all parameters. All three SBI algorithms suggest that ($M_r, M_y, M_g$) for cell density is non-identifiable. With a significantly larger number of model simulations, BSL outperforms the other SBI algorithms on the parameters ($R_r, R_y, R_g$) for both cell trajectory and cell density. Note that there may appear to be some bias in the parameter estimates as the posterior modes do not correspond exactly to the true parameter values.  However, when dealing with small sample sizes, due to natural variation, the optimal parameter values for a given dataset will not equal the true parameter values. However, we would expect the 95\% credible interval (CI) of the posterior to contain the true parameter value around 95\% of the time.  As we see from the plots, the true parameter values are contained within the 95\% CI. For cell density, BSL places a large mass of probability for ($M_r, M_g$)  on the boundary of the prior distribution. Following \textbf{Guideline 3}, we should use posterior predictive checks to verify if the estimated posteriors are reasonable and not overconfident.

\begin{figure}[H]
\begin{subfigure}{1\textwidth}
\caption{ }
    \begin{subfigure}{.32\textwidth}
        \includegraphics[width=1\linewidth]{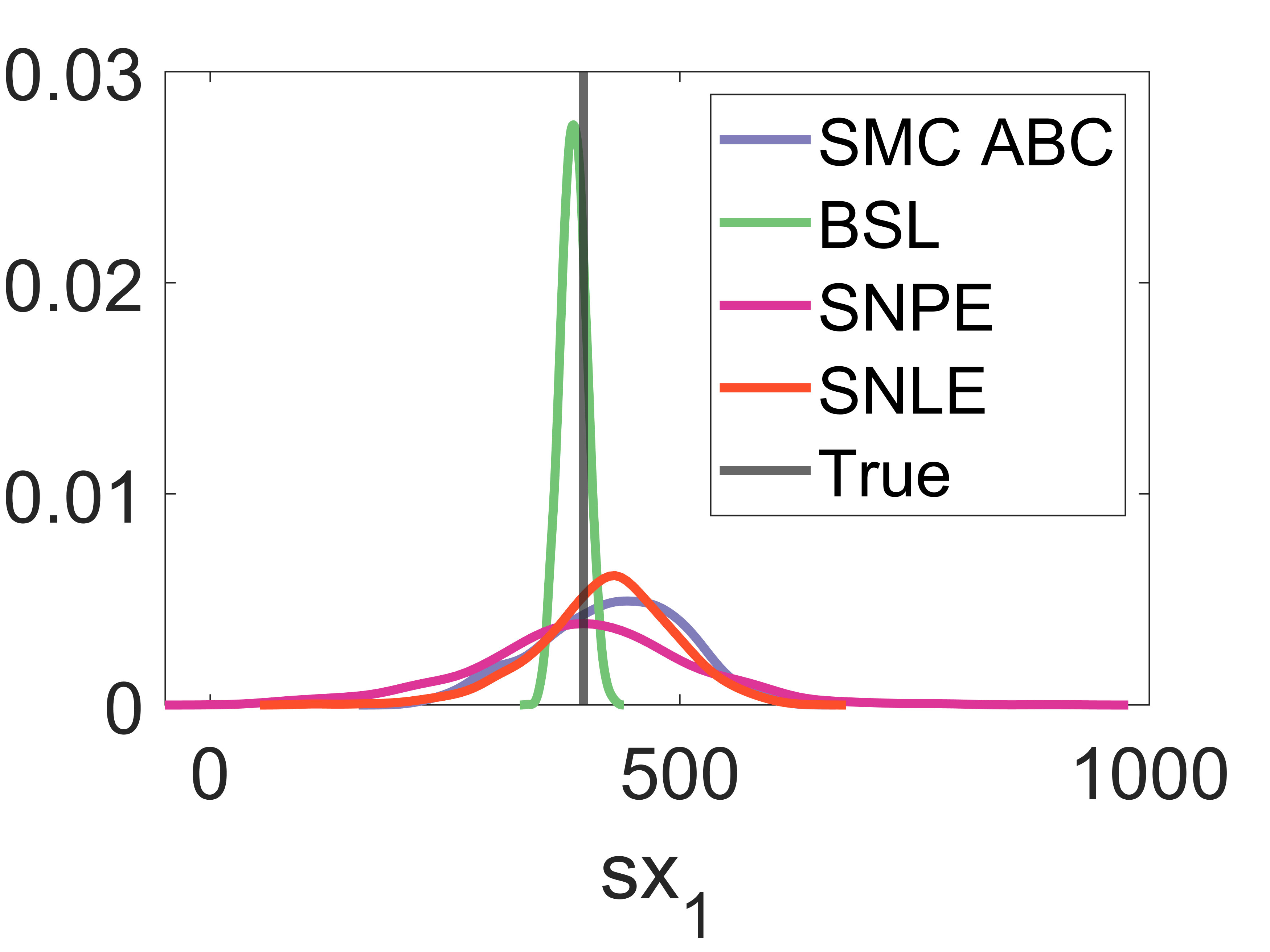}  
    \end{subfigure}
    \hfill
    \begin{subfigure}{.32\textwidth}
        \includegraphics[width=1\linewidth]{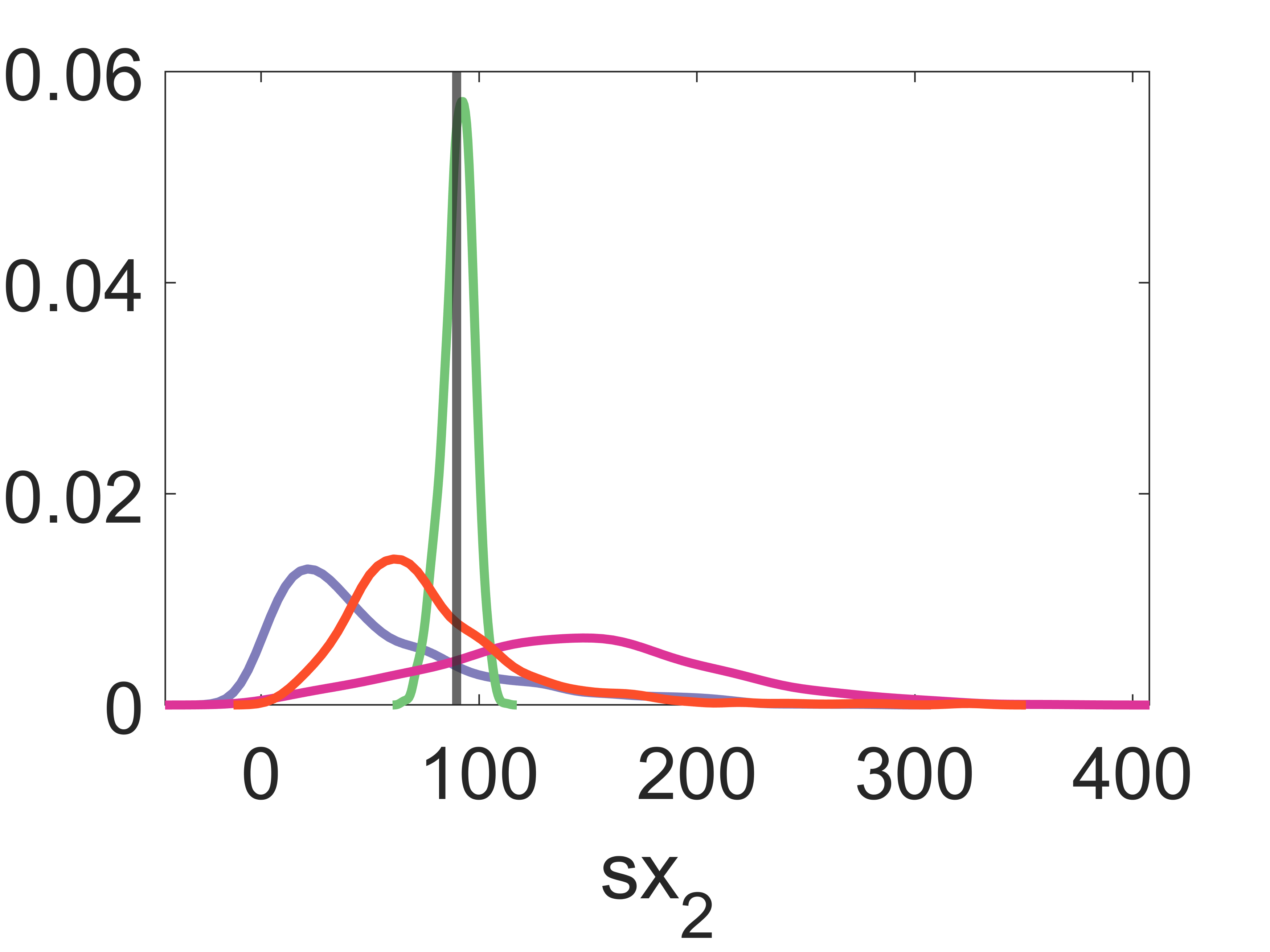}  
    \end{subfigure}
    \hfill
    \begin{subfigure}{.32\textwidth}
        \includegraphics[width=1\linewidth]{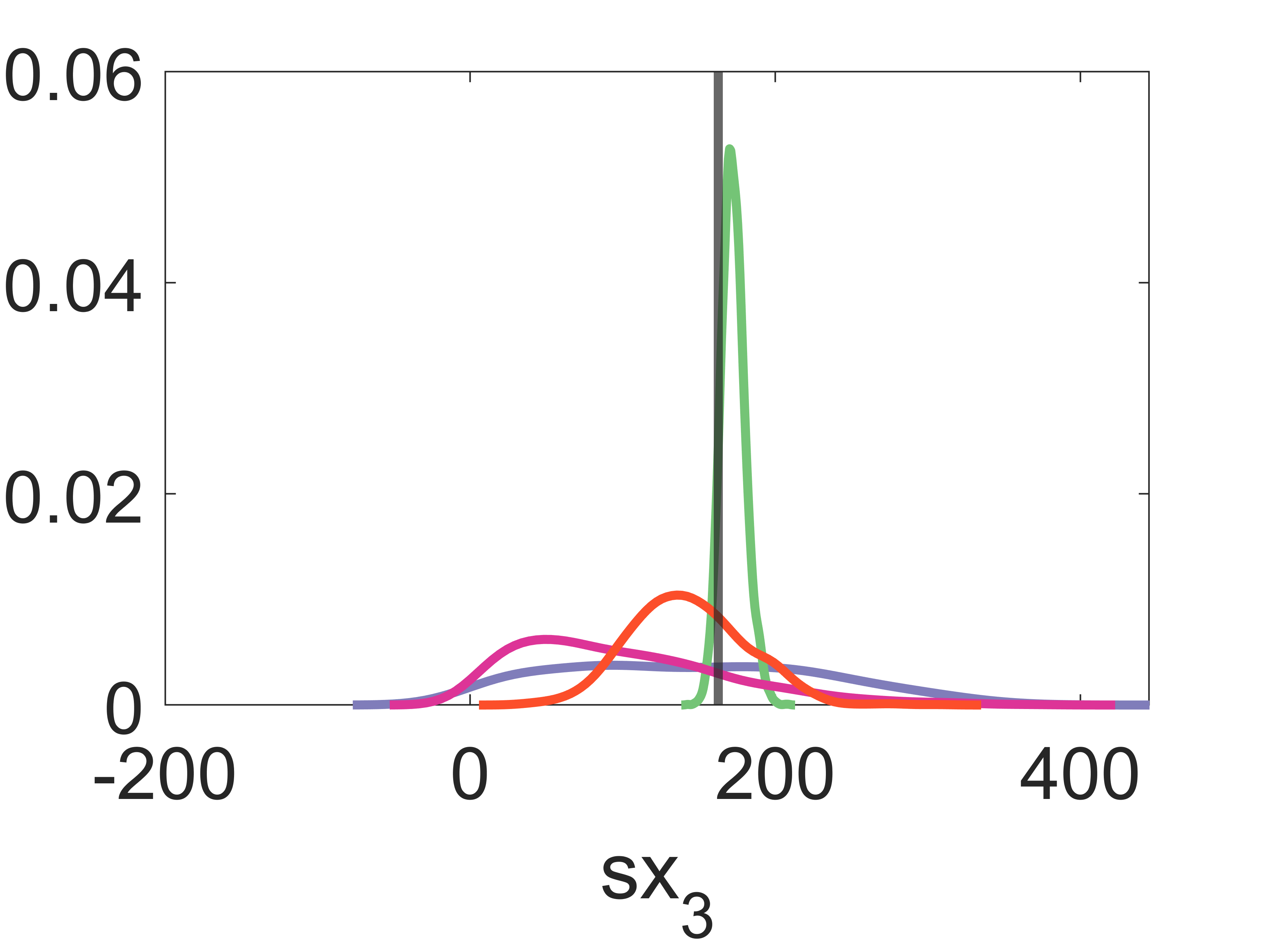}  
    \end{subfigure}
    \vfill
    \begin{subfigure}{.32\textwidth}
        \includegraphics[width=1\linewidth]{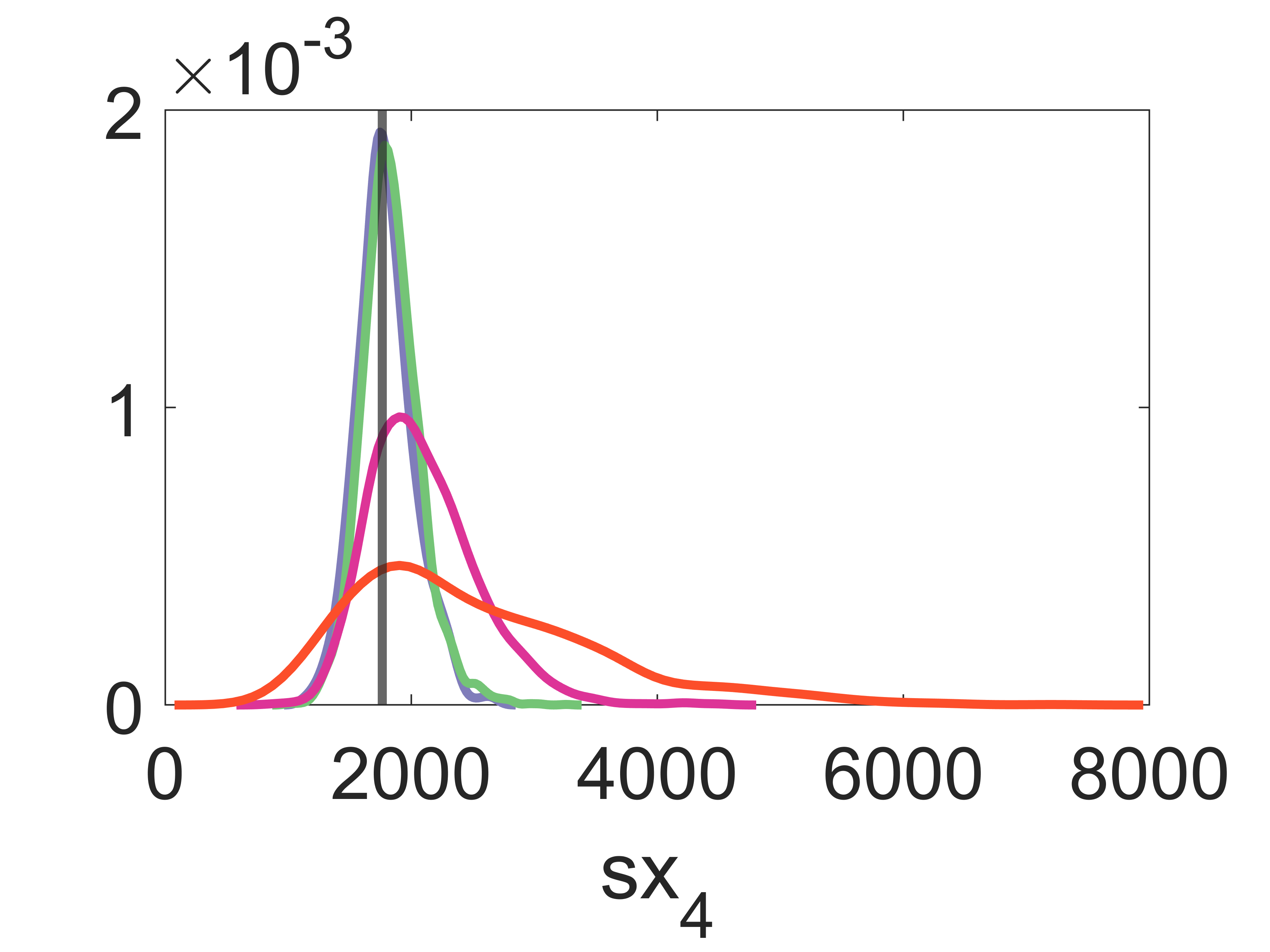}  
    \end{subfigure}
    \hfill
    \begin{subfigure}{.32\textwidth}
        \includegraphics[width=1\linewidth]{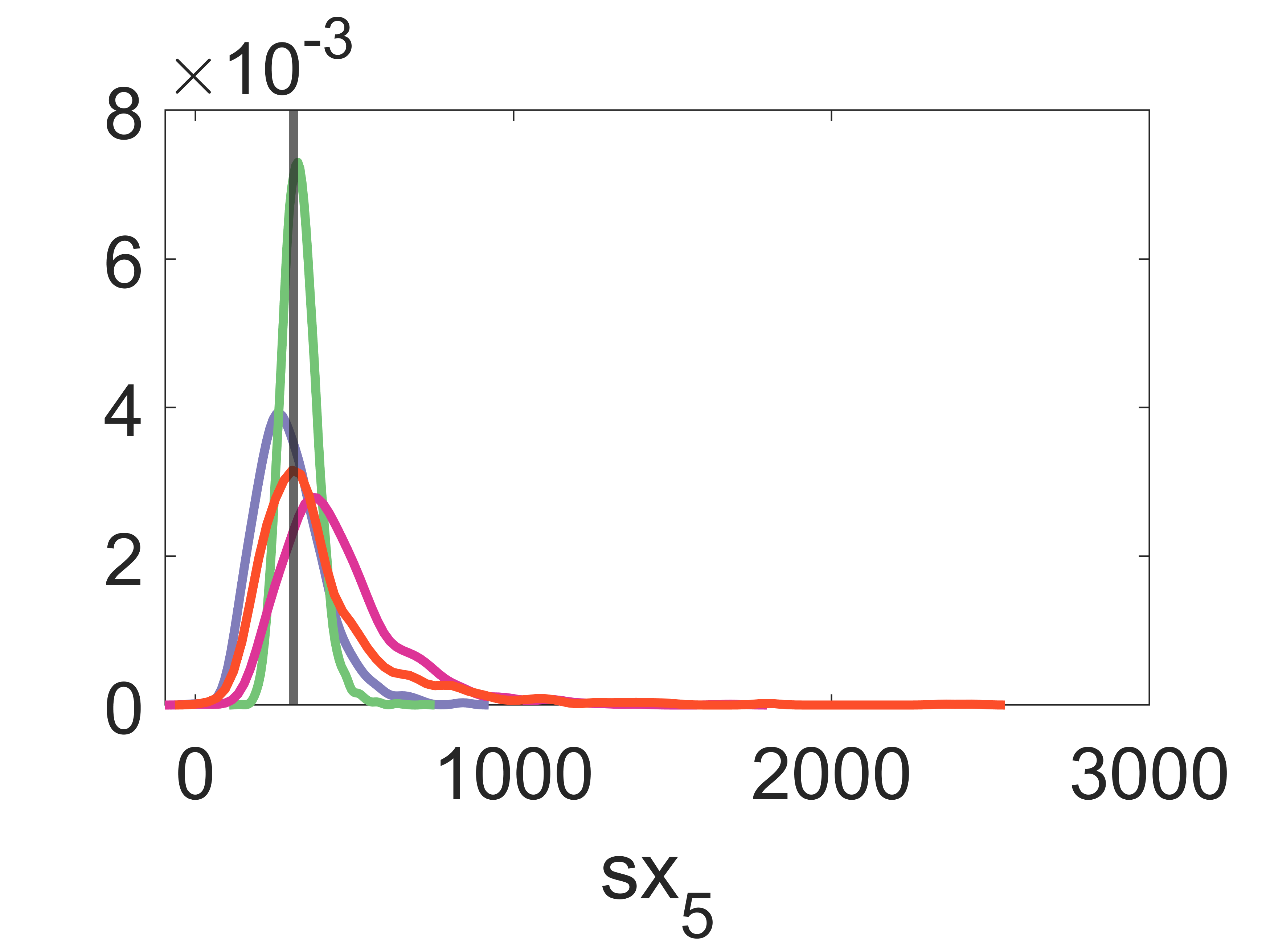}  
    \end{subfigure}
    \hfill
    \begin{subfigure}{.32\textwidth}
        \includegraphics[width=1\linewidth]{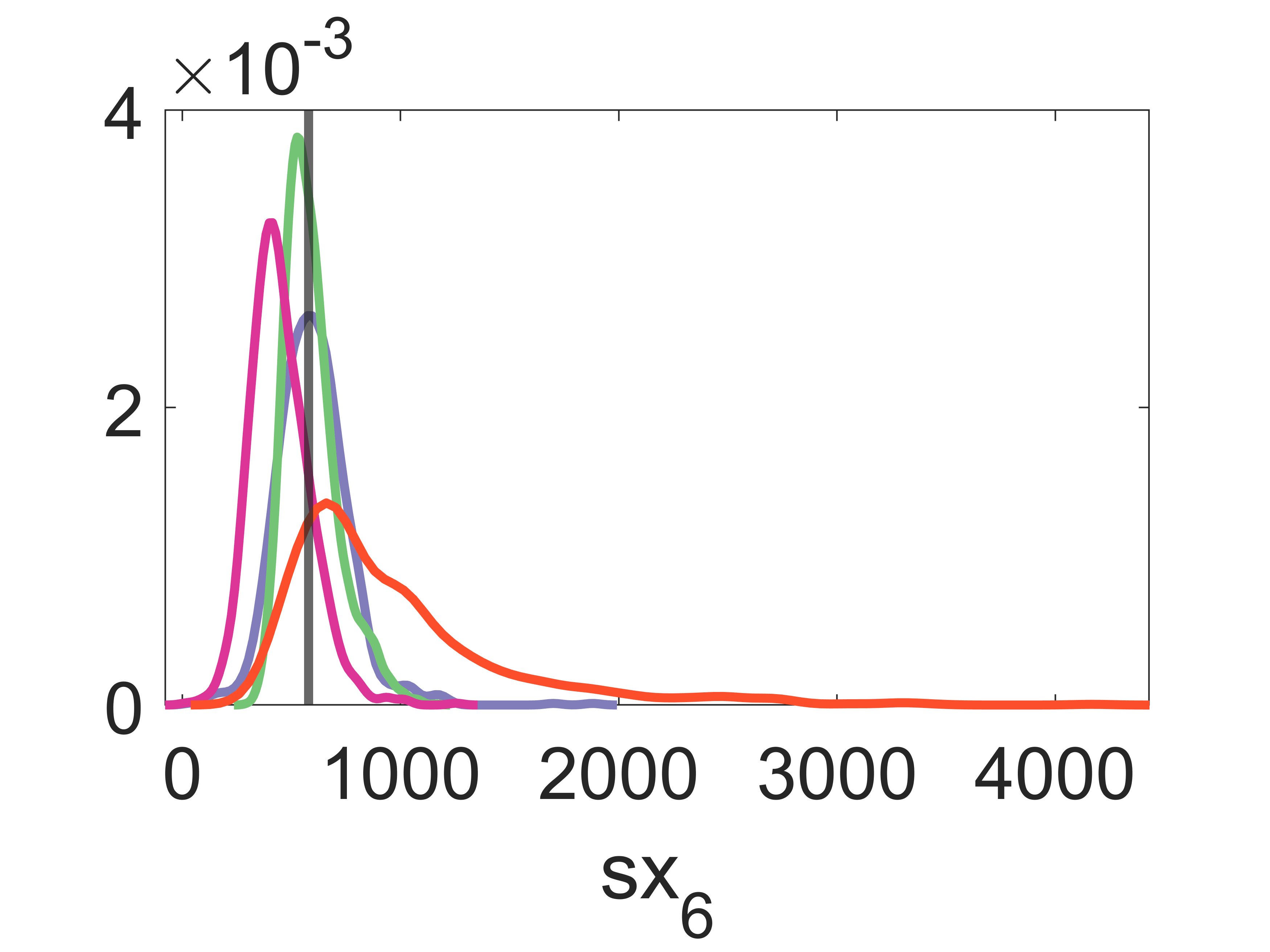}  
    \end{subfigure}
    \label{tracking synthetics data calibration}
\end{subfigure}
\vfill
\begin{subfigure}{1\textwidth}
\caption{ }
    \begin{subfigure}{.19\textwidth}
        \includegraphics[width=1\linewidth]{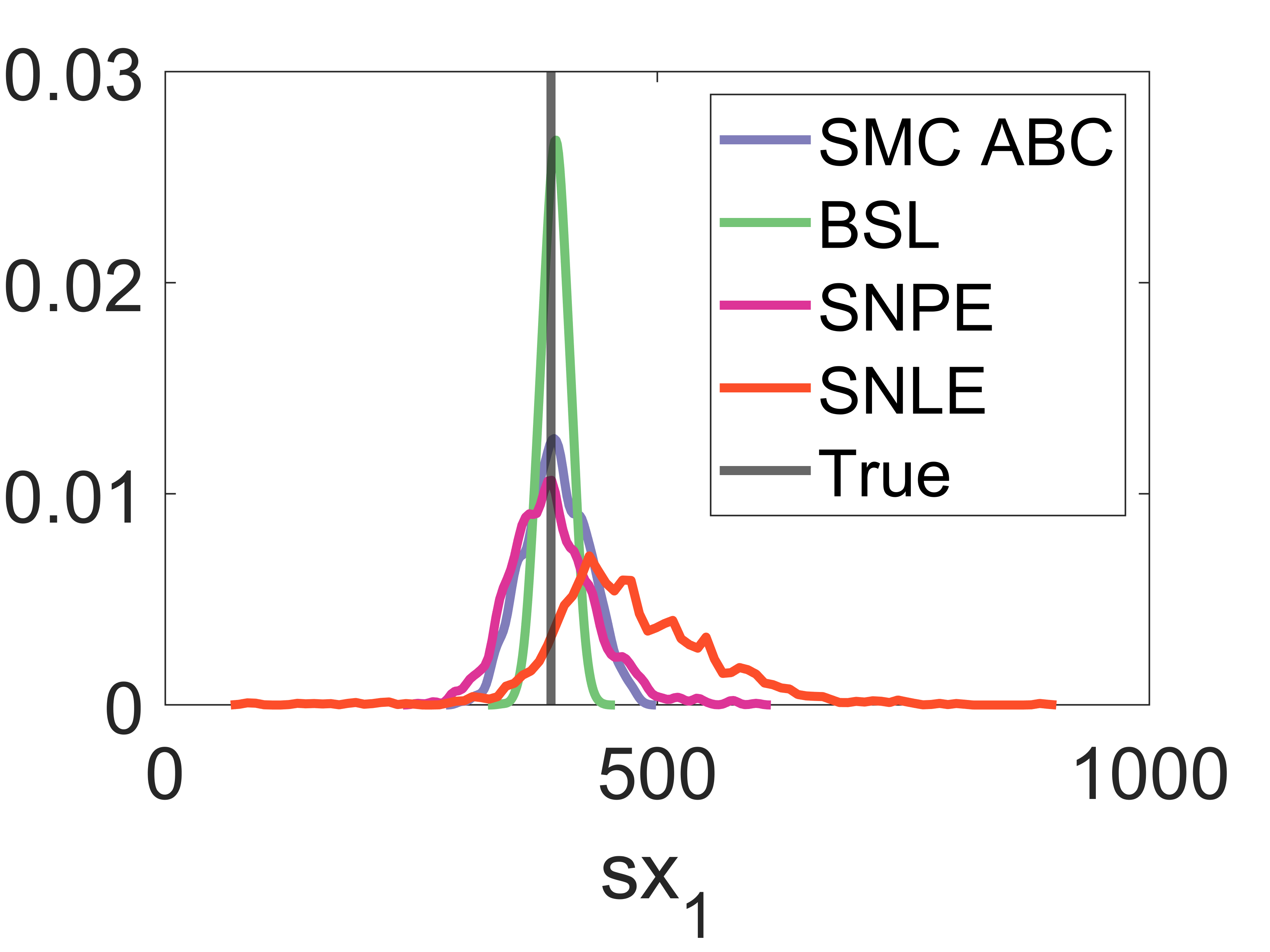}  
    \end{subfigure}
    \hfill
    \begin{subfigure}{.19\textwidth}
        \includegraphics[width=1\linewidth]{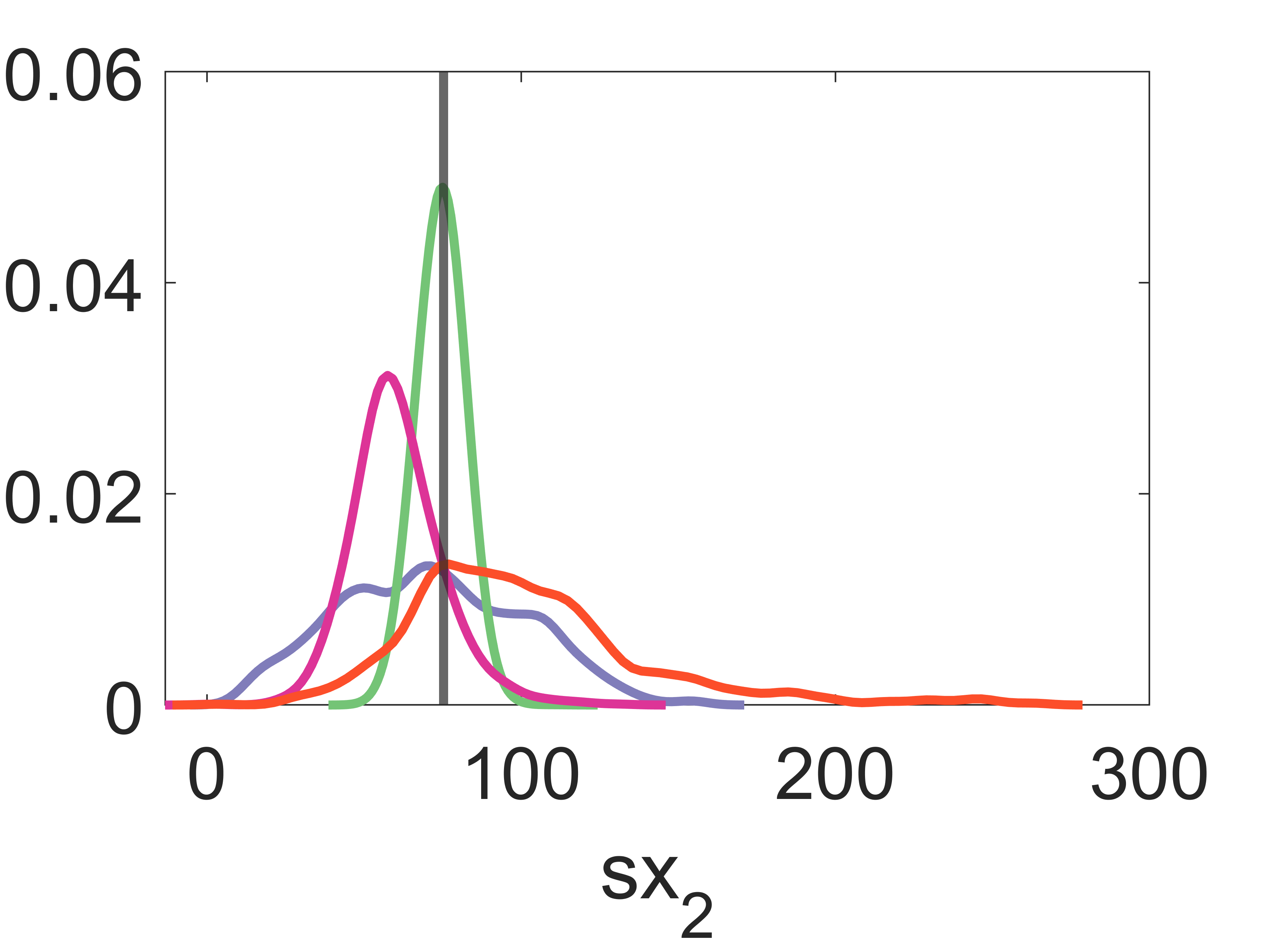}  
    \end{subfigure}
    \hfill
    \begin{subfigure}{.19\textwidth}
        \includegraphics[width=1\linewidth]{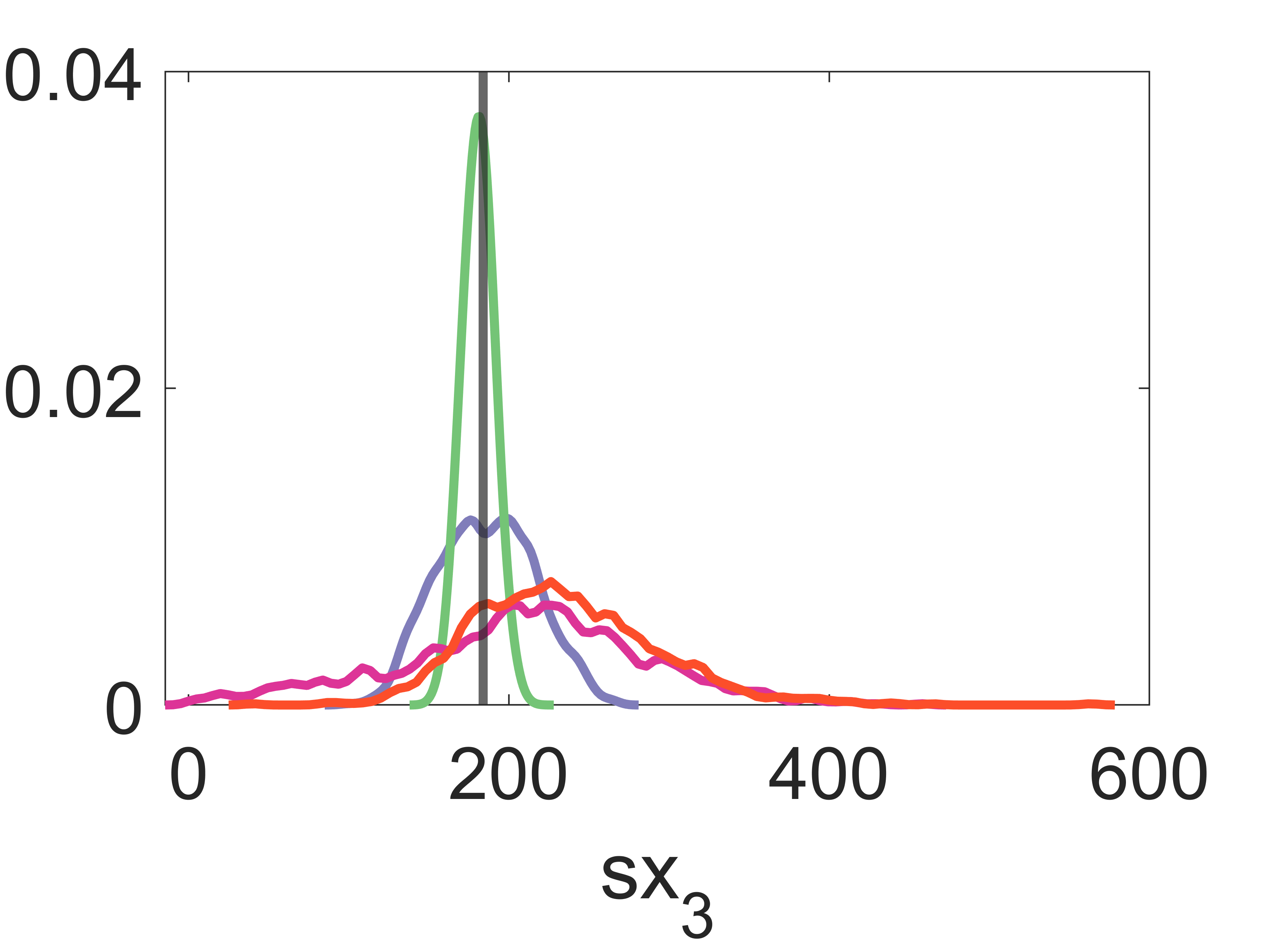}  
    \end{subfigure}
    \hfill
    \begin{subfigure}{.19\textwidth}
        \includegraphics[width=1\linewidth]{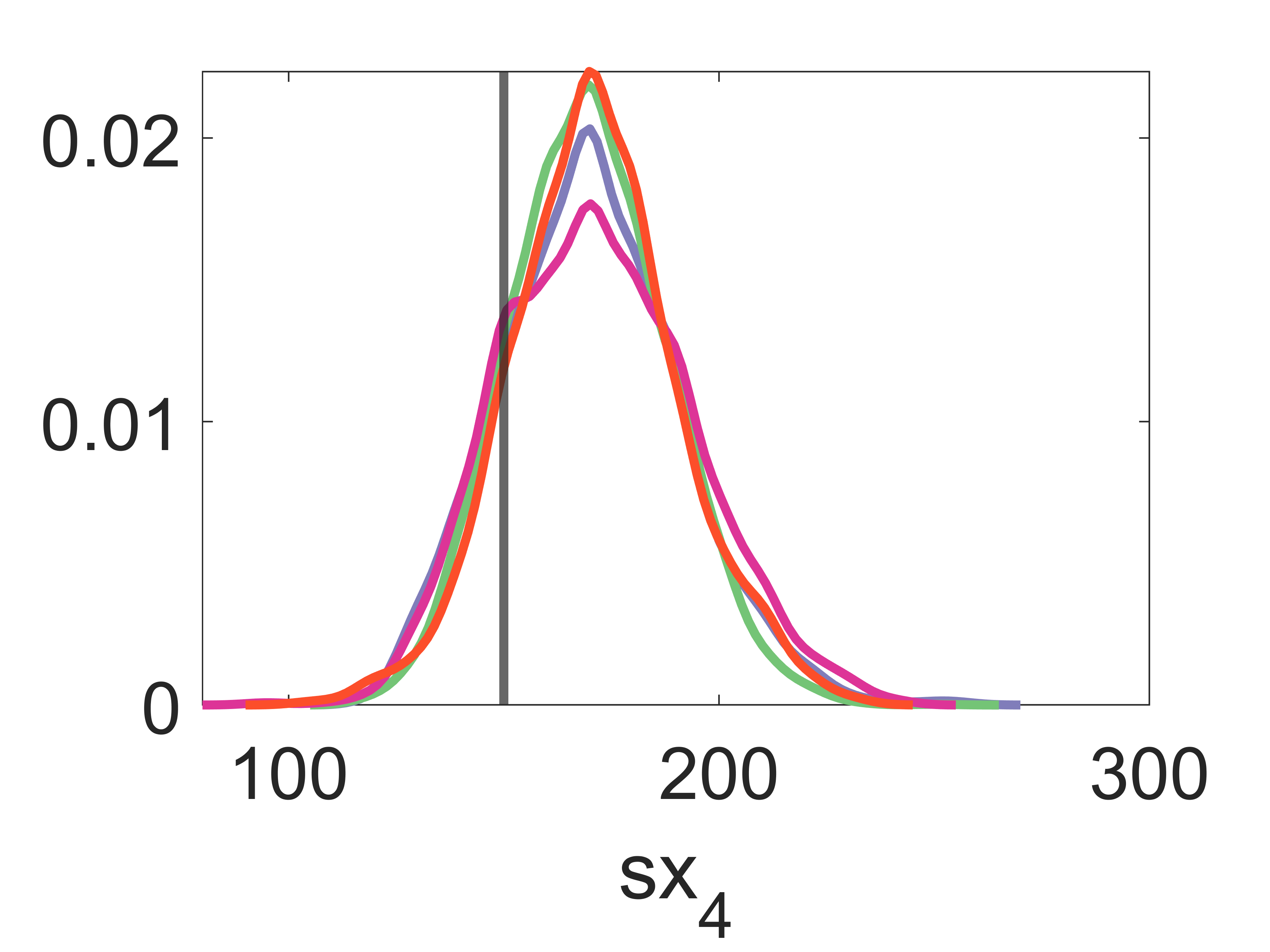}  
    \end{subfigure}
    \hfill
    \begin{subfigure}{.19\textwidth}
        \includegraphics[width=1\linewidth]{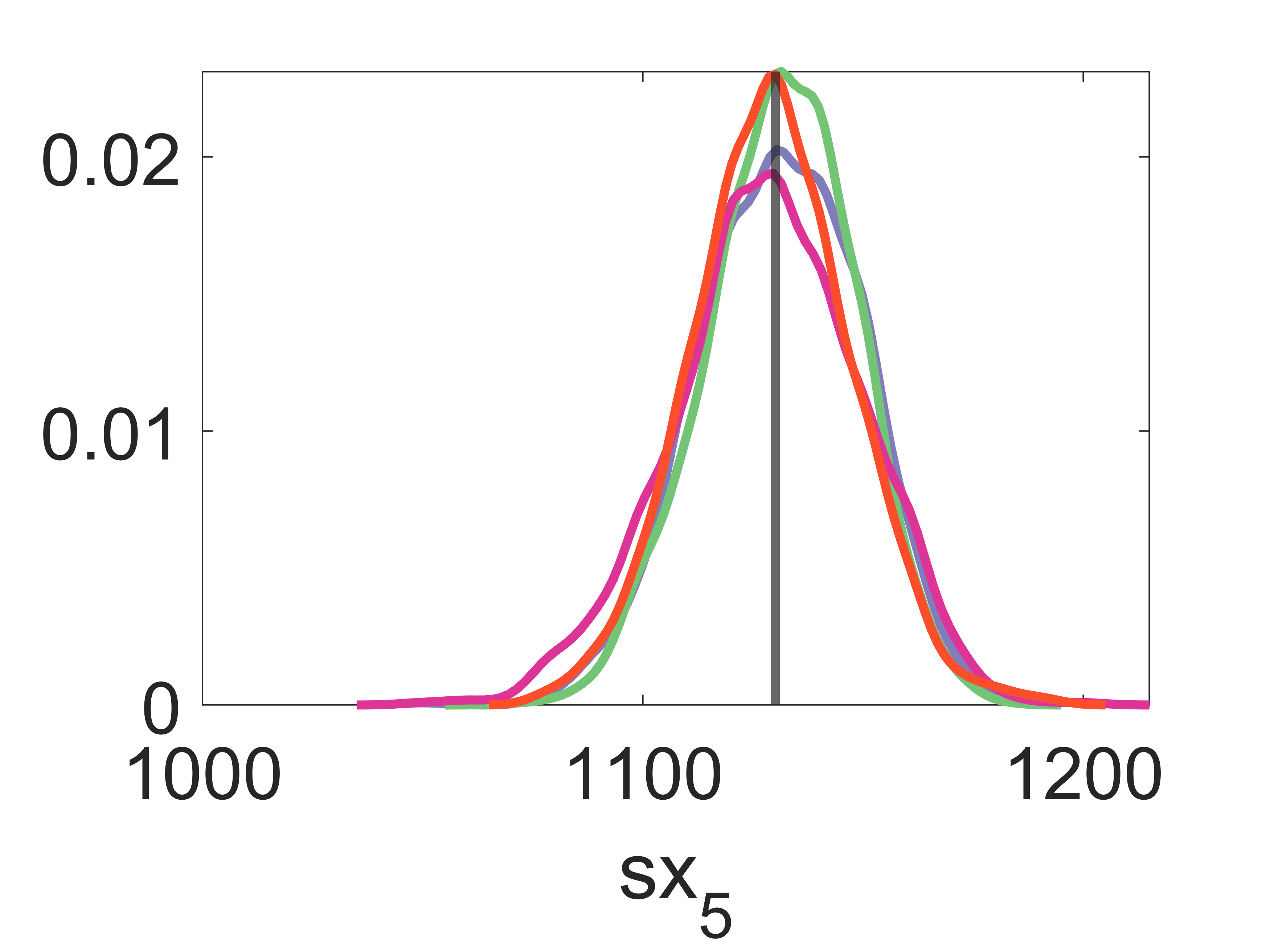}  
    \end{subfigure}
    \vfill
    \begin{subfigure}{.19\textwidth}
        \includegraphics[width=1\linewidth]{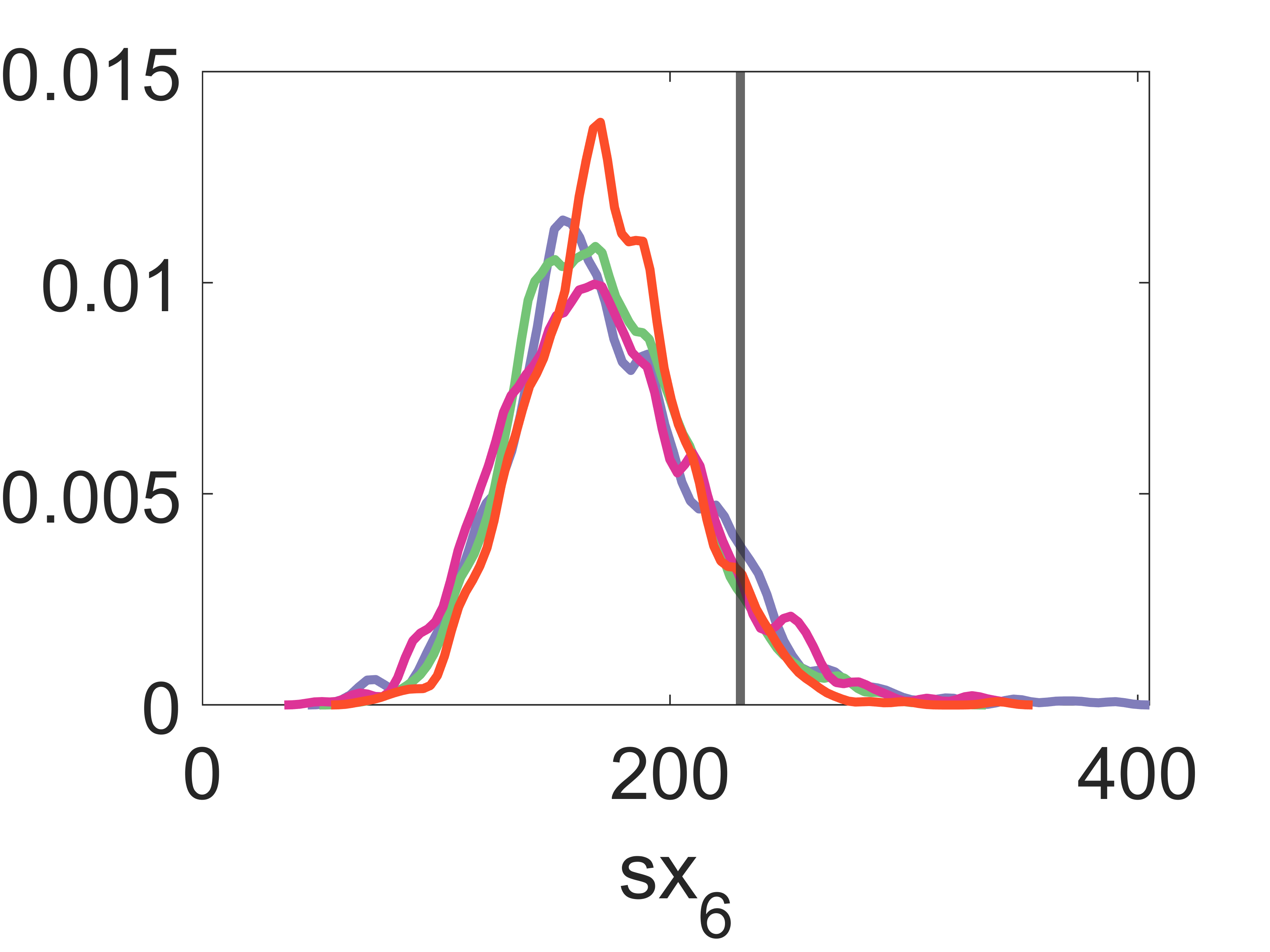}  
    \end{subfigure}
    \hfill
     \begin{subfigure}{.19\textwidth}
        \includegraphics[width=1\linewidth]{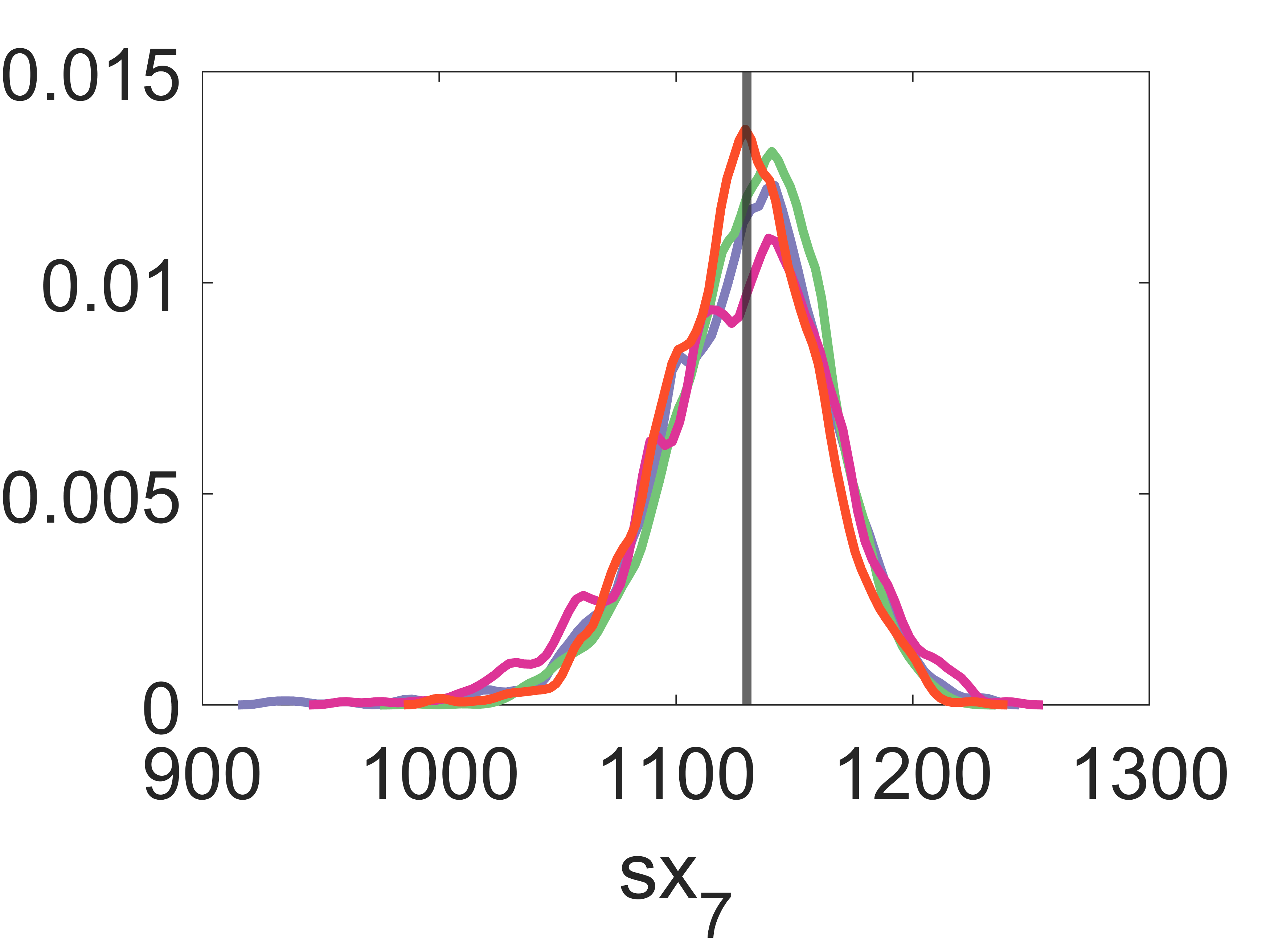}  
    \end{subfigure}
    \hfill
    \begin{subfigure}{.19\textwidth}
        \includegraphics[width=1\linewidth]{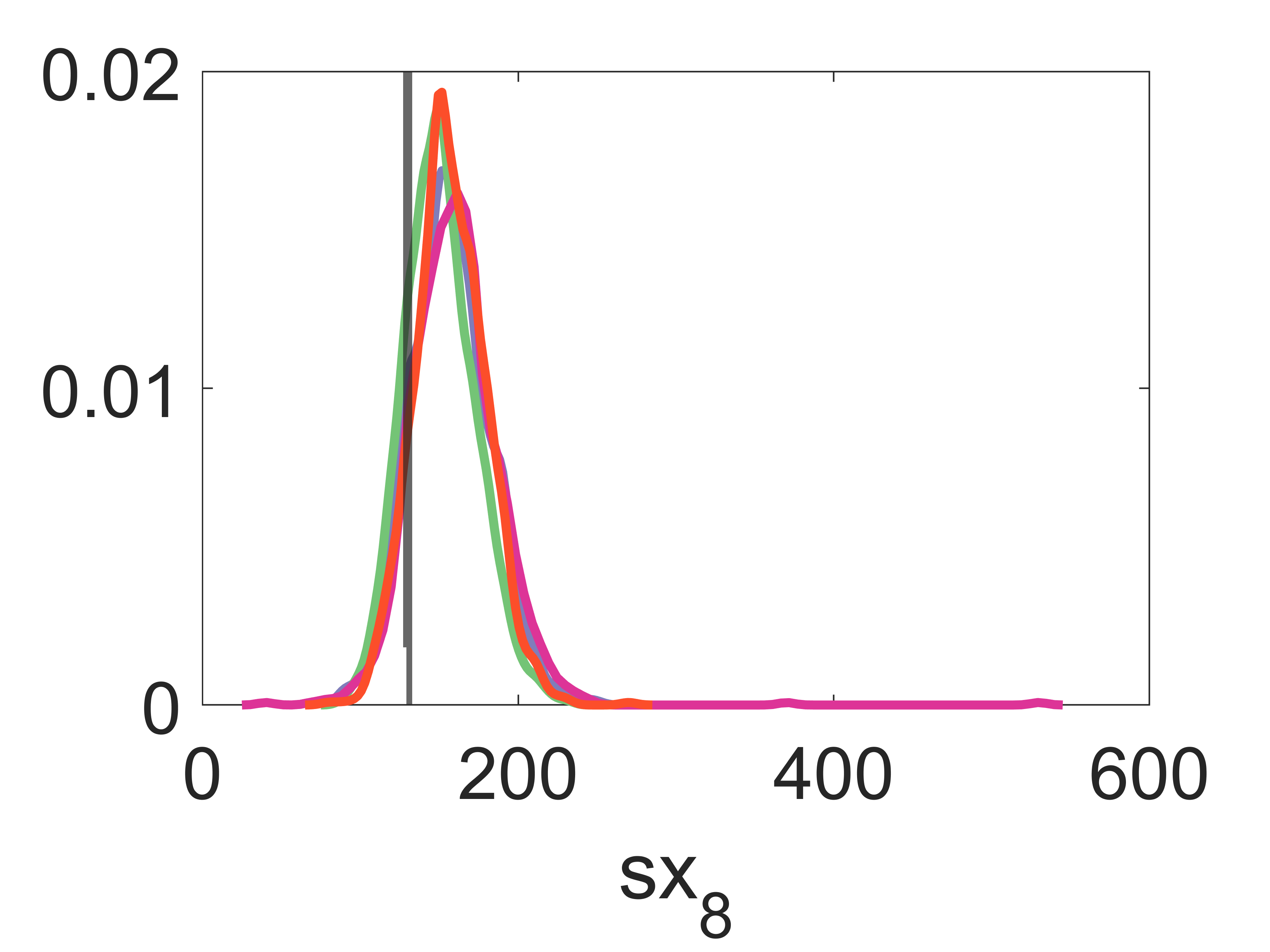}  
    \end{subfigure}
    \hfill
    \begin{subfigure}{.19\textwidth}
        \includegraphics[width=1\linewidth]{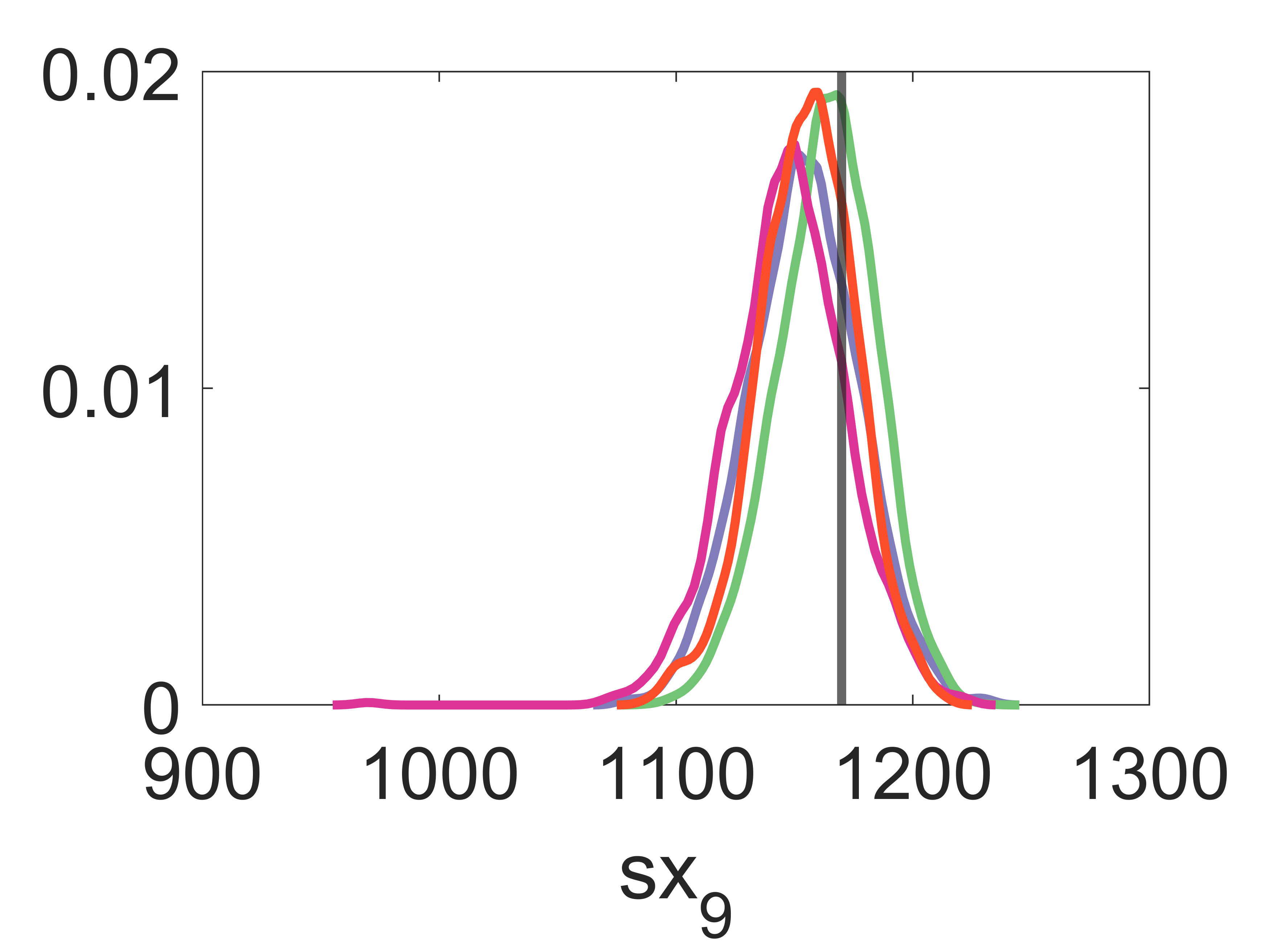}  
    \end{subfigure}
    \hfill
    \begin{subfigure}{.19\textwidth}
        \includegraphics[width=1\linewidth]{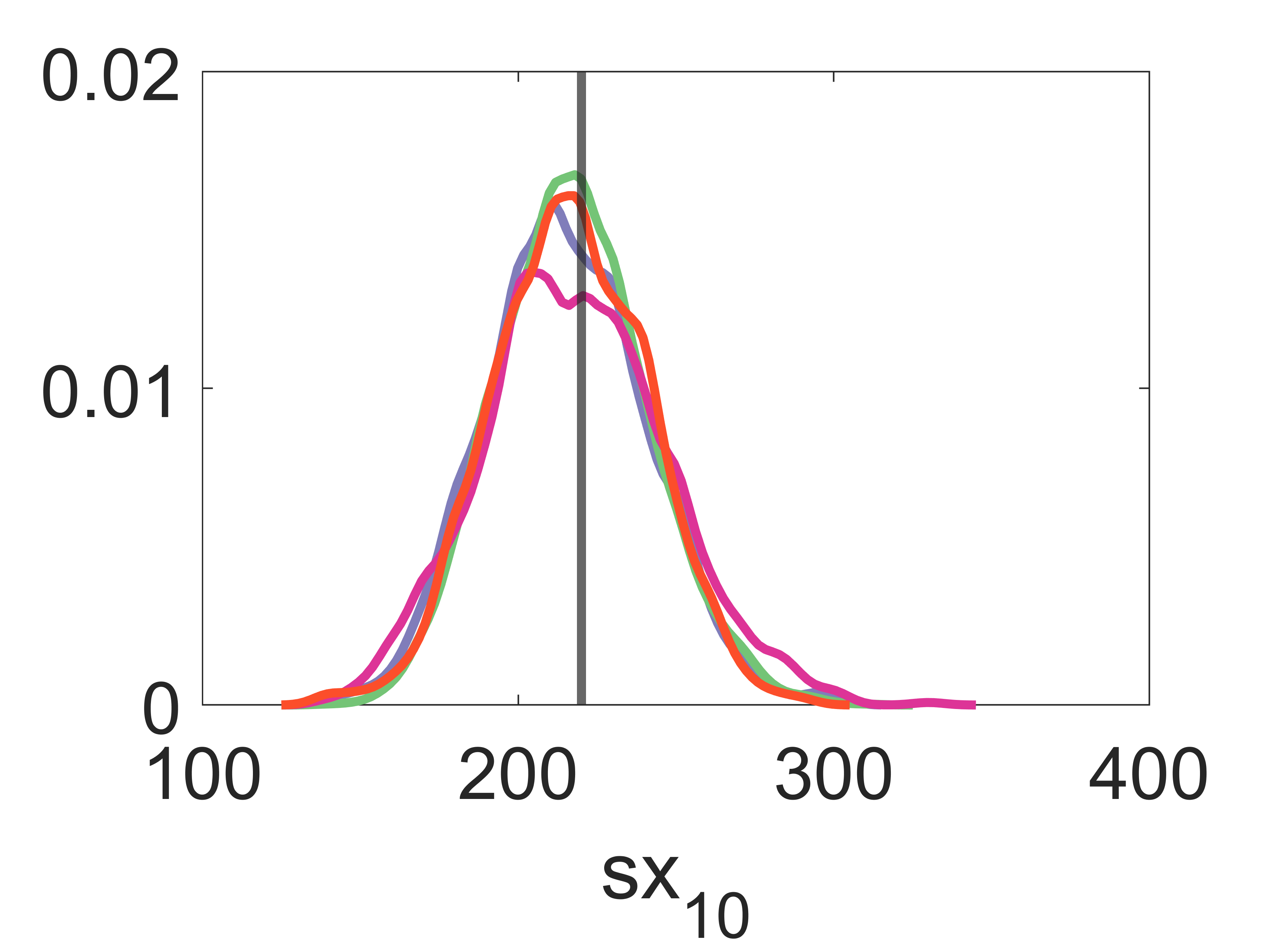}  
    \end{subfigure}
    \vfill
    \begin{subfigure}{.19\textwidth}
        \includegraphics[width=1\linewidth]{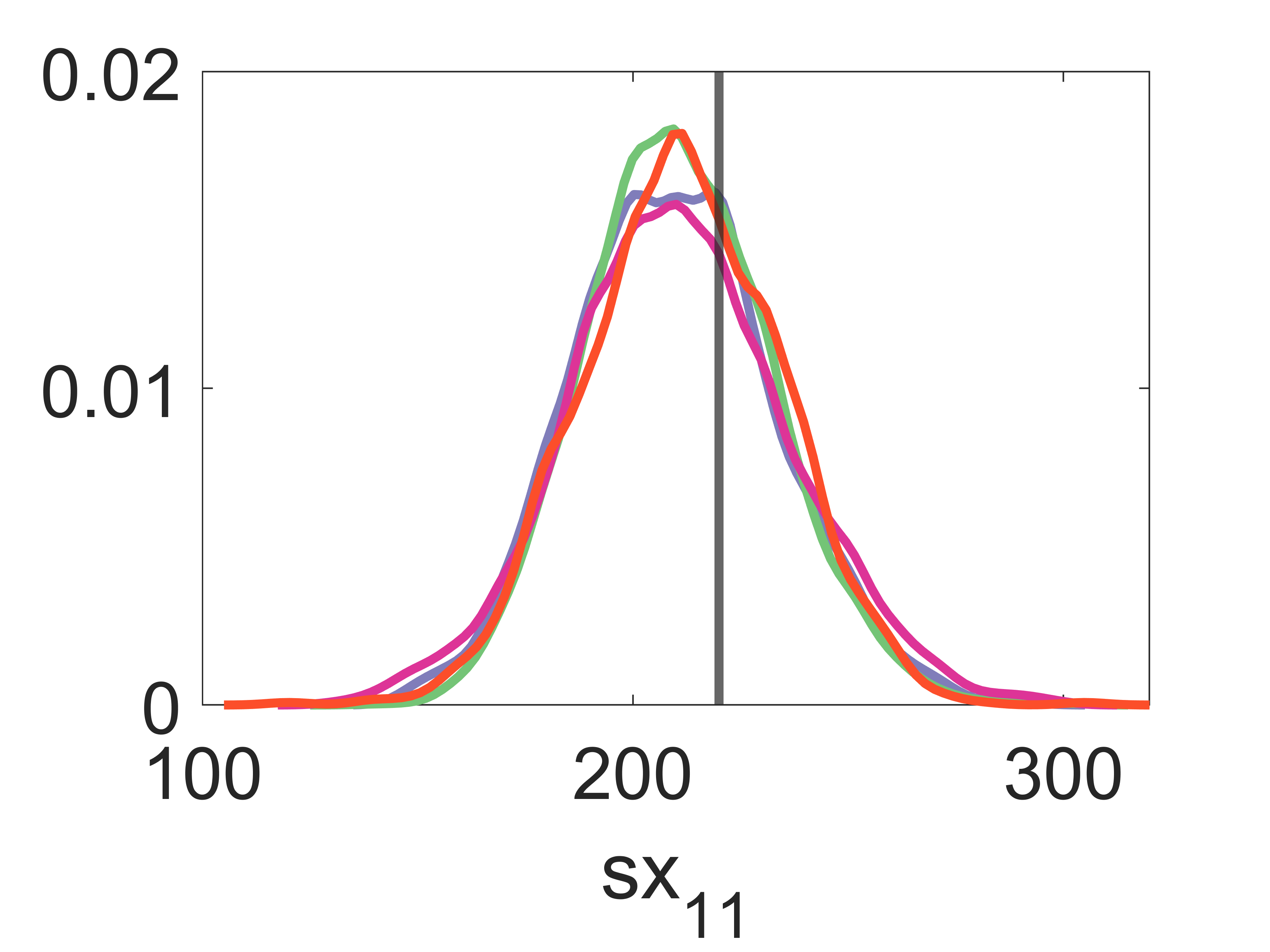}  
    \end{subfigure}
    \hfill
    \begin{subfigure}{.19\textwidth}
        \includegraphics[width=1\linewidth]{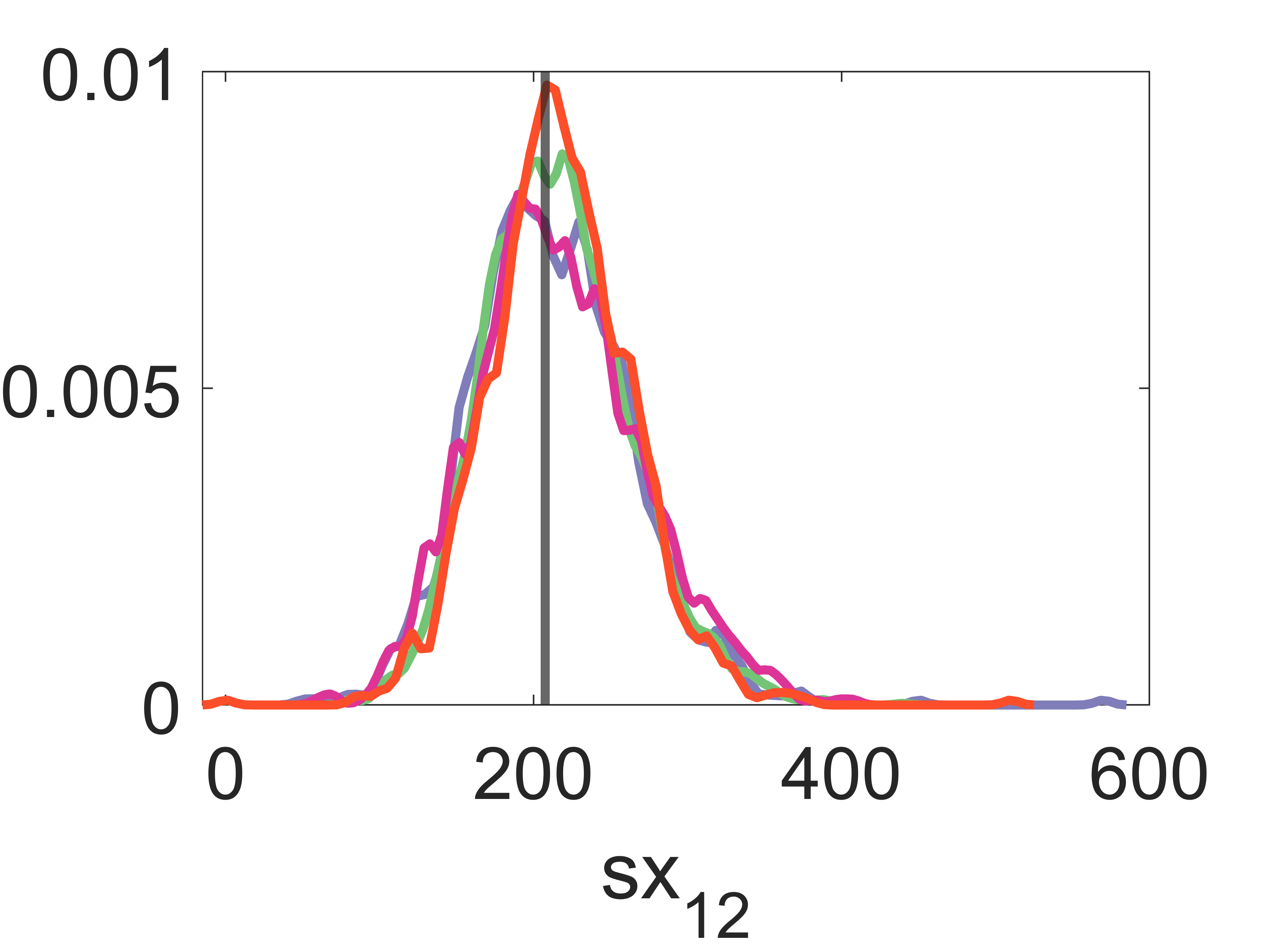}  
    \end{subfigure}
    \hfill
    \begin{subfigure}{.19\textwidth}
        \includegraphics[width=1\linewidth]{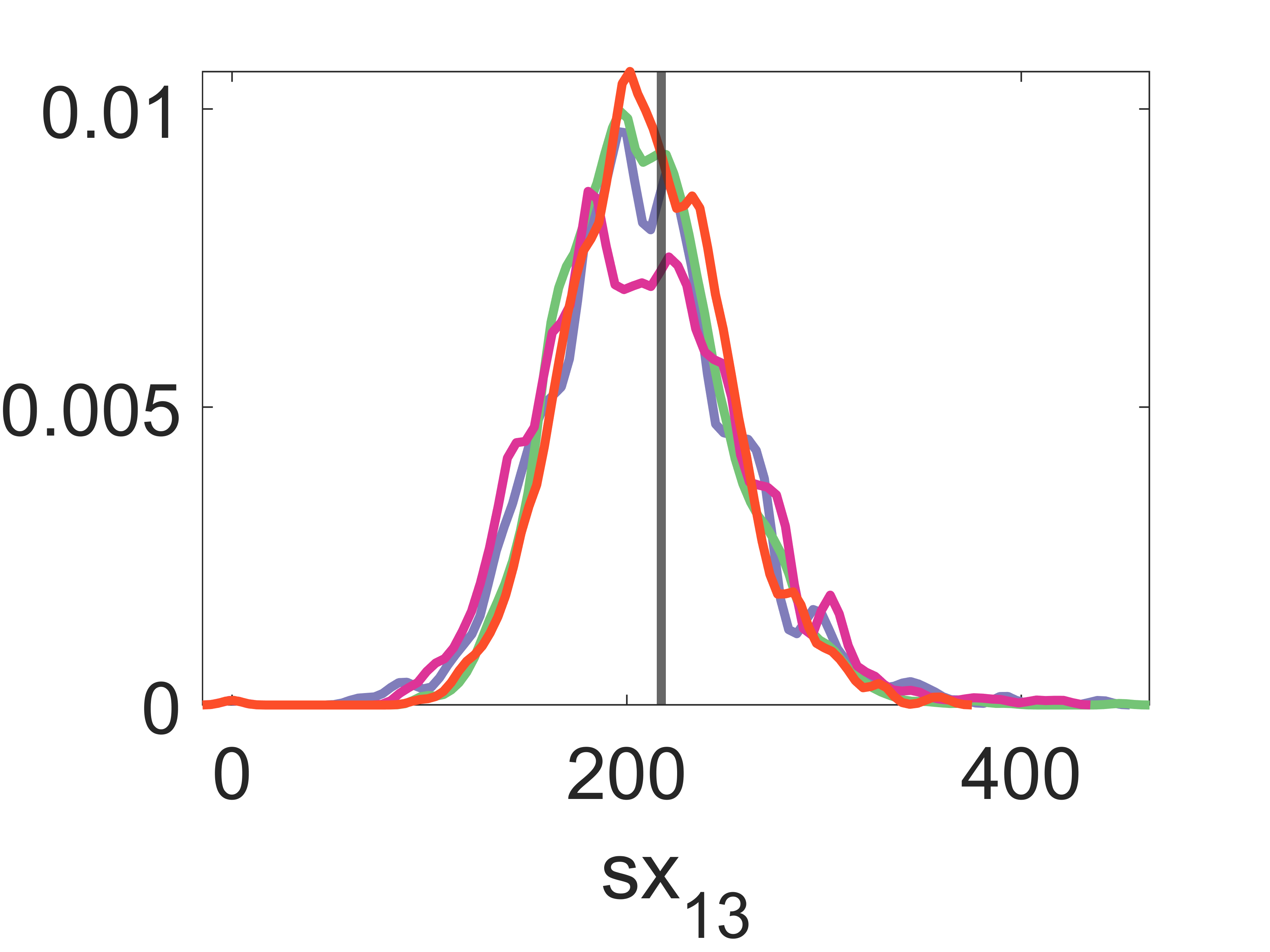}  
    \end{subfigure}
    \hfill
    \begin{subfigure}{.19\textwidth}
        \includegraphics[width=1\linewidth]{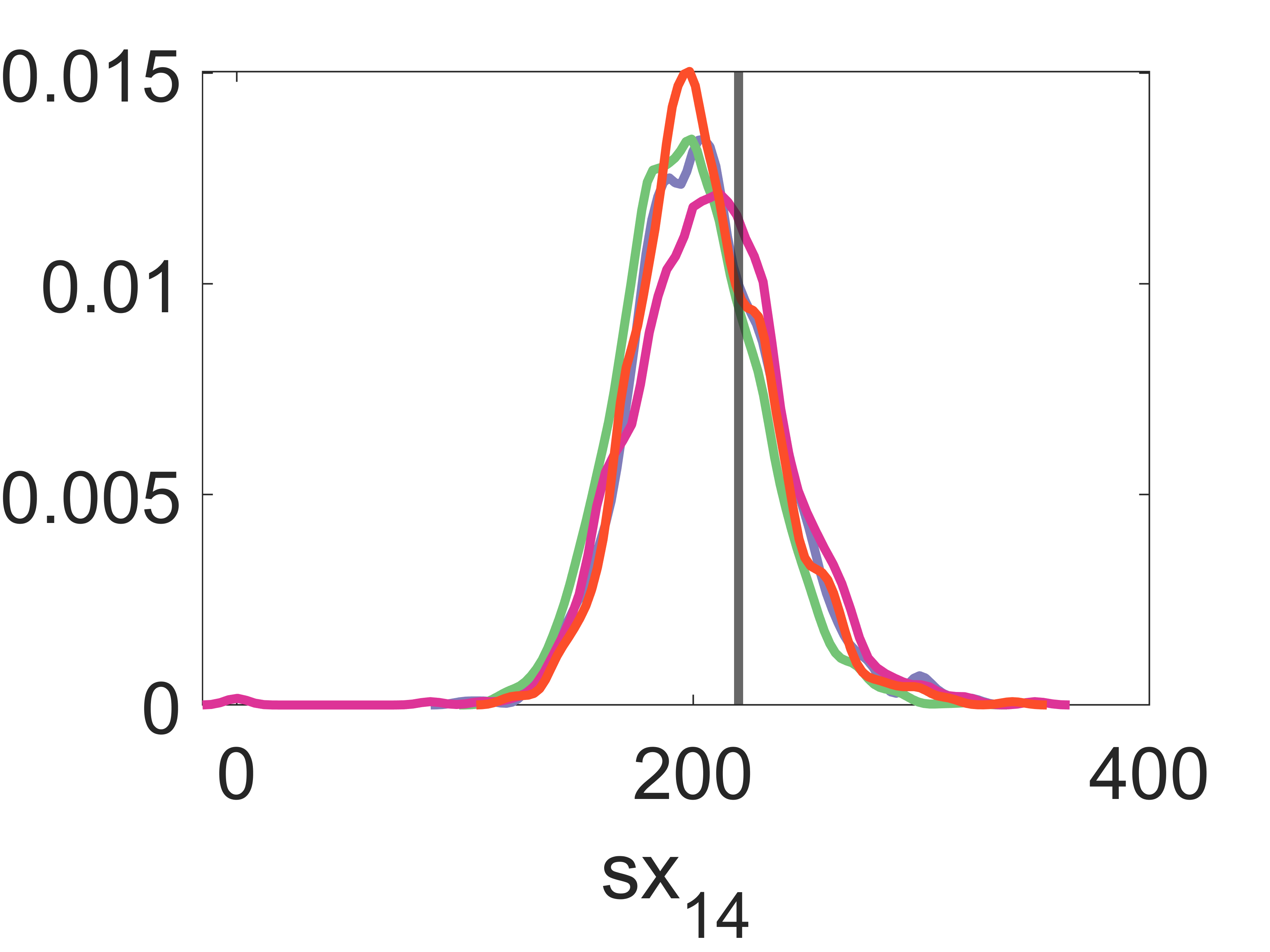}  
    \end{subfigure}
    \hfill
    \begin{subfigure}{.19\textwidth}
        \includegraphics[width=1\linewidth]{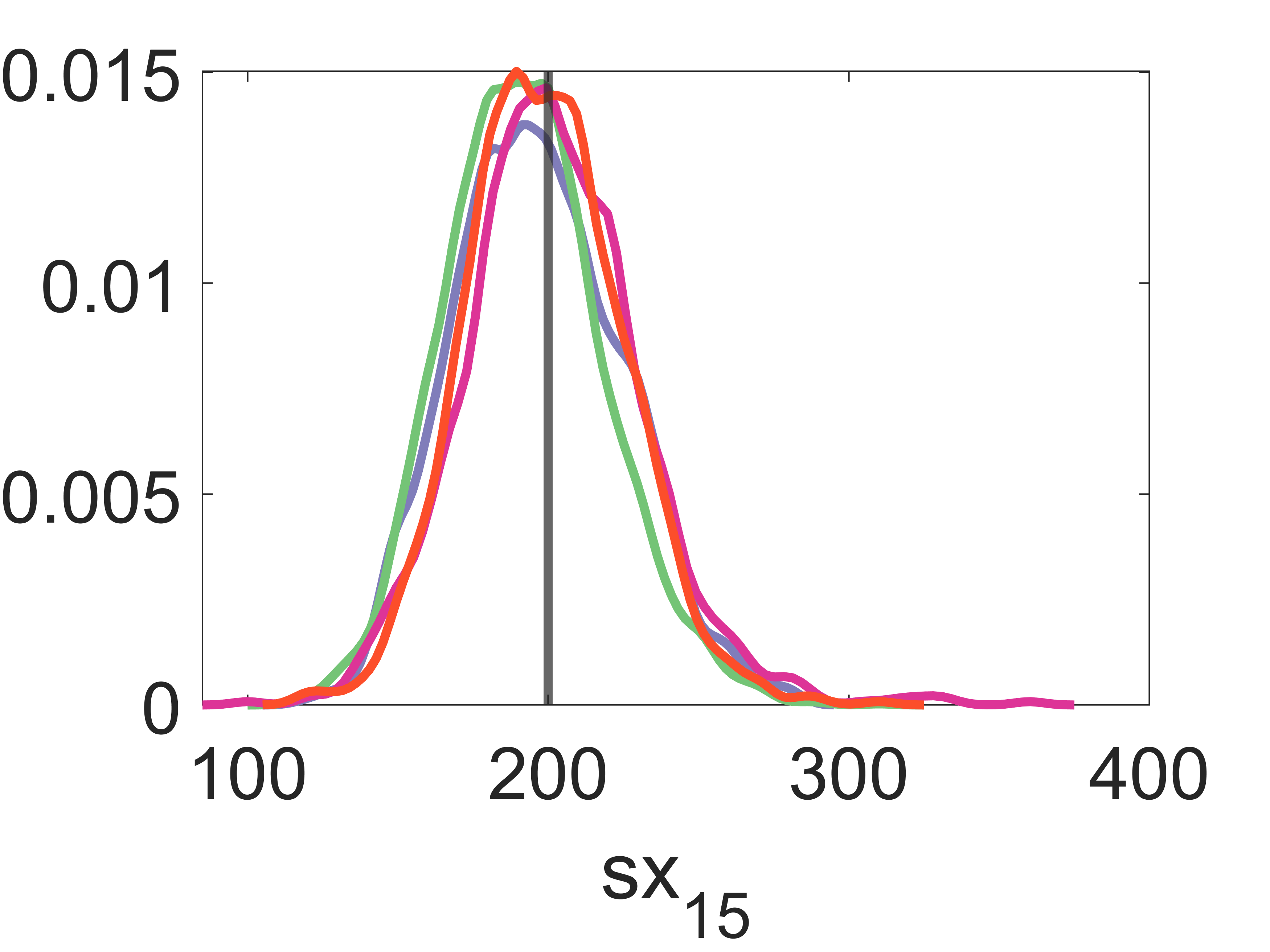}  
    \end{subfigure}
    \label{density synthetics data calibration}
\end{subfigure}
\caption{ \textbf{Posterior predictive distribution for the stochastic cell invasion model of summary statistics of synthetic datasets:} dataset with (a) cell trajectories and (b) cell density as summary statistics for cell movement across different methods. The violet solid lines show the SMC ABC's posterior predictive distributions, the green solid lines show those from the BSL method, the purple solid lines show the posterior predictive distribution by SNPE and the orange solid lines show those from SNLE. The black vertical lines are the summary statistics values for the synthetic dataset.}
\label{cell synthetics data calibration}
\end{figure}

We present the posterior predictive distribution in Figure \ref{cell synthetics data calibration}. It is evident that BSL outperforms other SBI algorithms on cell trajectory (Figure \ref{tracking synthetics data calibration}, green line). This indicates that obtaining estimation is acceptable since the summary statistics might not be sufficient for the parameters $(M_r, M_y, M_g)$. We also find that BSL outperforms others on cell density (Figure \ref{density synthetics data calibration}, green line) for $({sx}_1, {sx}_2, {sx}_3)$, which are the counts for cells in each cell cycle stage. These results verify that the estimated marginal posteriors for BSL are not overconfident. 

\subsubsection{Stage 2: SBI stage}
We recorded the computational cost for each algorithm on each synthetic dataset for both examples. All neural SBI algorithms (SNPE and SNLE) used 100k total simulations. SMC ABC required around 200k to 400k total simulations for the four synthetic datasets. BSL required 2 million and 3 million total simulations for cell trajectories and cell density, respectively. Additionally, we ran BSL with a 400k simulation budget (including the burn-in period) to check if BSL could still outperform the others. For cell trajectory, BSL performs similarly to SMC ABC under this computational budget, while for cell density, the burn-in period requires 450k simulations, making it difficult to compare the performance with others. If the computational time is more affordable, BSL might be a good choice, but that is not the case in this example. Considering both the posterior predictive distribution and computational cost, we chose SMC ABC, SNPE, and SNLE for the inference task on real datasets.

\subsubsection{Stage 3: uncertainty analysis stage}
\begin{figure}[H]
\begin{subfigure}{1\textwidth}
\caption{ }
    \begin{subfigure}{.32\textwidth}
        \includegraphics[width=1\linewidth]{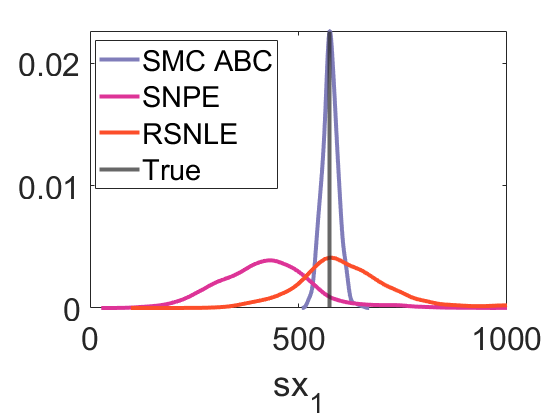}  
    \end{subfigure}
    \hfill
    \begin{subfigure}{.32\textwidth}
        \includegraphics[width=1\linewidth]{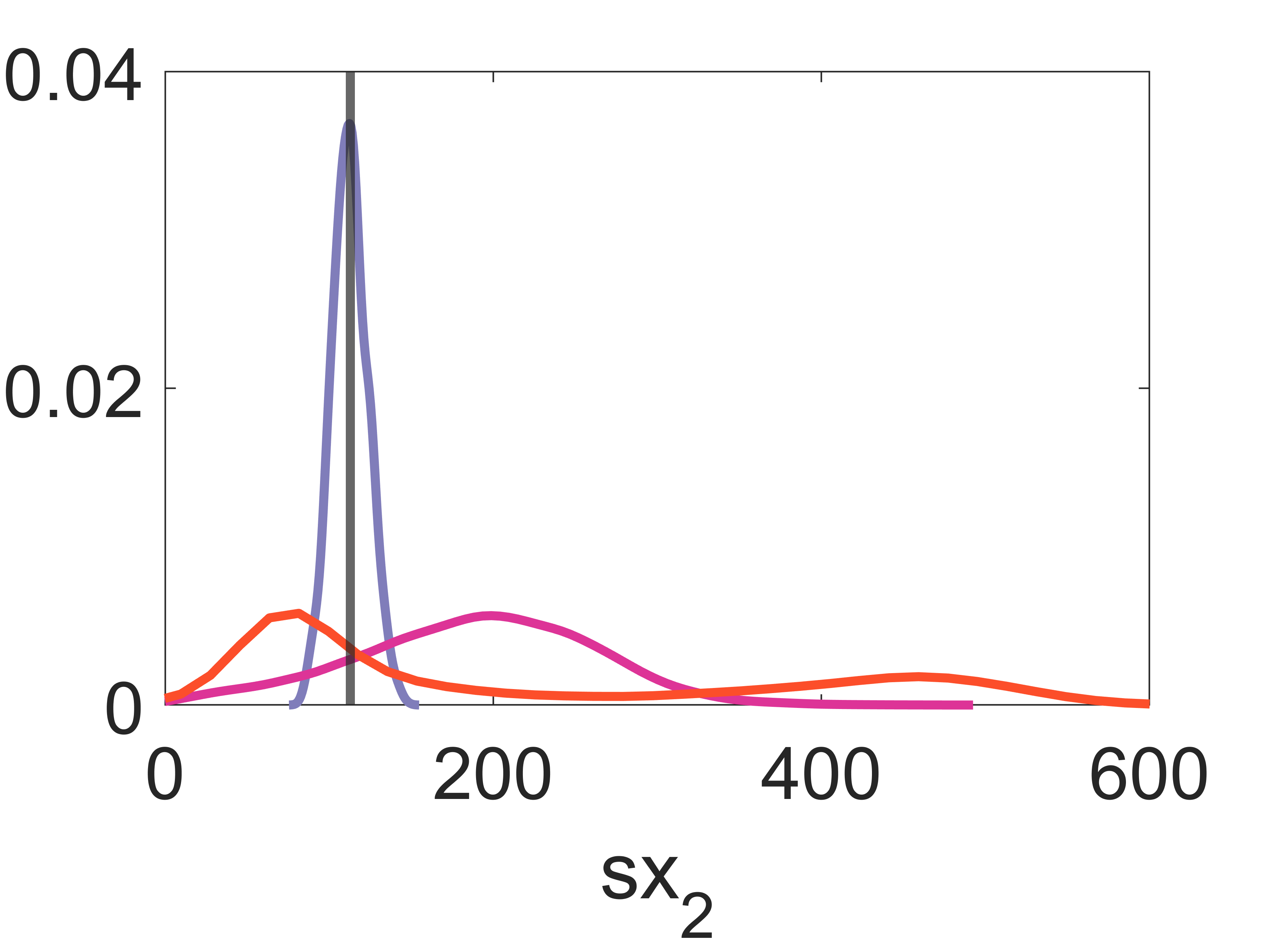}  
    \end{subfigure}
    \hfill
    \begin{subfigure}{.32\textwidth}
        \includegraphics[width=1\linewidth]{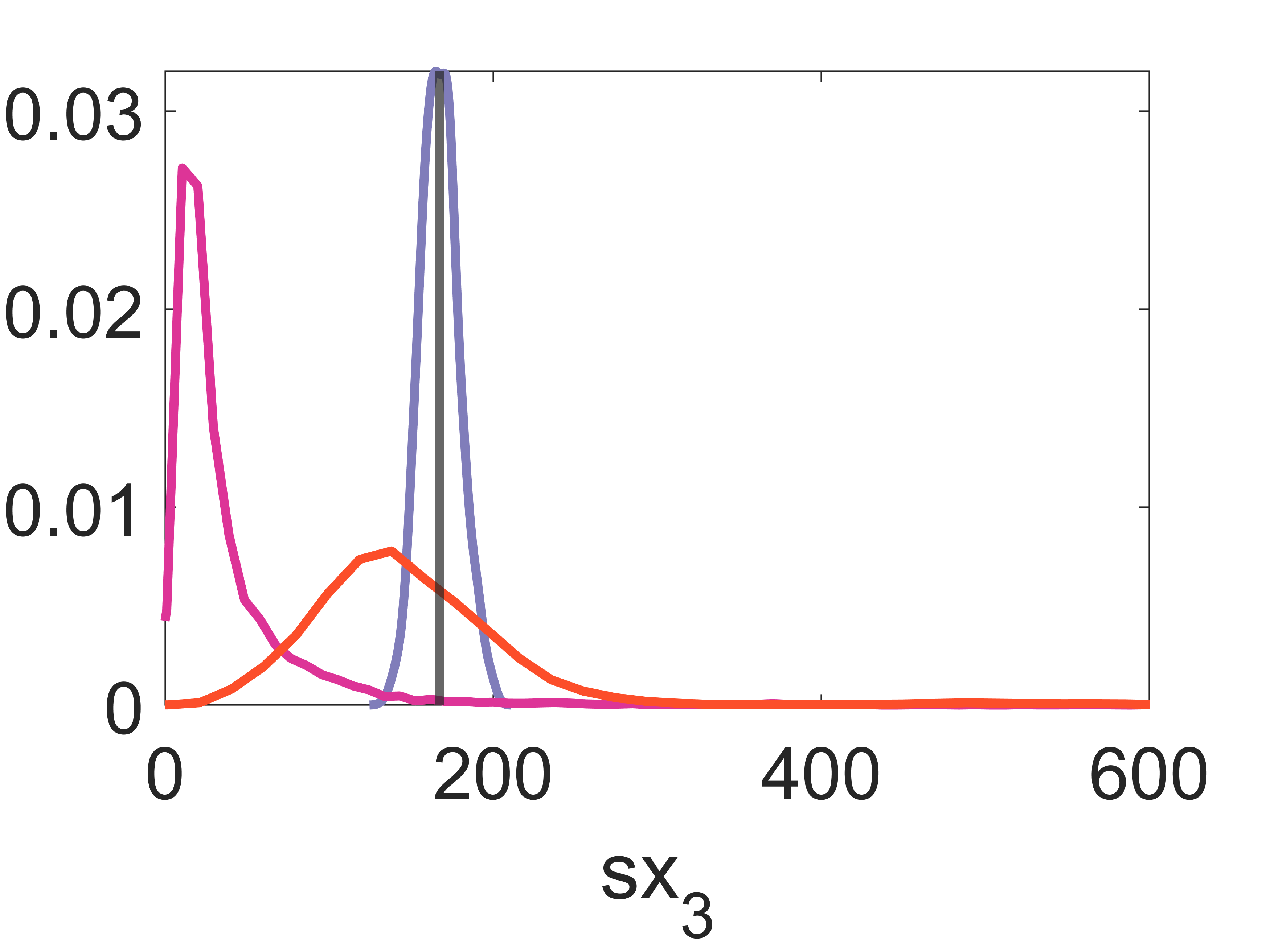}  
    \end{subfigure}
    \vfill
    \begin{subfigure}{.32\textwidth}
        \includegraphics[width=1\linewidth]{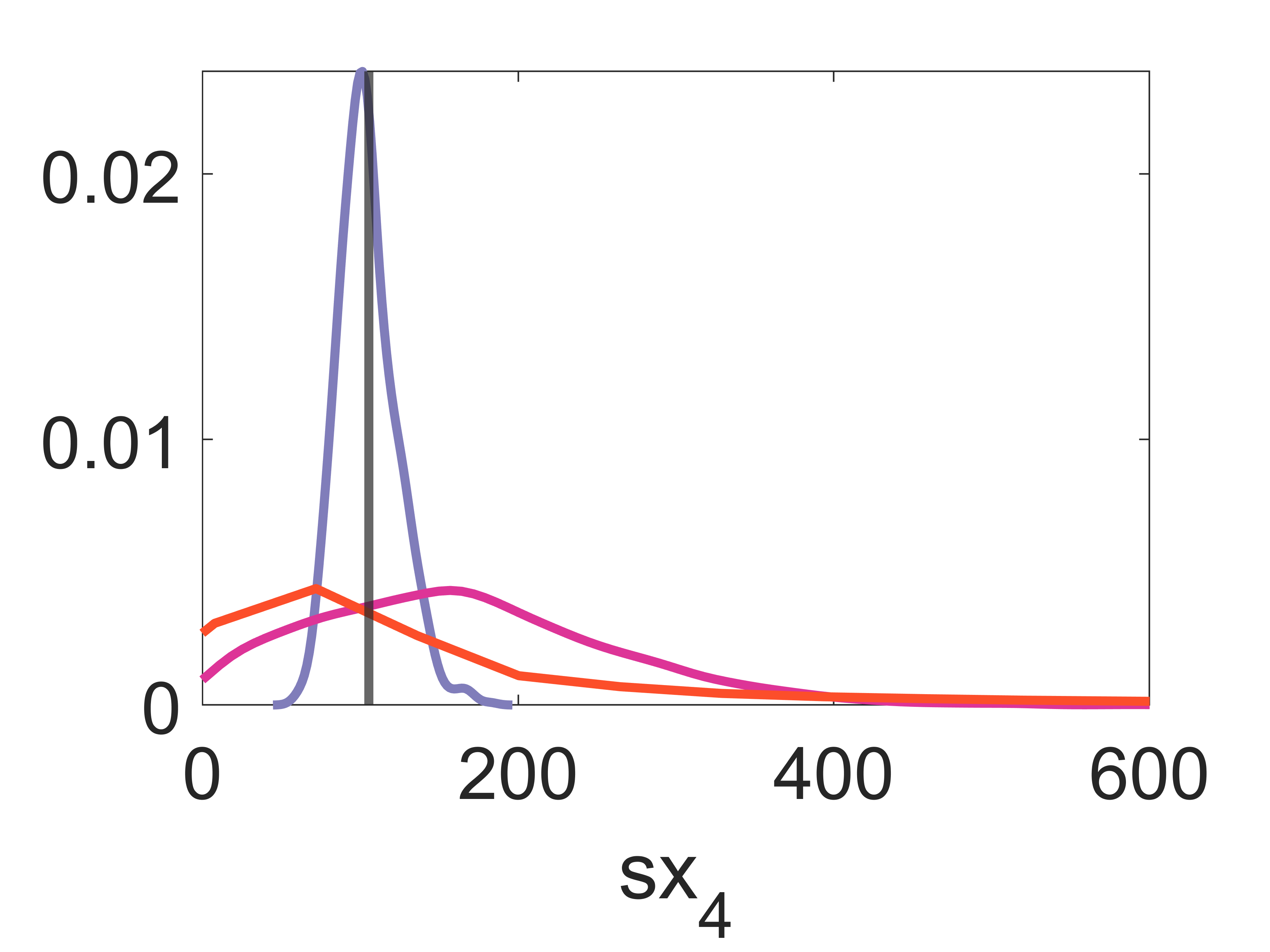}  
    \end{subfigure}
    \hfill
    \begin{subfigure}{.32\textwidth}
        \includegraphics[width=1\linewidth]{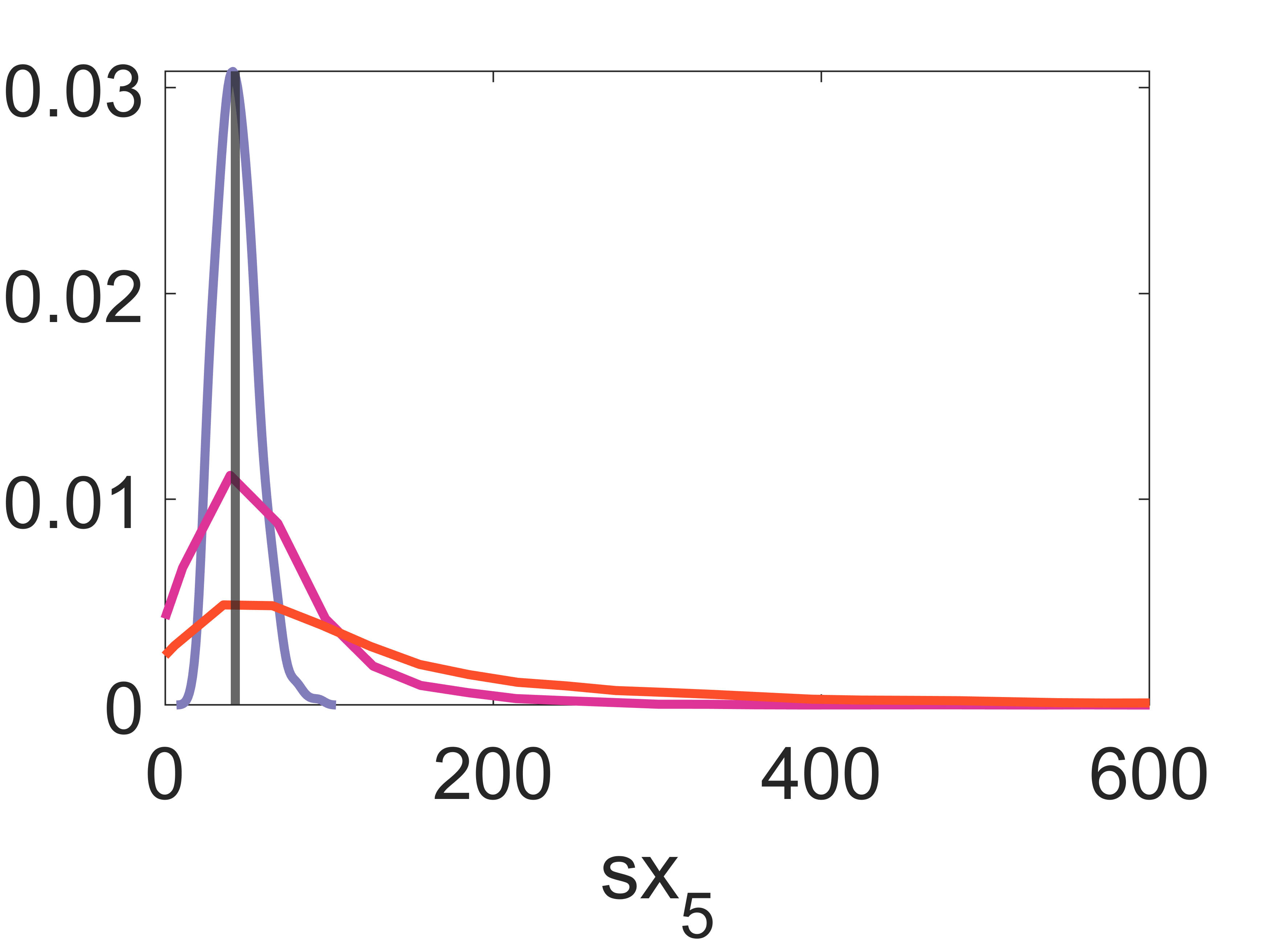}  
    \end{subfigure}
    \hfill
    \begin{subfigure}{.32\textwidth}
        \includegraphics[width=1\linewidth]{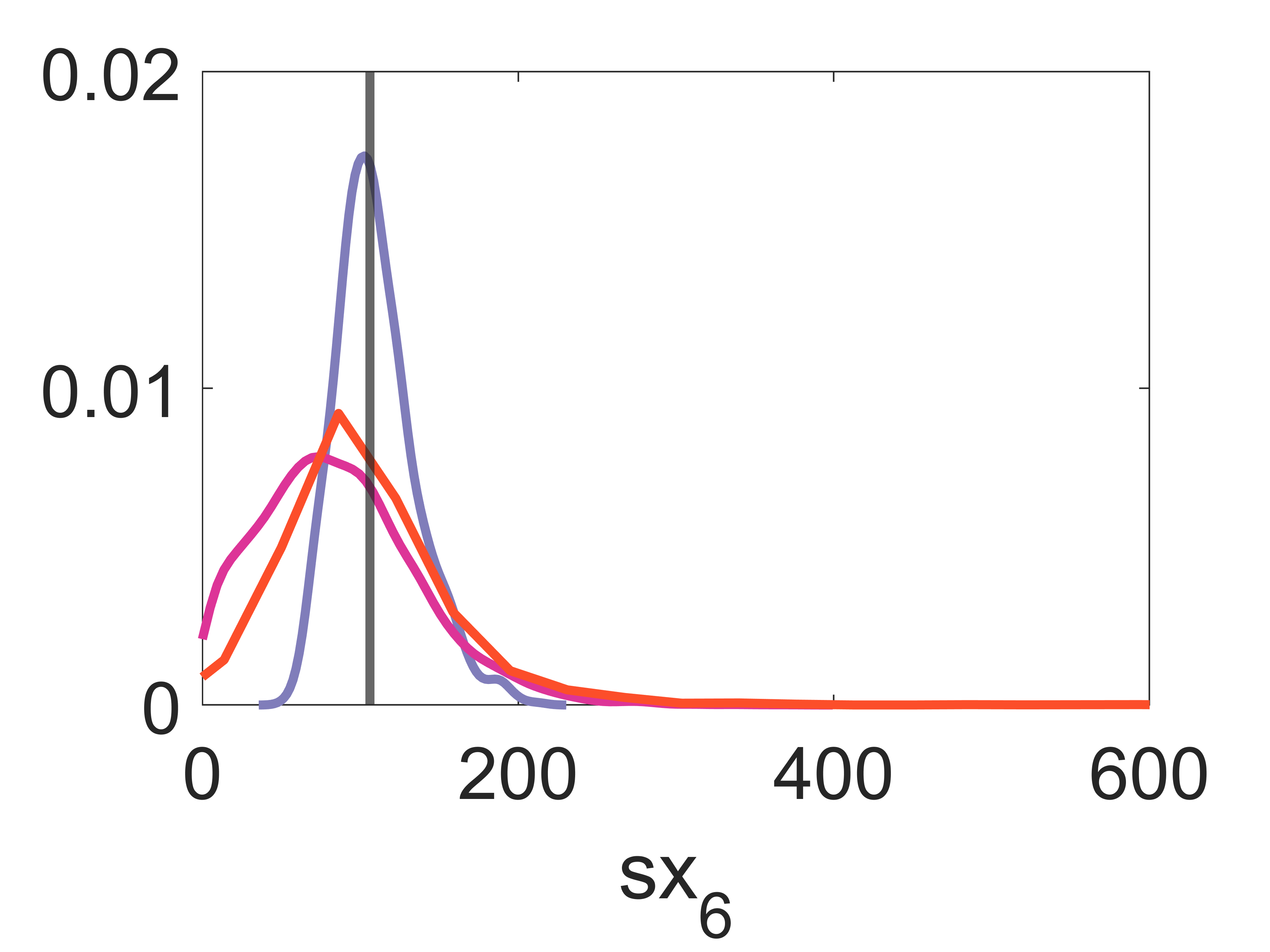}  
    \end{subfigure}
    \label{tracking real data calibration}
\end{subfigure}
\vfill
\begin{subfigure}{1\textwidth}
\caption{ }
    \begin{subfigure}{.19\textwidth}
        \includegraphics[width=1\linewidth]{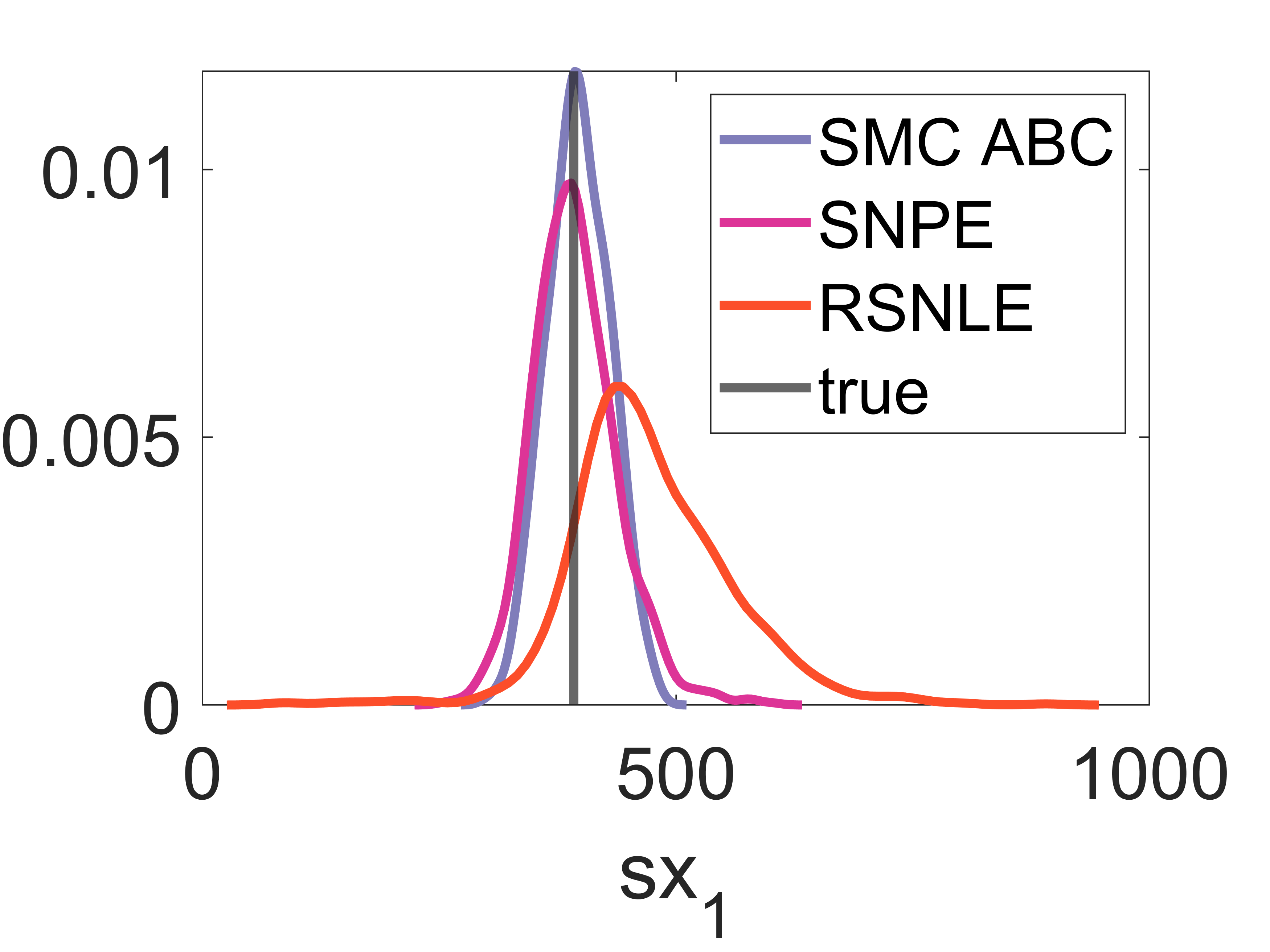}  
    \end{subfigure}
    \hfill
    \begin{subfigure}{.19\textwidth}
        \includegraphics[width=1\linewidth]{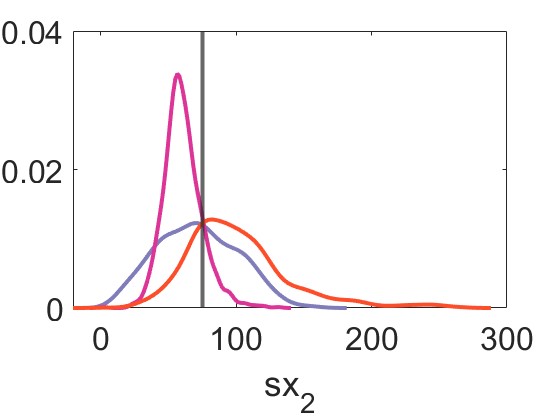}  
    \end{subfigure}
    \hfill
    \begin{subfigure}{.19\textwidth}
        \includegraphics[width=1\linewidth]{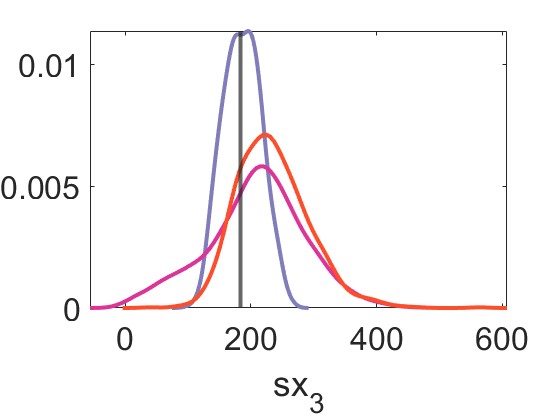}  
    \end{subfigure}
    \hfill
    \begin{subfigure}{.19\textwidth}
        \includegraphics[width=1\linewidth]{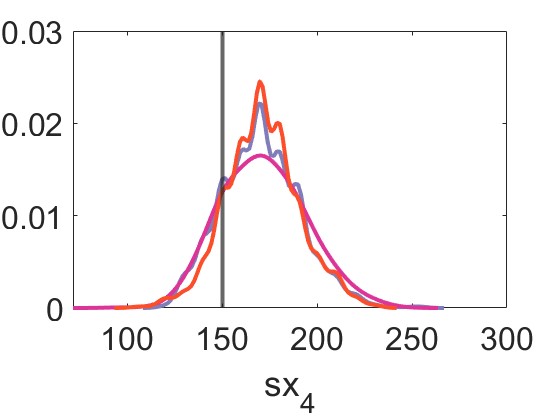}  
    \end{subfigure}
    \hfill
    \begin{subfigure}{.19\textwidth}
        \includegraphics[width=1\linewidth]{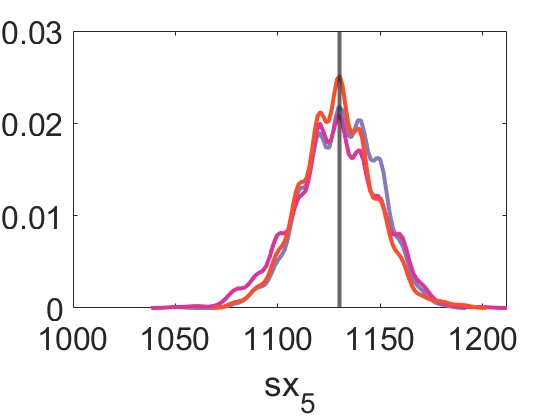}  
    \end{subfigure}
    \vfill
    \begin{subfigure}{.19\textwidth}
        \includegraphics[width=1\linewidth]{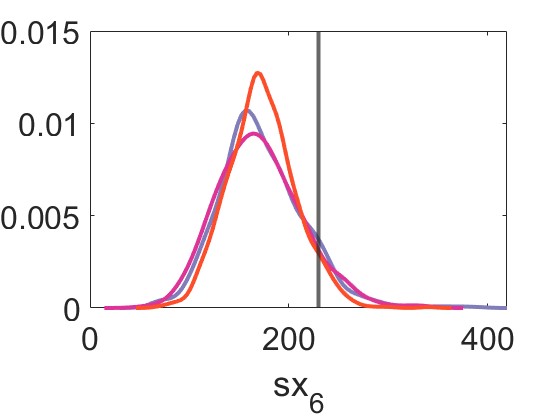}  
    \end{subfigure}
    \hfill
     \begin{subfigure}{.19\textwidth}
        \includegraphics[width=1\linewidth]{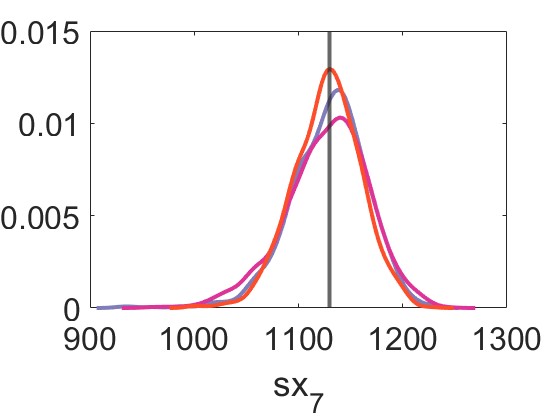}  
    \end{subfigure}
    \hfill
    \begin{subfigure}{.19\textwidth}
        \includegraphics[width=1\linewidth]{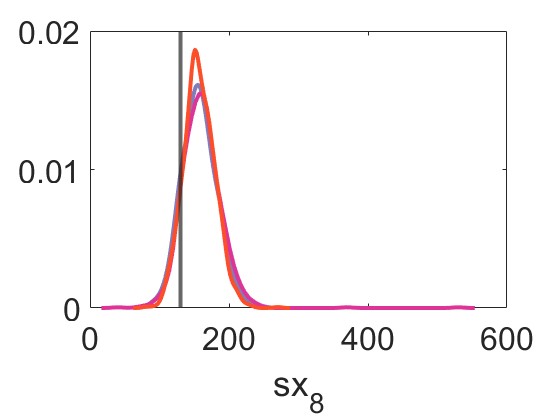}  
    \end{subfigure}
    \hfill
    \begin{subfigure}{.19\textwidth}
        \includegraphics[width=1\linewidth]{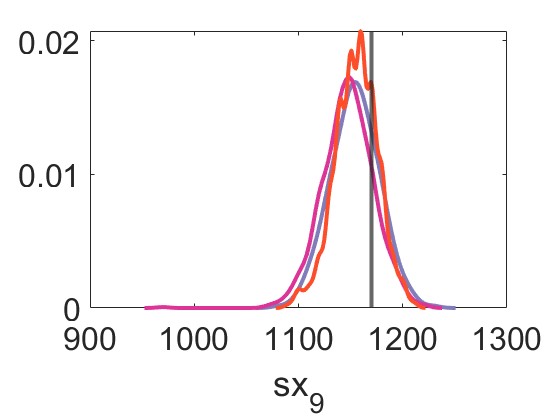}  
    \end{subfigure}
    \hfill
    \begin{subfigure}{.19\textwidth}
        \includegraphics[width=1\linewidth]{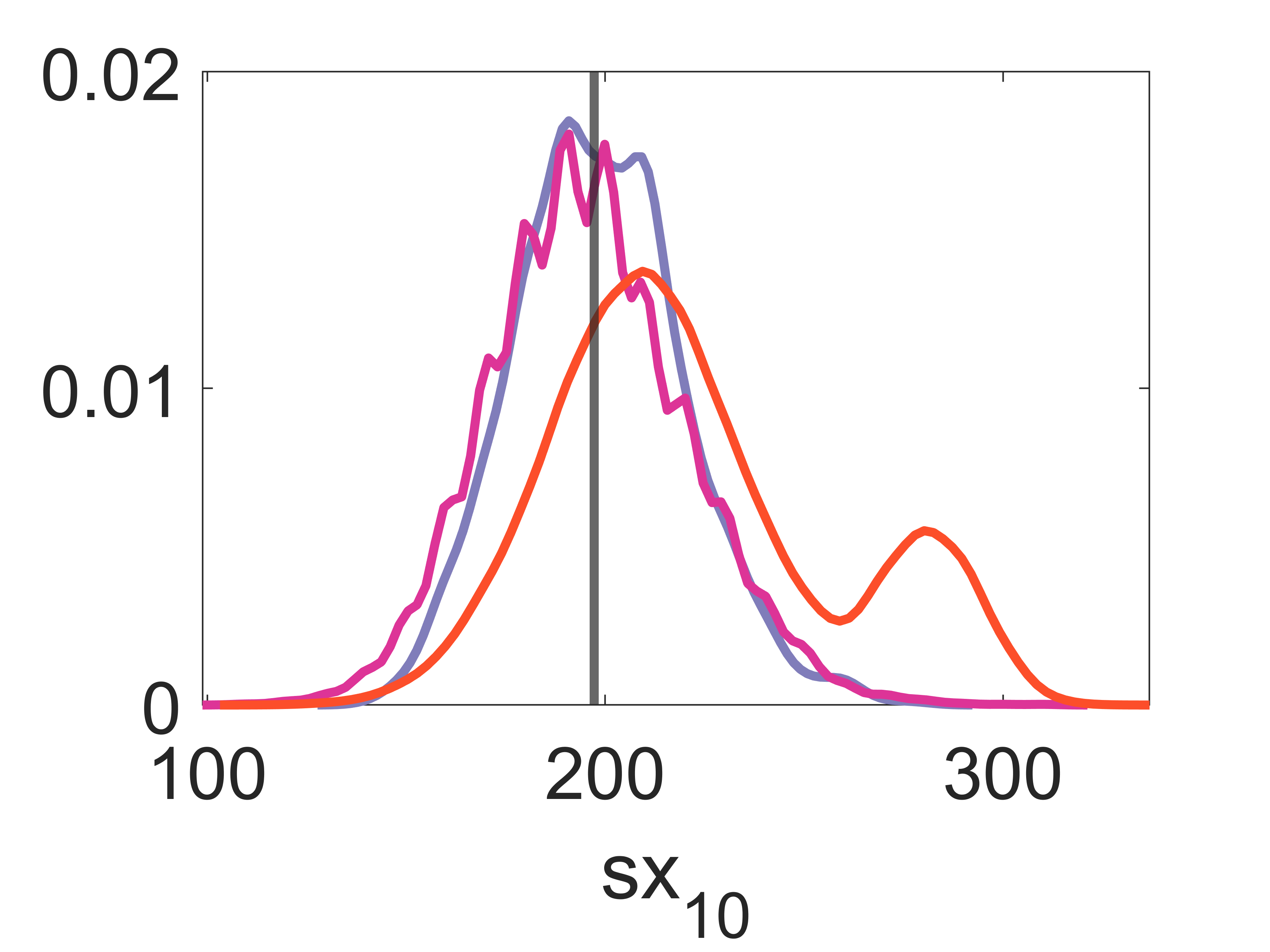}  
    \end{subfigure}
    \vfill
    \begin{subfigure}{.19\textwidth}
        \includegraphics[width=1\linewidth]{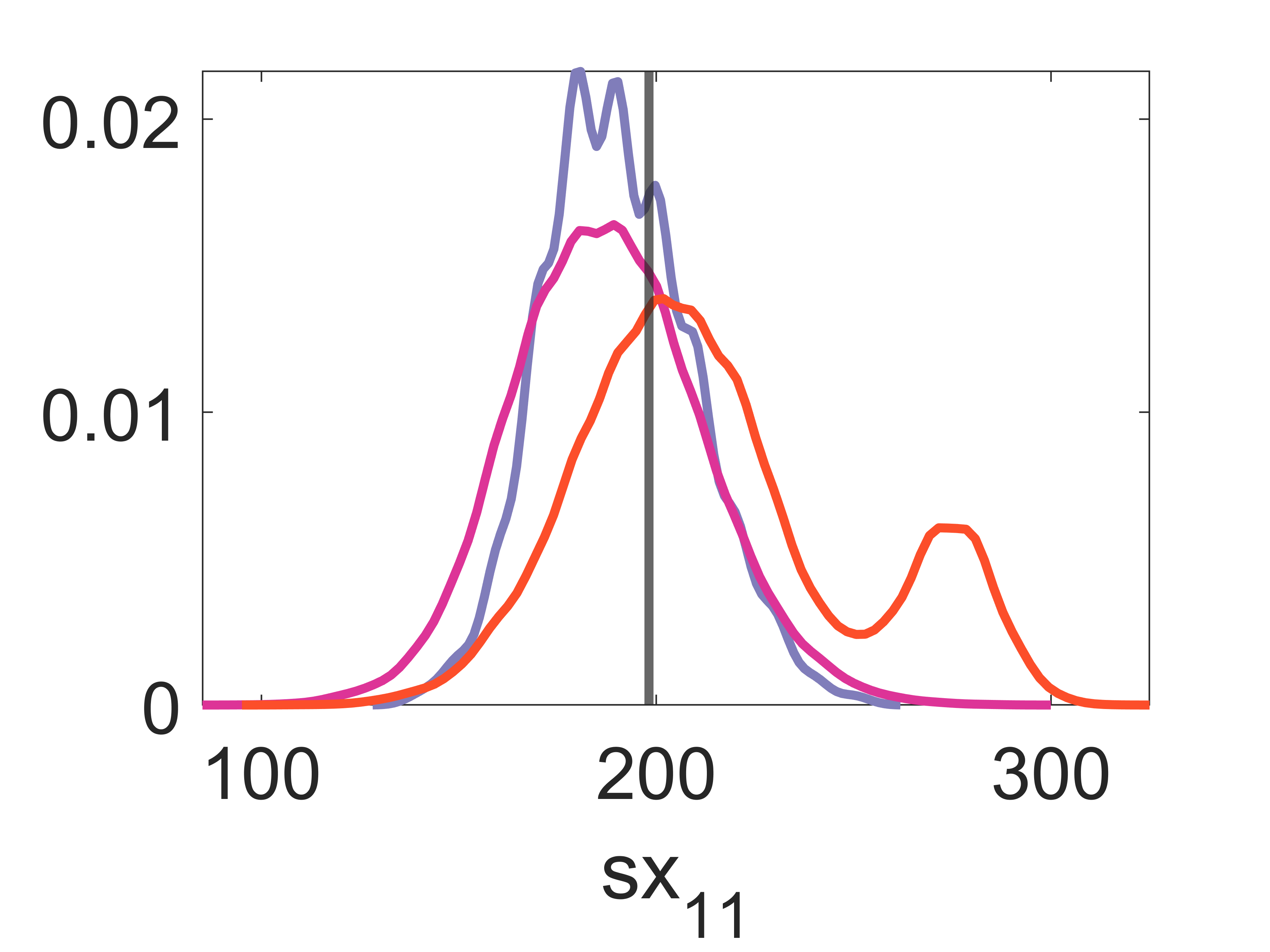}  
    \end{subfigure}
    \hfill
    \begin{subfigure}{.19\textwidth}
        \includegraphics[width=1\linewidth]{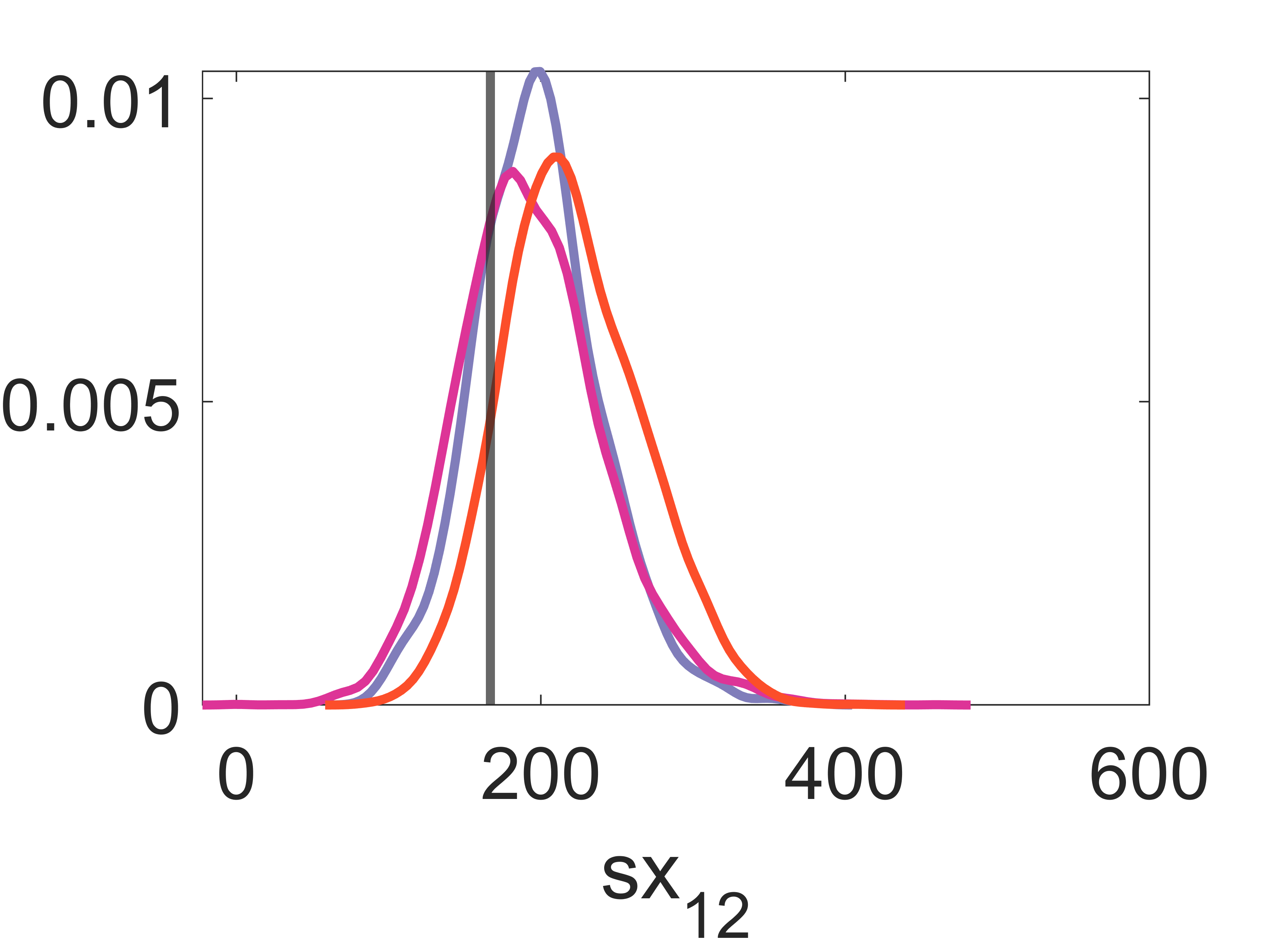}  
    \end{subfigure}
    \hfill
    \begin{subfigure}{.19\textwidth}
        \includegraphics[width=1\linewidth]{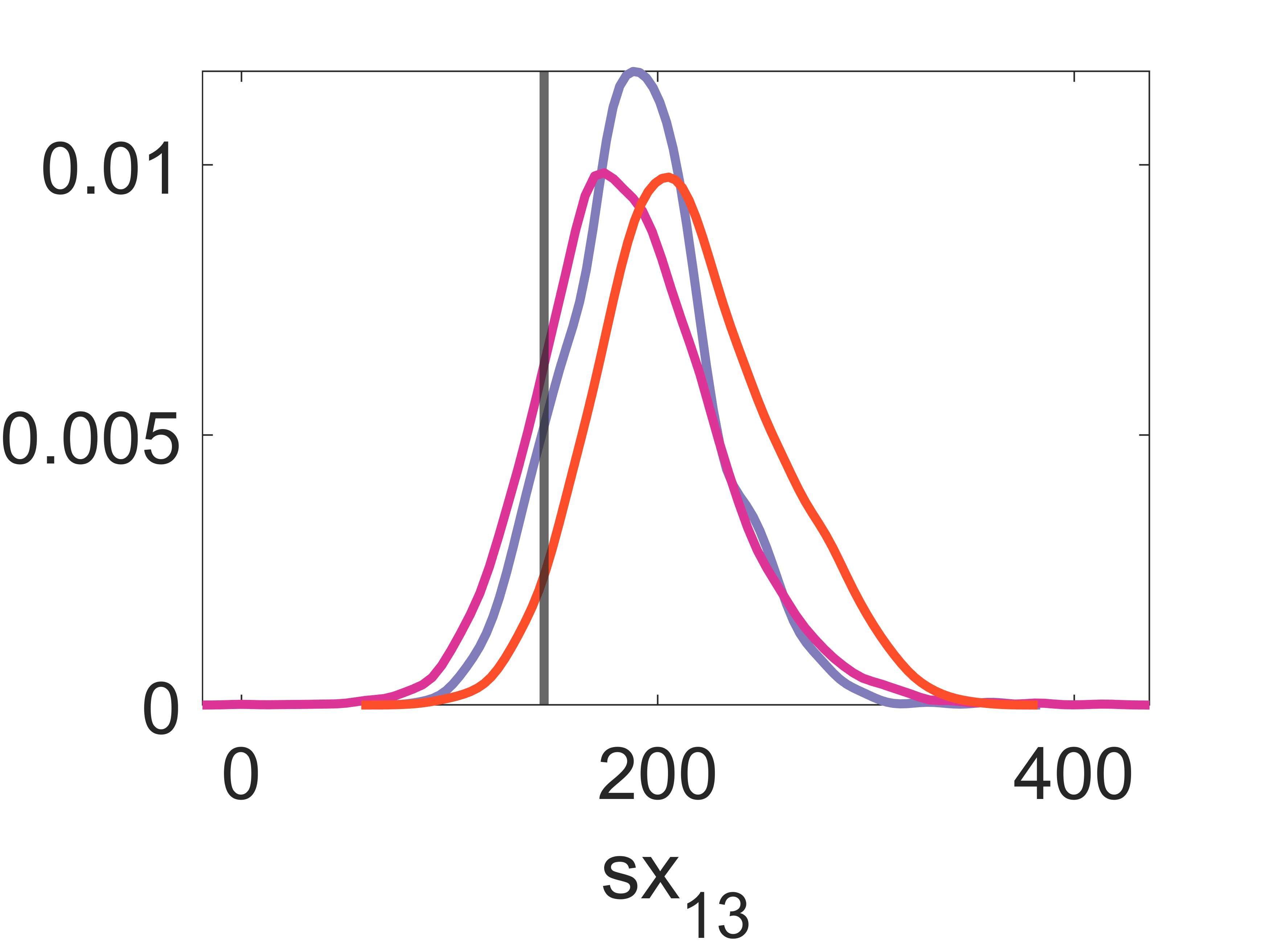}  
    \end{subfigure}
    \hfill
    \begin{subfigure}{.19\textwidth}
        \includegraphics[width=1\linewidth]{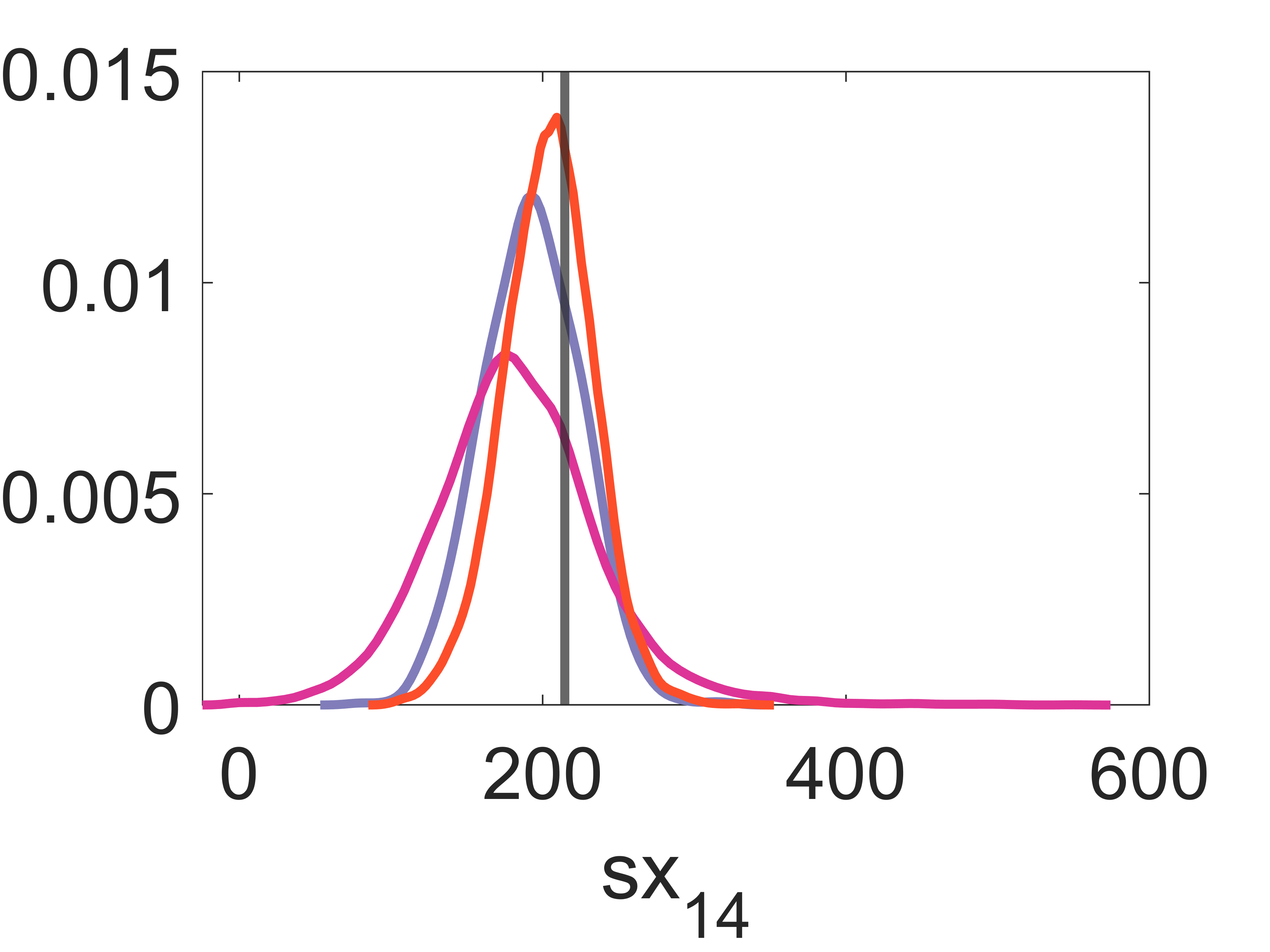}  
    \end{subfigure}
    \hfill
    \begin{subfigure}{.19\textwidth}
        \includegraphics[width=1\linewidth]{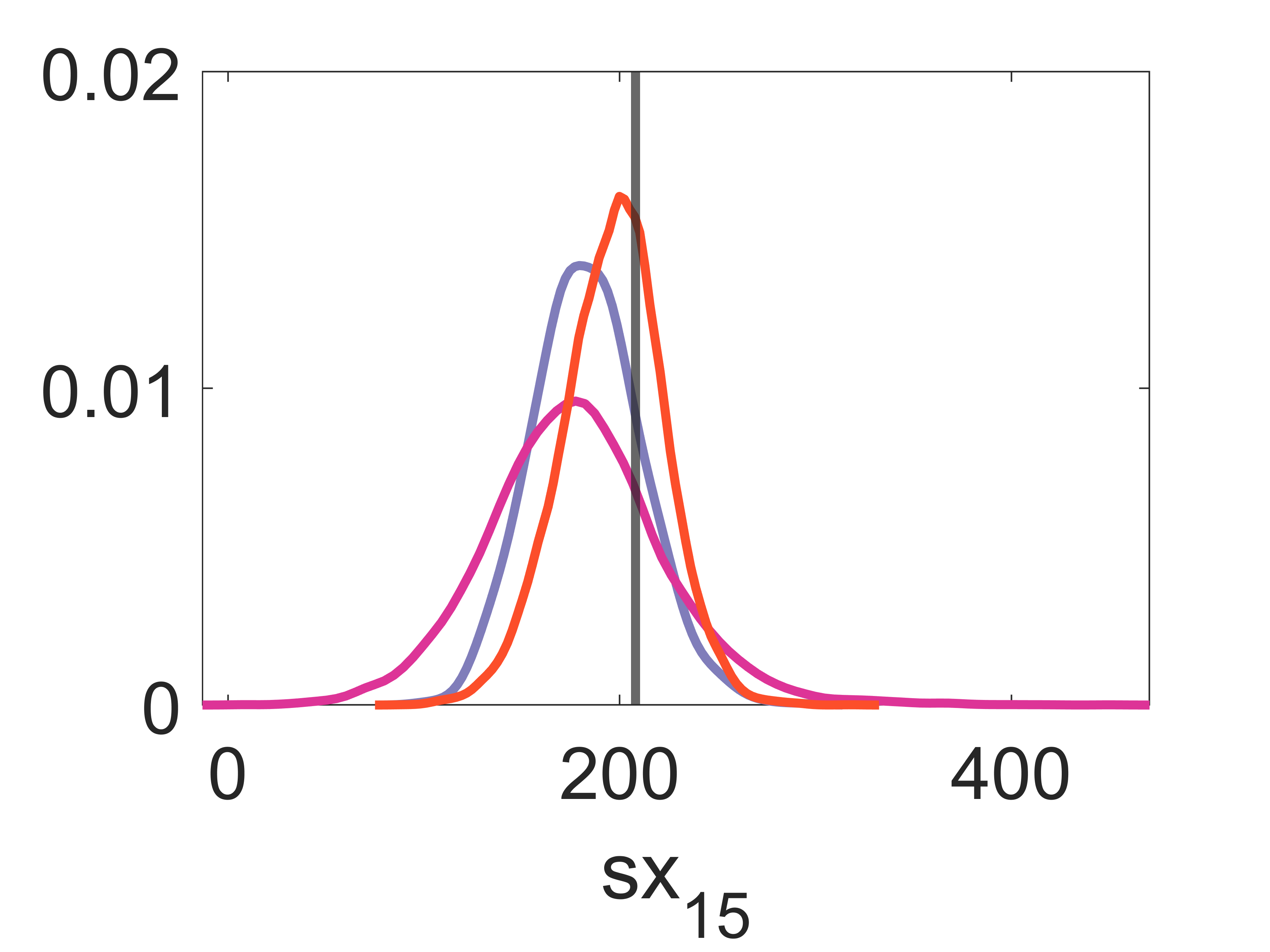}  
    \end{subfigure}
    \label{density real data calibration}
\end{subfigure}
\caption{ \textbf{Posterior predictive distribution for the stochastic cell invasion model of summary statistics of real datasets}: dataset with (a) cell trajectories and (b) cell density as summary statistics for cell movement across different methods. The violet solid lines show the SMC ABC's posterior predictive distributions, the purple solid lines show the posterior predictive distribution by SNPE and the orange solid lines show those from SNLE. The black vertical lines are the true summary statistics values.}
\label{cell real data calibration}
\end{figure}

We ran the selected candidate SBI algorithms on the real datasets for cell trajectory and cell density. As proposed in \textbf{Guideline 5}, the posterior predictive check is one of the few reliable tools to assess whether the model can adequately recover the real datasets. We present the posterior predictive distribution for both cell trajectory and cell density in Figure \ref{cell real data calibration}, and the estimated marginal posterior in Section \textcolor{blue}{C 2.2} of the Supplementary document. In Figure \ref{tracking real data calibration}, it is evident that SMC ABC outperforms on $({sx}_1, {sx}_2, {sx}_3)$, indicating that SMC ABC better captures cell proliferation. However, the neural SBI algorithms perform relatively poorly, especially for SNPE, it failed to recover ${sx}_3$. For cell density (Figure \ref{density real data calibration}), there is no significant difference between the posterior predictive distributions among all three algorithms, providing evidence that all three SBI algorithms are suitable.

As proposed in \textbf{Guideline 6}, we need to check for model misspecification based on the posterior predictive check. We present the predictive interval plots for both cell trajectory and cell density in Section \textcolor{blue}{C 2.2} of the Supplementary document for a more detailed description. In Figure \ref{cell real data calibration}, we can conclude that most of the real datasets lie within the high-density regions, indicating that the model is compatible. However, we noticed that the parameters ($M_r, M_y, M_g$) are non-identifiable for cell density, which suggests potential improvements to the experimental design so that it is possible to provide more information for the cell movement, as recommended by \textbf{Guideline 7}.

\section{Discussion}
In this study, we present a guideline with a set of stages to follow to determine which SBI algorithms are useful for a given computational model. We apply our guidelines to two agent-based models: the BVCBM and the cell invasion model. When we applied neural SBI algorithms to both examples, we found those can be more computational efficient than statistical SBI if the model is compatible. For the cell invasion model, summary statistics were used to mitigate the impact of highly noisy datasets, allowing the neural SBI methods to better balance the trade-off between computational cost and estimation accuracy. The neural networks perform strongly on datasets that are in-distribution (synthetic datasets), but can have difficulty generalising to actual observations that are out-of-distribution. In the BVCBM example, we confirmed that if the datasets are highly noisy, even robust neural SBI methods can fail to perform accurate inference.

Although neural SBI methods are more efficient than statistical SBI methods, the latter may still be preferred when real datasets are used. ABC methods, based on the comparison of data trajectories, find parameter sets that can generate data similar to real datasets if the model is compatible. As pointed out by \citet{frazier2020model}, the ABC method can converge to pseudo-true parameter values when the model is misspecified. BSL, if the multivariate Gaussian assumption holds, can provide highly accurate estimations. Secondly, statistical SBI methods are easier to analyse and understand compared to `black-box' neural SBI methods. Neural networks are sensitive to input data, which can lead to difficulties in interpretation. To make neural SBI methods work, it is necessary to apply many numerical techniques. However, if neural networks are successfully trained, such methods can often be more efficient than statistical SBI methods.

In this paper, we provide a comprehensive guideline for applying SBI methods to real datasets, which leads to unavoidable effects from model misspecification. As proposed in the guideline, the prior predictive check is a useful tool for providing initial insights into model misspecification. In this step, we only require a fixed number of model simulations (1k simulation for our demonstration). However, as discussed above, robust neural SBI algorithms might not perform well, and this can be due to factors other than model misspecification.

At the pre-analysis stage, neural SBI algorithms outperform statistical SBI algorithms not only in computational efficiency but also in estimation accuracy on synthetic datasets. There has been extensive investigation into the superior performance of neural SBI algorithms in well-specified settings, where the `true' datasets are generated from the DGP. It is not surprising to see neural SBI algorithms being the better candidates at this stage. However, as studied in \citet{cannon2022investigating}, neural SBI algorithms can be significantly affected by misspecification. Even when the model is compatible, neural SBI algorithms can still fail. For example, in the BVCBM case, SMC ABC produced posterior predictive distributions that covered most of the real pancreatic tumour growth datasets, indicating model compatibility. However, since the actual observations contain substantially more noise than the simulated datasets used to train the NCDE, even RSNL failed because the neural network fails to produce accurate inference on the out-of-distribution data.

For demonstration purposes, we focus on two SBI methods that are easy to implement and popular in the field of computational inference. The SMC ABC methods used in this paper can be replaced by more advanced methods, such as those proposed by \citet{hammer2024approximate, warne2022rapid, simola2021adaptive, raynal2019abc}. It is also possible to use full datasets instead of summary statistics by employing different distance metrics \citep{drovandi2022comparison}, such as Wasserstein ABC \citep{bernton2019approximate} and K2-ABC \citep{park2016k2}. For BSL, a semi-parametric approach \citep{an2020robust, priddle2023transformations} has been introduced to relax the restrictive normality assumption, and many approaches have been developed to accelerate the BSL method \citep{an2019accelerating, priddle2022efficient}. There is also a rich body of research on model misspecification in statistical SBI methods, such as \citet{frazier2020model, frazier2020robust, frazier2021robust, frazier2023bayesian, frazier2024synthetic}. Neural SBI methods, such as SNPE, have many variants, including the use of normalizing flows \citep{papamakarios2016fast, papamakarios2019sequential, lueckmann2017flexible, greenberg2019automatic, deistler2022energy} as CNDEs and the use of adversarial generative networks as classifiers to train a CNDE \citep{wang2022adversarial, glockler2023adversarial, ramesh2022gatsbi}. There is some research on model misspecification for neural SBI methods, such as \citet{kelly2024misspecification,huang2024learning,cannon2022investigating,dellaporta2022robust,lemos2023robust}.

Simulation-based inference is a rapidly growing research area in modern science, aimed at providing efficient and robust approaches to estimate parameters and quantify their uncertainty in increasingly complex and expensive biological models, where the likelihood is often intractable. The comprehensive guidelines provided can help readers better calibrate models with real datasets to build more realistic models. From the demonstrated examples, we can see that no single algorithm outperforms all others in every case, leading to a need to balance the trade-off between computational cost and estimation accuracy when selecting an SBI algorithm for real datasets. 

\section*{Acknowledgement}
We thank the computational resources provided by QUT’s High Performance Computing and Research Support Group (HPC). Xiaoyu Wang and Christopher Drovandi were supported by an Australian Research Council Future Fellowship (FT210100260). Ryan P. Kelly is supported by PhD Research Training Program scholarship from the Australian Government and a QUT Centre for Data Science top-up scholarship. Adrianne L Jenner was supported by an Australian Research Council Discovery Project (DP230100025)

\bibliography{ref}

\begin{thebibliography}{97}
\providecommand{\natexlab}[1]{#1}
\providecommand{\url}[1]{\texttt{#1}}
\expandafter\ifx\csname urlstyle\endcsname\relax
  \providecommand{\doi}[1]{doi: #1}\else
  \providecommand{\doi}{doi: \begingroup \urlstyle{rm}\Url}\fi

\bibitem[An et~al.(2009)An, Mi, Dutta-Moscato, and Vodovotz]{an2009agent}
G.~An, Q.~Mi, J.~Dutta-Moscato, and Y.~Vodovotz.
\newblock Agent-based models in translational systems biology.
\newblock \emph{{Wiley Interdisciplinary Reviews: Systems Biology and
  Medicine}}, 1\penalty0 (2):\penalty0 159--171, 2009.

\bibitem[An et~al.(2019)An, South, Nott, and Drovandi]{an2019accelerating}
Z.~An, L.~F. South, D.~J. Nott, and C.~C. Drovandi.
\newblock Accelerating {B}ayesian synthetic likelihood with the graphical
  lasso.
\newblock \emph{Journal of Computational and Graphical Statistics}, 28\penalty0
  (2):\penalty0 471--475, 2019.

\bibitem[An et~al.(2020)An, Nott, and Drovandi]{an2020robust}
Z.~An, D.~J. Nott, and C.~Drovandi.
\newblock Robust {B}ayesian synthetic likelihood via a semi-parametric
  approach.
\newblock \emph{Statistics and {C}omputing}, 30\penalty0 (3):\penalty0
  543--557, 2020.

\bibitem[Beaumont(2010)]{beaumont2010approximate}
M.~A. Beaumont.
\newblock Approximate {B}ayesian computation in evolution and ecology.
\newblock \emph{Annual {R}eview of {E}cology, {E}volution, and {S}ystematics},
  41:\penalty0 379--406, 2010.

\bibitem[Beaumont(2019)]{beaumont2019approximate}
M.~A. Beaumont.
\newblock Approximate {B}ayesian computation.
\newblock \emph{Annual {R}eview of {S}tatistics and its {A}pplication},
  6:\penalty0 379--403, 2019.

\bibitem[Beaumont et~al.(2002)Beaumont, Zhang, and
  Balding]{beaumont2002approximate}
M.~A. Beaumont, W.~Zhang, and D.~J. Balding.
\newblock Approximate {B}ayesian computation in population genetics.
\newblock \emph{Genetics}, 162\penalty0 (4):\penalty0 2025--2035, 2002.

\bibitem[Beaumont et~al.(2009)Beaumont, Cornuet, Marin, and
  Robert]{beaumont2009adaptive}
M.~A. Beaumont, J.-M. Cornuet, J.-M. Marin, and C.~P. Robert.
\newblock Adaptive approximate {B}ayesian computation.
\newblock \emph{Biometrika}, 96\penalty0 (4):\penalty0 983--990, 2009.

\bibitem[Bernton et~al.(2019)Bernton, Jacob, Gerber, and
  Robert]{bernton2019approximate}
E.~Bernton, P.~E. Jacob, M.~Gerber, and C.~P. Robert.
\newblock Approximate {B}ayesian computation with the {W}asserstein distance.
\newblock \emph{{Journal of the Royal Statistical Society Series B: Statistical
  Methodology}}, 81\penalty0 (2):\penalty0 235--269, 2019.

\bibitem[Biswas et~al.(2022)Biswas, Chen, Karslake, Farhat, Ames, Raiymbek,
  Freddolino, Biteen, and Ragunathan]{biswas2022hp1}
S.~Biswas, Z.~Chen, J.~D. Karslake, A.~Farhat, A.~Ames, G.~Raiymbek, P.~L.
  Freddolino, J.~S. Biteen, and K.~Ragunathan.
\newblock {HP1 oligomerization compensates for low-affinity H3K9me recognition
  and provides a tunable mechanism for heterochromatin-specific localization}.
\newblock \emph{Science {A}dvances}, 8\penalty0 (27):\penalty0 eabk0793, 2022.

\bibitem[Black et~al.(2011)Black, Harel, and Betsy~McCoach]{black2011missing}
A.~C. Black, O.~Harel, and D.~Betsy~McCoach.
\newblock Missing data techniques for multilevel data: Implications of model
  misspecification.
\newblock \emph{{Journal of Applied Statistics}}, 38\penalty0 (9):\penalty0
  1845--1865, 2011.

\bibitem[Box(1979)]{box1979all}
G.~Box.
\newblock All models are wrong, but some are useful.
\newblock \emph{Robustness in {S}tatistics}, 202\penalty0 (1979):\penalty0 549,
  1979.

\bibitem[Box and Tiao(2011)]{box2011bayesian}
G.~E. Box and G.~C. Tiao.
\newblock \emph{Bayesian inference in statistical analysis}.
\newblock John Wiley \& Sons, 2011.

\bibitem[Brodland(2015)]{brodland2015computational}
G.~W. Brodland.
\newblock How computational models can help unlock biological systems.
\newblock In \emph{Seminars in {C}ell \& {D}evelopmental {B}iology}, volume~47,
  pages 62--73. Elsevier, 2015.

\bibitem[Browning et~al.(2019)Browning, Haridas, and
  Simpson]{browning2019bayesian}
A.~P. Browning, P.~Haridas, and M.~J. Simpson.
\newblock {A Bayesian sequential learning framework to parameterise continuum
  models of melanoma invasion into human skin}.
\newblock \emph{{Bulletin of Mathematical Biology}}, 81\penalty0 (3):\penalty0
  676--698, 2019.

\bibitem[Browning et~al.(2022)Browning, Drovandi, Turner, Jenner, and
  Simpson]{browning2022efficient}
A.~P. Browning, C.~Drovandi, I.~W. Turner, A.~L. Jenner, and M.~J. Simpson.
\newblock Efficient inference and identifiability analysis for differential
  equation models with random parameters.
\newblock \emph{{PLOS Computational Biology}}, 18\penalty0 (11):\penalty0
  e1010734, 2022.

\bibitem[Cannon et~al.(2022)Cannon, Ward, and Schmon]{cannon2022investigating}
P.~Cannon, D.~Ward, and S.~M. Schmon.
\newblock Investigating the impact of model misspecification in neural
  simulation-based inference.
\newblock \emph{arXiv preprint arXiv:2209.01845}, 2022.

\bibitem[Carr et~al.(2021)Carr, Simpson, and Drovandi]{carr2021estimating}
M.~J. Carr, M.~J. Simpson, and C.~Drovandi.
\newblock {Estimating parameters of a stochastic cell invasion model with
  fluorescent cell cycle labelling using approximate Bayesian computation}.
\newblock \emph{{Journal of the Royal Society Interface}}, 18\penalty0
  (182):\penalty0 20210362, 2021.

\bibitem[Cranmer et~al.(2020)Cranmer, Brehmer, and Louppe]{cranmer2020frontier}
K.~Cranmer, J.~Brehmer, and G.~Louppe.
\newblock The frontier of simulation-based inference.
\newblock \emph{{Proceedings of the National Academy of Sciences}},
  117\penalty0 (48):\penalty0 30055--30062, 2020.

\bibitem[Csill{\'e}ry et~al.(2010)Csill{\'e}ry, Blum, Gaggiotti, and
  Fran{\c{c}}ois]{csillery2010approximate}
K.~Csill{\'e}ry, M.~G. Blum, O.~E. Gaggiotti, and O.~Fran{\c{c}}ois.
\newblock {Approximate Bayesian computation (ABC) in practice}.
\newblock \emph{Trends in {E}cology \& {E}volution}, 25\penalty0 (7):\penalty0
  410--418, 2010.

\bibitem[Deistler et~al.(2022{\natexlab{a}})Deistler, Goncalves, and
  Macke]{deistler2022truncated}
M.~Deistler, P.~J. Goncalves, and J.~H. Macke.
\newblock Truncated proposals for scalable and hassle-free simulation-based
  inference.
\newblock \emph{{Advances in Neural Information Processing Systems}},
  35:\penalty0 23135--23149, 2022{\natexlab{a}}.

\bibitem[Deistler et~al.(2022{\natexlab{b}})Deistler, Macke, and
  Gon{\c{c}}alves]{deistler2022energy}
M.~Deistler, J.~H. Macke, and P.~J. Gon{\c{c}}alves.
\newblock Energy-efficient network activity from disparate circuit parameters.
\newblock \emph{{Proceedings of the National Academy of Sciences}},
  119\penalty0 (44):\penalty0 e2207632119, 2022{\natexlab{b}}.

\bibitem[Dellaporta et~al.(2022)Dellaporta, Knoblauch, Damoulas, and
  Briol]{dellaporta2022robust}
C.~Dellaporta, J.~Knoblauch, T.~Damoulas, and F.-X. Briol.
\newblock Robust {B}ayesian inference for simulator-based models via the {MMD}
  posterior bootstrap.
\newblock In \emph{International Conference on Artificial Intelligence and
  Statistics}, pages 943--970. PMLR, 2022.

\bibitem[Drovandi and Frazier(2022)]{drovandi2022comparison}
C.~Drovandi and D.~T. Frazier.
\newblock A comparison of likelihood-free methods with and without summary
  statistics.
\newblock \emph{Statistics and Computing}, 32\penalty0 (3):\penalty0 1--23,
  2022.

\bibitem[Drovandi and Pettitt(2011)]{drovandi2011estimation}
C.~C. Drovandi and A.~N. Pettitt.
\newblock Estimation of parameters for macroparasite population evolution using
  approximate {B}ayesian computation.
\newblock \emph{Biometrics}, 67\penalty0 (1):\penalty0 225--233, 2011.

\bibitem[Durkan et~al.(2019)Durkan, Bekasov, Murray, and
  Papamakarios]{durkan2019neural}
C.~Durkan, A.~Bekasov, I.~Murray, and G.~Papamakarios.
\newblock Neural spline flows.
\newblock \emph{Advances in {N}eural {I}nformation {P}rocessing {S}ystems}, 32,
  2019.

\bibitem[Durkan et~al.(2020)Durkan, Murray, and
  Papamakarios]{durkan2020contrastive}
C.~Durkan, I.~Murray, and G.~Papamakarios.
\newblock On contrastive learning for likelihood-free inference.
\newblock In \emph{International conference on machine learning}, pages
  2771--2781. PMLR, 2020.

\bibitem[Fasiolo et~al.(2018)Fasiolo, Wood, Hartig, and
  Bravington]{fasiolo2018extended}
M.~Fasiolo, S.~N. Wood, F.~Hartig, and M.~V. Bravington.
\newblock An extended empirical saddlepoint approximation for intractable
  likelihoods.
\newblock \emph{{Electronic Journal of Statistics}}, 12(1):\penalty0
  1544--1578, 2018.

\bibitem[Fisher et~al.(2019)Fisher, Rudin, and Dominici]{fisher2019all}
A.~Fisher, C.~Rudin, and F.~Dominici.
\newblock {All Models are Wrong, but Many are Useful: Learning a Variable's
  Importance by Studying an Entire Class of Prediction Models Simultaneously.}
\newblock \emph{{Journal Machine Learning Research}}, 20\penalty0
  (177):\penalty0 1--81, 2019.

\bibitem[Frazier and Drovandi(2021)]{frazier2021robust}
D.~T. Frazier and C.~Drovandi.
\newblock Robust approximate {B}ayesian inference with synthetic likelihood.
\newblock \emph{{Journal of Computational and Graphical Statistics}},
  30\penalty0 (4):\penalty0 958--976, 2021.

\bibitem[Frazier et~al.(2020{\natexlab{a}})Frazier, Drovandi, and
  Loaiza-Maya]{frazier2020robust}
D.~T. Frazier, C.~Drovandi, and R.~Loaiza-Maya.
\newblock Robust approximate {B}ayesian computation: {A}n adjustment approach.
\newblock \emph{arXiv preprint arXiv:2008.04099}, 2020{\natexlab{a}}.

\bibitem[Frazier et~al.(2020{\natexlab{b}})Frazier, Robert, and
  Rousseau]{frazier2020model}
D.~T. Frazier, C.~P. Robert, and J.~Rousseau.
\newblock Model misspecification in approximate {B}ayesian computation:
  consequences and diagnostics.
\newblock \emph{{Journal of the Royal Statistical Society Series B: Statistical
  Methodology}}, 82\penalty0 (2):\penalty0 421--444, 2020{\natexlab{b}}.

\bibitem[Frazier et~al.(2023)Frazier, Nott, Drovandi, and
  Kohn]{frazier2023bayesian}
D.~T. Frazier, D.~J. Nott, C.~Drovandi, and R.~Kohn.
\newblock Bayesian inference using synthetic likelihood: asymptotics and
  adjustments.
\newblock \emph{Journal of the American Statistical Association}, 118\penalty0
  (544):\penalty0 2821--2832, 2023.

\bibitem[Frazier et~al.(2024)Frazier, Nott, and Drovandi]{frazier2024synthetic}
D.~T. Frazier, D.~J. Nott, and C.~Drovandi.
\newblock Synthetic likelihood in misspecified models.
\newblock \emph{{Journal of the American Statistical Association}}, \penalty0
  (In press):\penalty0 1--23, 2024.

\bibitem[Frostig et~al.(2018)Frostig, Johnson, and Leary]{frostig2018compiling}
R.~Frostig, M.~J. Johnson, and C.~Leary.
\newblock Compiling machine learning programs via high-level tracing.
\newblock \emph{Systems for {M}achine {L}earning}, 4\penalty0 (9), 2018.

\bibitem[Ghasemi et~al.(2011)Ghasemi, Lindsey, Yang, Nguyen, Huang, and
  Jin]{ghasemi2011bayesian}
O.~Ghasemi, M.~L. Lindsey, T.~Yang, N.~Nguyen, Y.~Huang, and Y.-F. Jin.
\newblock Bayesian parameter estimation for nonlinear modelling of biological
  pathways.
\newblock \emph{BMC {S}ystems {B}iology}, 5\penalty0 (3):\penalty0 1--10, 2011.

\bibitem[Gloeckler et~al.(2023)Gloeckler, Deistler, and
  Macke]{glockler2023adversarial}
M.~Gloeckler, M.~Deistler, and J.~H. Macke.
\newblock Adversarial robustness of amortized {B}ayesian inference.
\newblock \emph{{Proceedings of the 40th International Conference on Machine
  Learning}}, 2023.

\bibitem[Greenberg et~al.(2019)Greenberg, Nonnenmacher, and
  Macke]{greenberg2019automatic}
D.~Greenberg, M.~Nonnenmacher, and J.~Macke.
\newblock Automatic posterior transformation for likelihood-free inference.
\newblock In \emph{International {C}onference on {M}achine {L}earning}, pages
  2404--2414. PMLR, 2019.

\bibitem[Hammer et~al.(2024)Hammer, Riegler, and
  Tjelmeland]{hammer2024approximate}
H.~L. Hammer, M.~A. Riegler, and H.~Tjelmeland.
\newblock Approximate {B}ayesian inference based on expected evaluation.
\newblock \emph{Bayesian Analysis}, 19\penalty0 (3):\penalty0 677--698, 2024.

\bibitem[Hinkelmann et~al.(2011)Hinkelmann, Murrugarra, Jarrah, and
  Laubenbacher]{hinkelmann2011mathematical}
F.~Hinkelmann, D.~Murrugarra, A.~S. Jarrah, and R.~Laubenbacher.
\newblock A mathematical framework for agent based models of complex biological
  networks.
\newblock \emph{Bulletin of {M}athematical {B}iology}, 73\penalty0
  (7):\penalty0 1583--1602, 2011.

\bibitem[Huang et~al.(2024)Huang, Bharti, Souza, Acerbi, and
  Kaski]{huang2024learning}
D.~Huang, A.~Bharti, A.~Souza, L.~Acerbi, and S.~Kaski.
\newblock Learning robust statistics for simulation-based inference under model
  misspecification.
\newblock \emph{Advances in Neural Information Processing Systems}, 36, 2024.

\bibitem[Jenner et~al.(2020)Jenner, Frascoli, Coster, and
  Kim]{jenner2020enhancing}
A.~L. Jenner, F.~Frascoli, A.~C. Coster, and P.~S. Kim.
\newblock Enhancing oncolytic virotherapy: {O}bservations from a {V}oronoi
  {C}ell-{B}ased model.
\newblock \emph{Journal of {T}heoretical {B}iology}, 485:\penalty0 110052,
  2020.

\bibitem[Jenner et~al.(2023)Jenner, Kelly, Dallaston, Araujo, Parfitt,
  Steinitz, Pooladvand, Kim, Wade, and Vine]{jenner2023examining}
A.~L. Jenner, W.~Kelly, M.~Dallaston, R.~Araujo, I.~Parfitt, D.~Steinitz,
  P.~Pooladvand, P.~S. Kim, S.~J. Wade, and K.~L. Vine.
\newblock Examining the efficacy of localised gemcitabine therapy for the
  treatment of pancreatic cancer using a hybrid agent-based model.
\newblock \emph{{PLOS Computational Biology}}, 19\penalty0 (1):\penalty0
  e1010104, 2023.

\bibitem[J{\o}rgensen et~al.(2022)J{\o}rgensen, Ghosh, Sturrock, and
  Shahrezaei]{jorgensen2022efficient}
A.~C.~S. J{\o}rgensen, A.~Ghosh, M.~Sturrock, and V.~Shahrezaei.
\newblock Efficient {B}ayesian inference for stochastic agent-based models.
\newblock \emph{{PLoS Computational Biology}}, 18\penalty0 (10):\penalty0
  e1009508, 2022.

\bibitem[Kelly et~al.(2024)Kelly, Nott, Frazier, Warne, and
  Drovandi]{kelly2024misspecification}
R.~Kelly, D.~J. Nott, D.~T. Frazier, D.~Warne, and C.~Drovandi.
\newblock Misspecification-robust sequential neural likelihood for
  simulation-based inference.
\newblock \emph{{Transactions on Machine Learning Research}}, 2024\penalty0
  (June), 2024.

\bibitem[Klowss et~al.(2022)Klowss, Browning, Murphy, Carr, Plank, Gunasingh,
  Haass, and Simpson]{klowss2022stochastic}
J.~J. Klowss, A.~P. Browning, R.~J. Murphy, E.~J. Carr, M.~J. Plank,
  G.~Gunasingh, N.~K. Haass, and M.~J. Simpson.
\newblock A stochastic mathematical model of 4d tumour spheroids with real-time
  fluorescent cell cycle labelling.
\newblock \emph{{Journal of the Royal Society Interface}}, 19\penalty0
  (189):\penalty0 20210903, 2022.

\bibitem[Lemos et~al.(2023)Lemos, Cranmer, Abidi, Hahn, Eickenberg, Massara,
  Yallup, and Ho]{lemos2023robust}
P.~Lemos, M.~Cranmer, M.~Abidi, C.~Hahn, M.~Eickenberg, E.~Massara, D.~Yallup,
  and S.~Ho.
\newblock Robust simulation-based inference in cosmology with {B}ayesian neural
  networks.
\newblock \emph{Machine Learning: Science and Technology}, 4\penalty0
  (1):\penalty0 01LT01, 2023.

\bibitem[Leung(2013)]{leung2013systems}
A.~W. Leung.
\newblock \emph{Systems of nonlinear partial differential equations:
  applications to biology and engineering}, volume~49.
\newblock Springer Science \& Business Media, 2013.

\bibitem[Liepe et~al.(2014)Liepe, Kirk, Filippi, Toni, Barnes, and
  Stumpf]{liepe2014framework}
J.~Liepe, P.~Kirk, S.~Filippi, T.~Toni, C.~P. Barnes, and M.~P. Stumpf.
\newblock A framework for parameter estimation and model selection from
  experimental data in systems biology using approximate {B}ayesian
  computation.
\newblock \emph{Nature {P}rotocols}, 9\penalty0 (2):\penalty0 439--456, 2014.

\bibitem[Lillacci and Khammash(2010)]{lillacci2010parameter}
G.~Lillacci and M.~Khammash.
\newblock Parameter estimation and model selection in computational biology.
\newblock \emph{{PLoS Computational Biology}}, 6\penalty0 (3):\penalty0
  e1000696, 2010.

\bibitem[Lueckmann et~al.(2017)Lueckmann, Goncalves, Bassetto, {\"O}cal,
  Nonnenmacher, and Macke]{lueckmann2017flexible}
J.-M. Lueckmann, P.~J. Goncalves, G.~Bassetto, K.~{\"O}cal, M.~Nonnenmacher,
  and J.~H. Macke.
\newblock Flexible statistical inference for mechanistic models of neural
  dynamics.
\newblock \emph{Advances in {N}eural {I}nformation {P}rocessing {S}ystems}, 30,
  2017.

\bibitem[Lueckmann et~al.(2021)Lueckmann, Boelts, Greenberg, Goncalves, and
  Macke]{lueckmann2021benchmarking}
J.-M. Lueckmann, J.~Boelts, D.~Greenberg, P.~Goncalves, and J.~Macke.
\newblock Benchmarking simulation-based inference.
\newblock In \emph{International {C}onference on {A}rtificial {I}ntelligence
  and {S}tatistics}, pages 343--351. PMLR, 2021.

\bibitem[Marin et~al.(2014)Marin, Pillai, Robert, and
  Rousseau]{marin2014relevant}
J.-M. Marin, N.~S. Pillai, C.~P. Robert, and J.~Rousseau.
\newblock Relevant statistics for {B}ayesian model choice.
\newblock \emph{{Journal of the Royal Statistical Society Series B: Statistical
  Methodology}}, 76\penalty0 (5):\penalty0 833--859, 2014.

\bibitem[Meineke et~al.(2001)Meineke, Potten, and Loeffler]{meineke2001cell}
F.~A. Meineke, C.~S. Potten, and M.~Loeffler.
\newblock Cell migration and organization in the intestinal crypt using a
  lattice-free model.
\newblock \emph{Cell {P}roliferation}, 34\penalty0 (4):\penalty0 253--266,
  2001.

\bibitem[Metzcar et~al.(2019)Metzcar, Wang, Heiland, and
  Macklin]{metzcar2019review}
J.~Metzcar, Y.~Wang, R.~Heiland, and P.~Macklin.
\newblock A review of cell-based computational modeling in cancer biology.
\newblock \emph{{JCO Clinical Cancer Informatics}}, 2:\penalty0 1--13, 2019.

\bibitem[Mori{\~n}a et~al.(2023)Mori{\~n}a, Fern{\'a}ndez-Fontelo, Caba{\~n}a,
  Arratia, and Puig]{morina2023estimated}
D.~Mori{\~n}a, A.~Fern{\'a}ndez-Fontelo, A.~Caba{\~n}a, A.~Arratia, and
  P.~Puig.
\newblock {Estimated Covid-19 burden in Spain: ARCH underreported
  non-stationary time series}.
\newblock \emph{BMC {M}edical {R}esearch {M}ethodology}, 23\penalty0
  (1):\penalty0 75, 2023.

\bibitem[N{\ae}vdal and Evje(2023)]{naevdal2023can}
G.~N{\ae}vdal and S.~Evje.
\newblock Can cancer cells inform us about the tumor microenvironment?
\newblock \emph{{Journal of Computational Physics}}, 492:\penalty0 112449,
  2023.

\bibitem[Nott et~al.(2023)Nott, Drovandi, and Frazier]{nott2023bayesian}
D.~J. Nott, C.~Drovandi, and D.~T. Frazier.
\newblock Bayesian inference for misspecified generative models.
\newblock \emph{{Annual Review of Statistics and Its Application}}, 11, 2023.

\bibitem[Papamakarios and Murray(2016)]{papamakarios2016fast}
G.~Papamakarios and I.~Murray.
\newblock Fast $\varepsilon$-free inference of simulation models with
  {B}ayesian conditional density estimation.
\newblock \emph{Advances in {N}eural {I}nformation {P}rocessing {S}ystems}, 29,
  2016.

\bibitem[Papamakarios et~al.(2019)Papamakarios, Sterratt, and
  Murray]{papamakarios2019sequential}
G.~Papamakarios, D.~Sterratt, and I.~Murray.
\newblock Sequential neural likelihood: Fast likelihood-free inference with
  autoregressive flows.
\newblock In \emph{The 22nd International Conference on Artificial Intelligence
  and Statistics}, pages 837--848. PMLR, 2019.

\bibitem[Papamakarios et~al.(2021)Papamakarios, Nalisnick, Rezende, Mohamed,
  and Lakshminarayanan]{papamakarios2021normalizing}
G.~Papamakarios, E.~Nalisnick, D.~J. Rezende, S.~Mohamed, and
  B.~Lakshminarayanan.
\newblock Normalizing flows for probabilistic modeling and inference.
\newblock \emph{{The Journal of Machine Learning Research}}, 22\penalty0
  (1):\penalty0 2617--2680, 2021.

\bibitem[Park et~al.(2016)Park, Jitkrittum, and Sejdinovic]{park2016k2}
M.~Park, W.~Jitkrittum, and D.~Sejdinovic.
\newblock {K2-ABC: Approximate Bayesian computation with kernel embeddings}.
\newblock In \emph{{Artificial intelligence and statistics}}, pages 398--407.
  PMLR, 2016.

\bibitem[Picchini and Forman(2019)]{picchini2019bayesian}
U.~Picchini and J.~L. Forman.
\newblock Bayesian inference for stochastic differential equation mixed effects
  models of a tumour xenography study.
\newblock \emph{{Journal of the Royal Statistical Society Series C: Applied
  Statistics}}, 68\penalty0 (4):\penalty0 887--913, 2019.

\bibitem[Prescott et~al.(2021)Prescott, Zhu, Zhao, and
  Baker]{prescott2021quantifying}
T.~P. Prescott, K.~Zhu, M.~Zhao, and R.~E. Baker.
\newblock Quantifying the impact of electric fields on single-cell motility.
\newblock \emph{Biophysical {J}ournal}, 120\penalty0 (16):\penalty0 3363--3373,
  2021.

\bibitem[Price et~al.(2018)Price, Drovandi, Lee, and Nott]{price2018bayesian}
L.~F. Price, C.~C. Drovandi, A.~Lee, and D.~J. Nott.
\newblock Bayesian synthetic likelihood.
\newblock \emph{{Journal of Computational and Graphical Statistics}},
  27\penalty0 (1):\penalty0 1--11, 2018.

\bibitem[Priddle and Drovandi(2023)]{priddle2023transformations}
J.~W. Priddle and C.~Drovandi.
\newblock Transformations in semi-parametric {B}ayesian synthetic likelihood.
\newblock \emph{{Computational Statistics \& Data Analysis}}, 187:\penalty0
  107797, 2023.

\bibitem[Priddle et~al.(2022)Priddle, Sisson, Frazier, Turner, and
  Drovandi]{priddle2022efficient}
J.~W. Priddle, S.~A. Sisson, D.~T. Frazier, I.~Turner, and C.~Drovandi.
\newblock Efficient {B}ayesian synthetic likelihood with whitening
  transformations.
\newblock \emph{{Journal of Computational and Graphical Statistics}},
  31\penalty0 (1):\penalty0 50--63, 2022.

\bibitem[Ramesh et~al.(2022)Ramesh, Lueckmann, Boelts, Tejero-Cantero,
  Greenberg, Gon{\c{c}}alves, and Macke]{ramesh2022gatsbi}
P.~Ramesh, J.-M. Lueckmann, J.~Boelts, {\'A}.~Tejero-Cantero, D.~S. Greenberg,
  P.~J. Gon{\c{c}}alves, and J.~H. Macke.
\newblock {GATSBI}: Generative {A}dversarial {T}raining for
  {S}imulation-{B}ased {I}nference.
\newblock \emph{arXiv preprint arXiv:2203.06481}, 2022.

\bibitem[Ram{\'\i}rez-Hassan and Frazier(2024)]{ramirez2024testing}
A.~Ram{\'\i}rez-Hassan and D.~T. Frazier.
\newblock Testing model specification in approximate {B}ayesian computation
  using asymptotic properties.
\newblock \emph{Journal of Computational and Graphical Statistics}, \penalty0
  (just-accepted):\penalty0 1--14, 2024.

\bibitem[Raynal et~al.(2019)Raynal, Marin, Pudlo, Ribatet, Robert, and
  Estoup]{raynal2019abc}
L.~Raynal, J.-M. Marin, P.~Pudlo, M.~Ribatet, C.~P. Robert, and A.~Estoup.
\newblock {ABC random forests for Bayesian parameter inference}.
\newblock \emph{Bioinformatics}, 35\penalty0 (10):\penalty0 1720--1728, 2019.

\bibitem[Ross et~al.(2017)Ross, Baker, Parker, Ford, Mort, and
  Yates]{ross2017using}
R.~J. Ross, R.~E. Baker, A.~Parker, M.~Ford, R.~Mort, and C.~Yates.
\newblock Using approximate {B}ayesian computation to quantify cell-cell
  adhesion parameters in a cell migratory process.
\newblock \emph{{NPJ Systems Biology and Applications}}, 3\penalty0
  (1):\penalty0 9, 2017.

\bibitem[Sakaue-Sawano et~al.(2008)Sakaue-Sawano, Kurokawa, Morimura, Hanyu,
  Hama, Osawa, Kashiwagi, Fukami, Miyata, Miyoshi,
  et~al.]{sakaue2008visualizing}
A.~Sakaue-Sawano, H.~Kurokawa, T.~Morimura, A.~Hanyu, H.~Hama, H.~Osawa,
  S.~Kashiwagi, K.~Fukami, T.~Miyata, H.~Miyoshi, et~al.
\newblock Visualizing spatiotemporal dynamics of multicellular cell-cycle
  progression.
\newblock \emph{Cell}, 132\penalty0 (3):\penalty0 487--498, 2008.

\bibitem[Simola et~al.(2021)Simola, Cisewski-Kehe, Gutmann, and
  Corander]{simola2021adaptive}
U.~Simola, J.~Cisewski-Kehe, M.~U. Gutmann, and J.~Corander.
\newblock {Adaptive approximate Bayesian computation tolerance selection}.
\newblock \emph{Bayesian analysis}, 16\penalty0 (2):\penalty0 397--423, 2021.

\bibitem[Simpson et~al.(2018)Simpson, Jin, Vittadello, Tambyah, Ryan,
  Gunasingh, Haass, and McCue]{simpson2018stochastic}
M.~J. Simpson, W.~Jin, S.~T. Vittadello, T.~A. Tambyah, J.~M. Ryan,
  G.~Gunasingh, N.~K. Haass, and S.~W. McCue.
\newblock Stochastic models of cell invasion with fluorescent cell cycle
  indicators.
\newblock \emph{{Physica A: Statistical Mechanics and its Applications}},
  510:\penalty0 375--386, 2018.

\bibitem[Sisson et~al.(2018)Sisson, Fan, and Beaumont]{sisson2018handbook}
S.~A. Sisson, Y.~Fan, and M.~Beaumont.
\newblock \emph{Handbook of approximate Bayesian computation}.
\newblock CRC Press, 2018.

\bibitem[Sorensen et~al.(2002)Sorensen, Gianola, and
  Gianola]{sorensen2002likelihood}
D.~Sorensen, D.~Gianola, and D.~Gianola.
\newblock {Likelihood, Bayesian and MCMC methods in quantitative genetics}.
\newblock 2002.

\bibitem[Spencer et~al.(2004)Spencer, Berryman, Garc{\'\i}a, and
  Abbott]{spencer2004ordinary}
S.~L. Spencer, M.~J. Berryman, J.~A. Garc{\'\i}a, and D.~Abbott.
\newblock An ordinary differential equation model for the multistep
  transformation to cancer.
\newblock \emph{{Journal of Theoretical Biology}}, 231\penalty0 (4):\penalty0
  515--524, 2004.

\bibitem[Sun et~al.(2011)Sun, Garibaldi, and Hodgman]{sun2011parameter}
J.~Sun, J.~M. Garibaldi, and C.~Hodgman.
\newblock Parameter estimation using metaheuristics in systems biology: a
  comprehensive review.
\newblock \emph{{IEEE/ACM transactions on computational biology and
  bioinformatics}}, 9\penalty0 (1):\penalty0 185--202, 2011.

\bibitem[Sunn{\aa}ker et~al.(2013)Sunn{\aa}ker, Busetto, Numminen, Corander,
  Foll, and Dessimoz]{sunnaaker2013approximate}
M.~Sunn{\aa}ker, A.~G. Busetto, E.~Numminen, J.~Corander, M.~Foll, and
  C.~Dessimoz.
\newblock Approximate {B}ayesian computation.
\newblock \emph{{PLoS Computational Biology}}, 9\penalty0 (1):\penalty0
  e1002803, 2013.

\bibitem[Tejero-Cantero et~al.(2020)Tejero-Cantero, Boelts, Deistler,
  Lueckmann, Durkan, Gon{\c{c}}alves, Greenberg, and Macke]{tejero2020sbi}
A.~Tejero-Cantero, J.~Boelts, M.~Deistler, J.-M. Lueckmann, C.~Durkan, P.~J.
  Gon{\c{c}}alves, D.~S. Greenberg, and J.~H. Macke.
\newblock {SBI--A toolkit for simulation-based inference}.
\newblock \emph{arXiv preprint arXiv:2007.09114}, 2020.

\bibitem[Toni et~al.(2009)Toni, Welch, Strelkowa, Ipsen, and
  Stumpf]{toni2009approximate}
T.~Toni, D.~Welch, N.~Strelkowa, A.~Ipsen, and M.~P. Stumpf.
\newblock Approximate {B}ayesian computation scheme for parameter inference and
  model selection in dynamical systems.
\newblock \emph{{Journal of the Royal Society Interface}}, 6\penalty0
  (31):\penalty0 187--202, 2009.

\bibitem[Tsimring(2014)]{tsimring2014noise}
L.~S. Tsimring.
\newblock Noise in biology.
\newblock \emph{{Reports on Progress in Physics}}, 77\penalty0 (2):\penalty0
  026601, 2014.

\bibitem[Valderrama-Baham{\'o}ndez and Fr{\"o}hlich(2019)]{valderrama2019mcmc}
G.~I. Valderrama-Baham{\'o}ndez and H.~Fr{\"o}hlich.
\newblock {MCMC techniques for parameter estimation of ODE based models in
  systems biology}.
\newblock \emph{{Frontiers in Applied Mathematics and Statistics}}, 5:\penalty0
  55, 2019.

\bibitem[Vanlier et~al.(2013)Vanlier, Tiemann, Hilbers, and
  Van~Riel]{vanlier2013parameter}
J.~Vanlier, C.~Tiemann, P.~Hilbers, and N.~Van~Riel.
\newblock Parameter uncertainty in biochemical models described by ordinary
  differential equations.
\newblock \emph{Mathematical {B}iosciences}, 246\penalty0 (2):\penalty0
  305--314, 2013.

\bibitem[Vo et~al.(2015)Vo, Drovandi, Pettitt, and Simpson]{vo2015quantifying}
B.~N. Vo, C.~C. Drovandi, A.~N. Pettitt, and M.~J. Simpson.
\newblock Quantifying uncertainty in parameter estimates for stochastic models
  of collective cell spreading using approximate {B}ayesian computation.
\newblock \emph{Mathematical {B}iosciences}, 263:\penalty0 133--142, 2015.

\bibitem[Wade(2019)]{wade2019fabrication}
S.~J. Wade.
\newblock Fabrication and preclinical assessment of drug eluting wet spun
  fibres for pancreatic cancer treatment.
\newblock 2019.

\bibitem[Wade et~al.(2020)Wade, Sahin, Piper, Talebian, Aghmesheh, Foroughi,
  Wallace, Moulton, and Vine]{wade2020dual}
S.~J. Wade, Z.~Sahin, A.-K. Piper, S.~Talebian, M.~Aghmesheh, J.~Foroughi,
  G.~G. Wallace, S.~E. Moulton, and K.~L. Vine.
\newblock Dual delivery of gemcitabine and paclitaxel by wet-spun coaxial
  fibers induces pancreatic ductal adenocarcinoma cell death, reduces tumor
  volume, and sensitizes cells to radiation.
\newblock \emph{{Advanced Healthcare Materials}}, 9\penalty0 (21):\penalty0
  2001115, 2020.

\bibitem[Walpole et~al.(2013)Walpole, Papin, and Peirce]{walpole2013multiscale}
J.~Walpole, J.~A. Papin, and S.~M. Peirce.
\newblock Multiscale computational models of complex biological systems.
\newblock \emph{Annual {R}eview of {B}iomedical {E}ngineering}, 15:\penalty0
  137--154, 2013.

\bibitem[Wang et~al.(2024{\natexlab{a}})Wang, Jenner, Salomone, Warne, and
  Drovandi]{wang2024calibration}
X.~Wang, A.~L. Jenner, R.~Salomone, D.~J. Warne, and C.~Drovandi.
\newblock Calibration of agent based models for monophasic and biphasic tumour
  growth using approximate {B}ayesian computation.
\newblock \emph{{Journal of Mathematical Biology}}, 88\penalty0 (3):\penalty0
  28, 2024{\natexlab{a}}.

\bibitem[Wang et~al.(2024{\natexlab{b}})Wang, Kelly, Warne, and
  Drovandi]{wang2024preconditioned}
X.~Wang, R.~P. Kelly, D.~J. Warne, and C.~Drovandi.
\newblock Preconditioned neural posterior estimation for likelihood-free
  inference.
\newblock \emph{{Transactions on Machine Learning Research}}, 2024\penalty0
  (September), 2024{\natexlab{b}}.

\bibitem[Wang and Blei(2019)]{wang2019variational}
Y.~Wang and D.~Blei.
\newblock Variational {B}ayes under model misspecification.
\newblock \emph{{Advances in Neural Information Processing Systems}}, 32, 2019.

\bibitem[Wang and Ro{\v{c}}kov{\'a}(2022)]{wang2022adversarial}
Y.~Wang and V.~Ro{\v{c}}kov{\'a}.
\newblock Adversarial {B}ayesian simulation.
\newblock \emph{arXiv preprint arXiv:2208.12113}, 2022.

\bibitem[Ward et~al.(2022)Ward, Cannon, Beaumont, Fasiolo, and
  Schmon]{ward2022robust}
D.~Ward, P.~Cannon, M.~Beaumont, M.~Fasiolo, and S.~Schmon.
\newblock Robust neural posterior estimation and statistical model criticism.
\newblock \emph{{Advances in Neural Information Processing Systems}},
  35:\penalty0 33845--33859, 2022.

\bibitem[Warne et~al.(2019{\natexlab{a}})Warne, Baker, and
  Simpson]{warne2019simulation}
D.~J. Warne, R.~E. Baker, and M.~J. Simpson.
\newblock Simulation and inference algorithms for stochastic biochemical
  reaction networks: from basic concepts to state-of-the-art.
\newblock \emph{{Journal of the Royal Society Interface}}, 16\penalty0
  (151):\penalty0 20180943, 2019{\natexlab{a}}.

\bibitem[Warne et~al.(2019{\natexlab{b}})Warne, Baker, and
  Simpson]{warne2019using}
D.~J. Warne, R.~E. Baker, and M.~J. Simpson.
\newblock Using experimental data and information criteria to guide model
  selection for reaction--diffusion problems in mathematical biology.
\newblock \emph{{Bulletin of Mathematical Biology}}, 81\penalty0 (6):\penalty0
  1760--1804, 2019{\natexlab{b}}.

\bibitem[Warne et~al.(2022)Warne, Baker, and Simpson]{warne2022rapid}
D.~J. Warne, R.~E. Baker, and M.~J. Simpson.
\newblock Rapid {B}ayesian inference for expensive stochastic models.
\newblock \emph{{Journal of Computational and Graphical Statistics}},
  31\penalty0 (2):\penalty0 512--528, 2022.

\bibitem[Wood(2010)]{wood2010statistical}
S.~N. Wood.
\newblock Statistical inference for noisy nonlinear ecological dynamic systems.
\newblock \emph{Nature}, 466\penalty0 (7310):\penalty0 1102--1104, 2010.

\bibitem[Zucknick and Richardson(2014)]{zucknick2014mcmc}
M.~Zucknick and S.~Richardson.
\newblock {MCMC} algorithms for {B}ayesian variable selection in the logistic
  regression model for large-scale genomic applications.
\newblock \emph{arXiv preprint arXiv:1402.2713}, 2014.

\end{thebibliography}

\end{document}